\pdfoutput=1
\documentclass[12pt,a4paper]{article}

\usepackage[cp1250]{inputenc}
\usepackage[round]{natbib} 
\usepackage{gensymb}
\usepackage{amsfonts,amsmath} 
\usepackage{amsthm}
\usepackage{enumerate} 
\usepackage{graphicx} 
\usepackage{authblk} 
\usepackage{epstopdf} 
\usepackage{floatrow}
\usepackage{todonotes}
\usepackage{comment}
\usepackage{mathtools}
\usepackage{cancel}

\usepackage{bbm}
\usepackage[hyperfootnotes=false]{hyperref}

\usepackage[normalem]{ulem} 
\usepackage{adjustbox}

\usepackage{tikz}
\usetikzlibrary{shapes, arrows, positioning,decorations.markings, external}

\usepackage[tmargin=1in,bmargin=1in,lmargin=0.75in,rmargin=0.75in]{geometry}

\usepackage[font={small}]{caption}

\usepackage{textcomp}
\usepackage{ulem}
\usepackage{dsfont}
\usepackage{amssymb}

\newcommand{\R}{\mathbb{R}}

\newcommand{\zin}{\mathrm{v}}
\newcommand{\nzin}{\mathrm{nv}}
\newcommand{\czin}{\mathrm{c}}

\usepackage[]{algorithm2e}

\usepackage{bm}
\usepackage{siunitx}
\usepackage{subcaption}
\usepackage{array}
\usepackage{multirow}

\newcommand\MyBox[2]{
  \fbox{\lower0.75cm
    \vbox to 1.7cm{\vfil
      \hbox to 1.7cm{\hfil\parbox{1.4cm}{#1\\#2}\hfil}
      \vfil}%
  }%
}

\newcommand{\voxel}{x}

\newcommand{\predictionModel}{g}

\newcommand{\phaseSegmentation}{L^\mathrm{SEM}}

\newcommand{\composition}{VFVM}

\def\submission{0}
\if\submission1
\usepackage[doublespacing]{setspace}
\usepackage{lineno}
\linenumbers
\fi

{\end{figure}}
\newcommand\blfootnote[1]{%
  \begingroup
  \renewcommand\thefootnote{}\footnote{#1}%
  \addtocounter{footnote}{-1}%
  \endgroup
}

\newcommand{\window}{W}

\title{\bf Multidimensional characterization of particle morphology and mineralogical composition using CT data and R-vine copulas}

\author{Orkun Furat$^{a,1,\ast}$, Tom Kirstein$^{a,1}$, Thomas Lei\ss{}ner$^b$, Kai Bachmann$^c$, Jens~Gutzmer$^c$, Urs A. Peuker$^{b}$, Volker~Schmidt$^{a}$}
\affil{\small \it $^a$Institute of Stochastics, Ulm University, D-89069 Ulm, Germany \\ 
\small \it $^b$Institute of Mechanical Process Engineering and Mineral Processing, Technische Universit\"{a}t Bergakademie Freiberg, D-09599 Freiberg, Germany \\ 
\small \it $^c$Helmholtz Institute Freiberg for Resource Technology, Helmholtz Zentrum Dresden-Rossendorf, D-01328 Dresden, Germany}
\date{}

\begin{document}
\maketitle
\sloppy 

\blfootnote{{$^1$OF and TK contributed equally.}}
\blfootnote{$^\ast$Corresponding author. E-mail address: \texttt{orkun.furat@uni-ulm.de},
 Phone: \texttt{+49731/50 23526},
\textit{Fax:} \texttt{+49731/50 23649} }

\begin{abstract}
\noindent 
 Computed tomography (CT) can capture volumes large enough to measure a statistically meaningful number of micron-sized particles with a sufficiently good resolution to allow for the analysis of individual particles. However, the development of methods to efficiently investigate such image data and interpretably model the observed particle features is still an active field of research. When image data of particles exhibiting a wide range of shapes and sizes is considered, traditional image segmentation methods, such as the classic watershed algorithm, struggle to recognize particles with satisfying accuracy. Thus, more advanced methods of machine learning must be utilized for image segmentation to improve the validity of subsequent analyzes. Moreover, CT data does not include information about the mineralogical composition of particles and, therefore, additional SEM-EDS image data has to be acquired. 
In this paper, micro-CT image data of a particle system mostly consisting of zinnwaldite-quartz composites is considered.
First, an image segmentation method is applied which uses  deep convolutional neural networks, in particular an adaptation of the U-net architecture. This has the advantage of requiring less hand-labeling than other machine learning methods, while also being more flexible with
the possibility of transfer learning. Then,   fully parameterized models based on vine copulas are designed to determine multivariate probability distributions of descriptor vectors for the  size, shape, texture and composition of particles---allowing for the estimation and interpretable characterization of interdependencies between particle descriptors. For model  fitting, the segmented three-dimensional CT data and co-registered two-dimensional SEM-EDS data are used. The models are applied to predict the mineralogical composition of particles, solely on the basis of particle descriptors observed in CT data.


\end{abstract}

\section{Introduction}
The nature of individual particles within materials influences the quality and
behavior of application-specific particulate materials, such as crushed mineral ores
considered in the mining industry and those materials used for the production of
coatings, membranes and electrodes. For example, descriptors of particle size, flatness and sphericity as well as their mineralogical composition play an important role in this context. 
Moreover, the behavior of  particle systems during processing depends on the descriptors of individual particles mentioned above \citep{Tripathy2017, Zheng2017}.
Therefore, the quantitative characterization of particle systems by means of such descriptors is of great interest in order to study their influence on macroscopic physical properties of particulate materials and, in particular, on the behavior of the underlying particle systems during processing.   

Typically, particle systems are characterized by univariate probability distributions of single particle descriptors, e.g., the particle size distribution or the distribution of shape descriptors \citep{ditscherlein_furat_2020,furat2018,SYGUSCH2021150725}. To do so, imaging techniques like, for example, computed tomography (CT) can be deployed to obtain image data of the  particle system under consideration, followed by image processing to obtain a particle-wise segmentation from which descriptors for individual particles can be computed \citep{DITSCHERLEIN2020989,Guil2}. 
However, in many cases,  the segmentation of image data is a non-trivial task, requiring a careful calibration of image processing algorithms, see e.g. \cite{ditscherlein_furat_2020}. In \cite{furat2018} the particle-wise segmentation of CT image data and its characterization by means of  probability distributions of particle descriptors was achieved by combining conventional segmentation algorithms with methods from machine learning, which have been trained using manually labeled 3D data.

Since particle descriptors are in general correlated and univariate probability distributions of single particle descriptors are unable to capture such correlations, see \cite{furat2019},  multivariate probability distributions have to be used for a more holistic characterization of particle systems. On the other hand, it is clear that this leads to additional complexity for the characterization task. In   \cite{ditscherlein_furat_2020} and \cite{furat2019}  so-called Archimedean copulas \citep{Czado2019,Nel06} have been used for modeling the joint multivariate distribution of 2- and 6-dimensional descriptor vectors, respectively. Yet, for modeling the distribution of vectors with dimension larger than two, so-called vine copulas have shown to be a more flexible tool, see e.g. \cite{aigner2022}.

For some particle systems, additional difficulties in  the course of their characterization arise when the particles are composed of various materials, or when they are even composites consisting of different material components. Then, 
CT image data often does not provide sufficient information on the composition of particles, see \cite{furat2018}. On the other hand, combining different imaging techniques, like the combination of scanning electron microscopy (SEM) and energy-dispersive X-ray spectroscopy (EDS), allows for a mineralogical characterization of planar 2D sections of particle systems. Using both 3D CT and co-registered 2D SEM-EDS image data of a particle system, a characterization with respect to the 3D particle morphology, texture as well as the mineralogical composition can be achieved, see \cite{furat2018,Reyes2017}.

The general goal  of   the  present paper  is the development of a workflow for the multivariate characterization of particle systems consisting of (i)  the segmentation of particulate CT data based on machine learning techniques, such as convolutional neural networks (CNN), which require less manually labeled data than the method described in \cite{furat2018}, see Fig.~\ref{fig:modeling-scheme} (first row; left and center); (ii) the combination of segmented CT image data and SEM-EDS data to obtain 
multidimensional vectors of particle descriptors regarding size, shape, texture and composition, see Fig.~\ref{fig:modeling-scheme} (first row; center and right); (iii) modeling of multivariate probability distributions of descriptor vectors using vine copulas which allows for better fits than the Archimedean copulas considered in \cite{ditscherlein_furat_2020} and \cite{furat2019}, see Fig.~\ref{fig:modeling-scheme} (second row; right); (iv) the calibration of prediction model which allows for the estimation of the mineralogical composition of particles imaged by CT without requiring additional SEM-EDS data, see Fig.~\ref{fig:modeling-scheme} (second row; left). 
To illustrate this workflow, crushed ore particles consisting of various mineral components, mainly quartz, topaz, zinnwaldite and muscovite, are characterized on the basis of micro-CT data using multivariate probability distributions of their descriptor
vectors. Since such particle systems typically undergo separation processes in  applications within the mining industry, we restrict the mineralogical characterization of particles to the assessment of 
their volume fractions of valuable (in this case zinnwaldite) and non-valuable materials.
Nevertheless, the methods of the present paper can be applied for further particle systems with different characterization tasks, as well. In particular, in Section~\ref{sec.dis} we discuss how a modified version of the prediction model could be used if the volume fractions of more than two groups of minerals are of interest, in contrast to the assessment of just the volume fractions of valuable and non-valuable minerals.
Moreover, the  methodology described above does not depend  on the range of particle sizes and  image resolution, given that an adequate amount of particles is imaged in sufficient detail.

\if\submission0
\begin{figure}[ht]
\centering

	\input{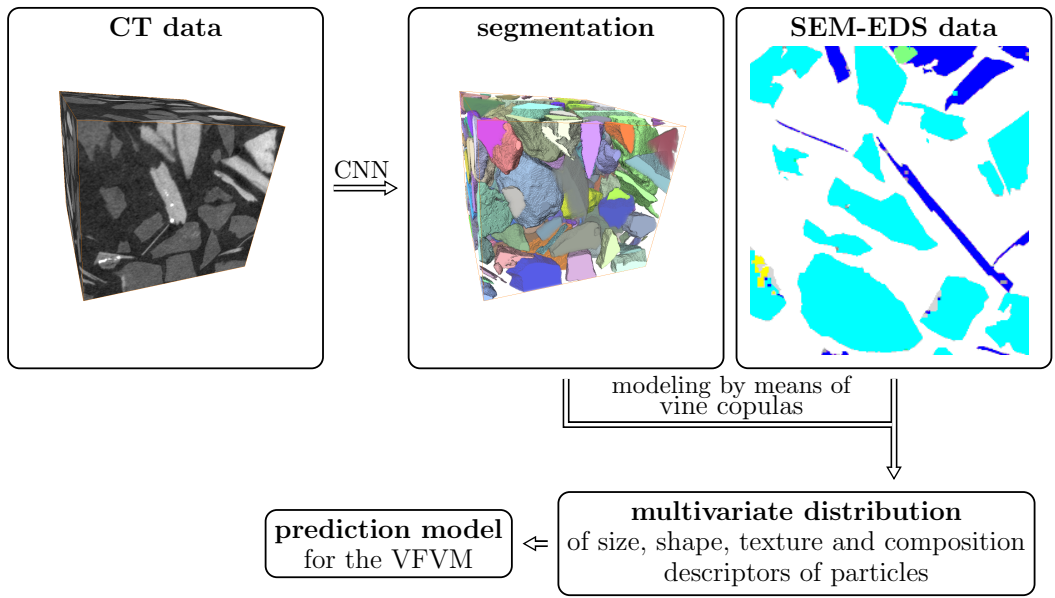}

\caption{Workflow for  multivariate stochastic modeling of particle descriptor vectors and  estimation of the particle-wise volume fraction of valuable materials (\composition{}) from CT data.} 
\label{fig:modeling-scheme}
\end{figure}

\fi

More precisely, the workflow for the characterization of particle systems, which is developed in the present paper, can be decomposed into the following steps. 
First,  the measured CT image data  undergoes a  pre-processing step based on machine learning,  where  suitable  neural networks
are considered. More precisely, we use a modified version of the network architectures described in \cite{3Dunet} and \cite{Furat2021}, which is a fully convolutional neural network, see~\cite{lecun2015deep}.
After the pre-processing step performed by the trained network, several binarization and segmentation steps are applied to the CT  data in order to extract individual particles   by means of a marker-based watershed algorithm, see e.g.~\cite{roerdink2001,spettl2015}.
Then, for each segmented particle,  the corresponding vector of size, shape and texture descriptors is computed.  Moreover, using  2D SEM-EDS data,  the mineralogical composition of those particles hitting some planar sections within the 3D sampling window of CT data is  correlated with corresponding particle descriptors computed from CT data. In this way, for such particles vectors of size, shape, texture and composition descriptors can be determined.
These descriptor vectors are then stochastically modeled by
 multivariate probability distributions, which are determined by means of an R-vine copula approach, see e.g. \cite{Joe15,kurowicka2010}.
These multivariate probability distributions can be used as a prediction model to estimate the mineralogical composition of particles, solely based on particle descriptors computed from CT data, assuming that the geometrical/textural descriptors and the mineralogical composition are correlated. In particular, using such multivariate probability distributions,  we extend the prediction model presented in \cite{furat2018} such that we can now estimate the mineralogical composition of particles quantitatively (i.e., the \composition{}) from CT data (without requiring additional SEM-EDS data) instead of just determining their predominant mineral component.

Besides the calibration of the prediction model for estimating the particle-wise \composition{} from CT data, multivariate probability distributions of particle descriptors have further useful applications. For example, as mentioned above, in the mining industry similar ore particle systems like that one considered in the present paper undergo various separation processes, e.g., magnetic separation \citep{LEISSNER2016}, for the extraction  of  zinnwaldite-rich  particles  from a crushed greisen-type ore. Typically, in the literature the quality of such separation processes is characterized by comparing univariate probability distributions of single particle descriptors before and after separation, using so-called partition curves \citep{LEISSNER2016}. Recently, in  \cite{SCHACH201978} a characterization of  separation processes has been proposed by considering bivariate probability distributions of two-dimensional descriptor vectors.
The workflow developed in the present paper, which can reliably model the multivariate distribution of  higher-dimensional vectors consisting of more than just two particle descriptors,  can be used as a basis for an even more holistic analysis of the quality of  separation processes. This will be discussed in detail in a forthcoming paper.

\section{Materials and methods}
\label{sec.sec.met}

\subsection{Description of materials and data acquisition}
\label{sec.des.mat}

The sample material used in this study is a crushed lithium ore, which consists of four main silicate components. These are quartz, topaz and the phyllosilicates zinnwaldite and muscovite (also grouped as mica), where quartz and zinnwaldite make up more than 90\% of the volume fraction of particles intersecting with a SEM-EDS slice. Mica is slightly paramagnetic and can be separated from quartz and topaz by magnetic separation, provided that there occurs no significant intergrowth of components within individual particles. In the case of composite particles, their particle magnetizability determines whether they are enriched in a magnetic or non-magnetic product, see \cite{LEISSNER2016}. 
Since the extraction of target particles with a defined composition is of great interest in the mining industry, the goal of the present paper is the  3D characterization of particles with respect to their size, shape, texture and  \composition{}.  Such particle data will be unbiased compared to data coming exclusively from 2D characterization methods even for anisotropic data, see \cite{Guil1}. 
For methods which deal with the characterization of 3D structures based on 2D image data, see~\citep{chiu2013,kench2021}.

The sample material  was prepared in such a way that SEM-EDS analyzes and micro-CT imaging could be performed on the same specimen, where
the particles were mixed with micron-sized graphite for better dispersion and embedded in epoxy blocks with a diameter of \SI{20}{\mm}. 
To avoid the influence of  segregation effects on the results of the 2D SEM-EDS analyzes, the grain mount was cut in the direction of sedimentation, rotated by \SI{90}{\degree} and re-embedded, see~\cite{Heinig2015}. 

The sample was analyzed by a Mineral Liberation Analyzer (MLA), 
a FEI Quanta 650F field emission scanning electron microscope (SEM) equipped with two Bruker Quantax X-Flash 5030 energy-dispersive X-ray detectors (EDX). Back-scattered electrons (BSE) are used for image segmentation and EDX-spectra for mineral classification. The GXMAP (grain-based X-ray mapping) measurement mode was selected for analyzing the sample. More detailed information about the functionality of MLA and offline processing of the data can be found in  \cite{Bachmann2017} and  \cite{Fandrich2007}. Measurement conditions were \SI{20}{\kilo\volt}, $500 \times 500$ px frame size and \SI{1000}{\micro\metre} horizontal field width (2 microns per pixel). EDX analyzes were performed every 6 px with an exposure time of \SI{7}{\milli\second}. BSE calibration was set on Au=252. Particle data processing was done with the software package MLA Dataview 3.1.4.686.

Micro-CT imaging was performed using a Zeiss Xradia 510 Versa X-ray microscope. Scanning was done using \SI{80}{\keV} acceleration voltage and \SI{7}{\watt} tube power. The polychromatic beam was filtered with the systems LE6 filter. The pixel size of the projections is \SI{12}{\micro\metre} (0.4X objective, source-sample distance: \SI{75}{\mm}, detector-sample distance: \SI{140}{\mm}, camera binning: 2). In this way, 1601 projections were acquired over \SI{360}{\degree} using \SI{25}{\second} exposure time to achieve an optimum signal-to-noise ratio.
Reconstruction was done using the Zeiss software XMReconstructor (version 11.1.8043) with the following parameters: automatic center shift correction, Gaussian smoothing (0.5), beam hardening correction (0.1), and no byte scaling. 

Note that the  methods described in the following sections, both for image segmentation as well as for  stochastic model-based characterization of particles, can also be applied  to image data resulting from other 3D imaging procedures, such as  nano-CT.

\subsection{Processing of CT image data via machine learning}\label{Sec:Seg}
In order to describe  the  particle system considered in this paper using multivariate probability densities of their size, shape and texture descriptors, we have to identify individual particles in the CT image. Therefore, an image segmentation procedure has to be deployed which partitions the CT image data into a background region and regions which correspond to individual particles, see Fig.~\ref{fig:modeling-scheme} (first row; center). However, the direct application of conventional methods like the watershed algorithm fails for the given  CT image data  since particles observed in the data often exhibit elongated, platelike shapes. More precisely, the watershed algorithm tends to split elongated particles into multiple regions, an issue which is referred to as oversegmentation, see e.g.
  \cite{soille2003}. 

Since the accuracy of the segmentation results strongly impacts the results of subsequent analyzes, i.e., the computation of descriptor vectors and their stochastic modeling in Sections~\ref{sec:stochastic-modeling-of-multidimensional-vectors-of-particle-characteristics} and \ref{Sec:Vine}, we utilize methods of machine learning to obtain an improved segmentation of the CT image data. 
More precisely, we consider a modified version of the network architectures described in \cite{3Dunet}, i.e.,
we use a CNN, namely the 3D U-net architecture, to achieve a reasonably good segmentation of the  3D image data.  The 3D~U-net is reported to only require annotated 2D cutouts for training, rather than annotated 3D voxel volumes, and thus reducing the amount of manually labeled data, see~\cite{3Dunet}. After training, the optimized network can be applied on the entire CT image data to produce a 3D segmentation of the entire data set.

  From here on, we assume that the CT image data can be described as a map  $I:\window\rightarrow\mathbb{R},$ where the voxel space $\window  \subset \mathbb{Z}^3$ is a three-dimensional interval and $I(\voxel)\in\mathbb{R}$ is the corresponding grayscale value of the voxel $\voxel=(\voxel_1,\voxel_2,\voxel_3)\in\window$.
 
\subsubsection{Description of the  network architecture}

The U-net architecture was first introduced in \cite{unet} with the goal of achieving accurate pixel-wise classification of 2D image data. Since then, in \cite{3Dunet}, the approach has been extended to allow for the segmentation of 3D image data. This network architecture, called 3D U-net, is a fully convolutional neural network with an encoder-decoder architecture. It has a downsampling path, followed by an upsampling path which are both comprising four levels. The levels on the downsampling path consist of two $3\times3\times3$~convolutional layers, each followed by a rectified linear unit (ReLu) activation layer, and a $2\times2\times2$ max-pooling layer with strides of two in each dimension. The levels on the upsampling path consist of a $2\times2\times2$~upsampling layer followed by two $3\times3\times3$ convolutional layers, each followed by an ReLu activation layer. Additionally, the number of feature channels is doubled in the first convolutional layer of each downsampling level and halved in the first convolutional layer of each upsampling level. The architecture also features residual connections between downsampling and upsampling levels as shown in Fig.~\ref{Fig:u-net}.  

We make the  following two adjustments with respect to the original architecture described in \cite{3Dunet}. 
First, the last two downsampling layers are extended by two dilated $3\times3\times3$ convolutional layers with dilation factors of 2 in each dimension, see \cite{yu2015multi}.
 The dilated convolutional layers allow the network to consider a larger region around each voxel. A similar effect could be achieved by increasing the kernel sizes in the convolutional layers which, contrary to the usage  of dilated convolutional layers, leads to an increase in trainable parameters and, thus, a decrease in computational feasibility.
 However, the  additional dilated convolutional layers still add depth to the network which can cause the problem of vanishing gradients during training. Second, in order to overcome this issue, we modified the network architecture described in~\cite{3Dunet} by adding further residual connections.  Specifically, in each  downsampling level, the input to the first convolutional layer is also  added to the output of the second convolutional layer. Similarly, the input to the first dilated convolutional layer in the last two downsampling levels is added to the output of the second dilated convolutional layer. In \cite{resnet_general} it was shown that such residual connections can improve learning in deep convolutional neural networks. The entire architecture is depicted in Fig.~\ref{Fig:u-net}.  
\if\submission0
\begin{figure}[ht]
\centering

\includegraphics[width=0.95\textwidth]{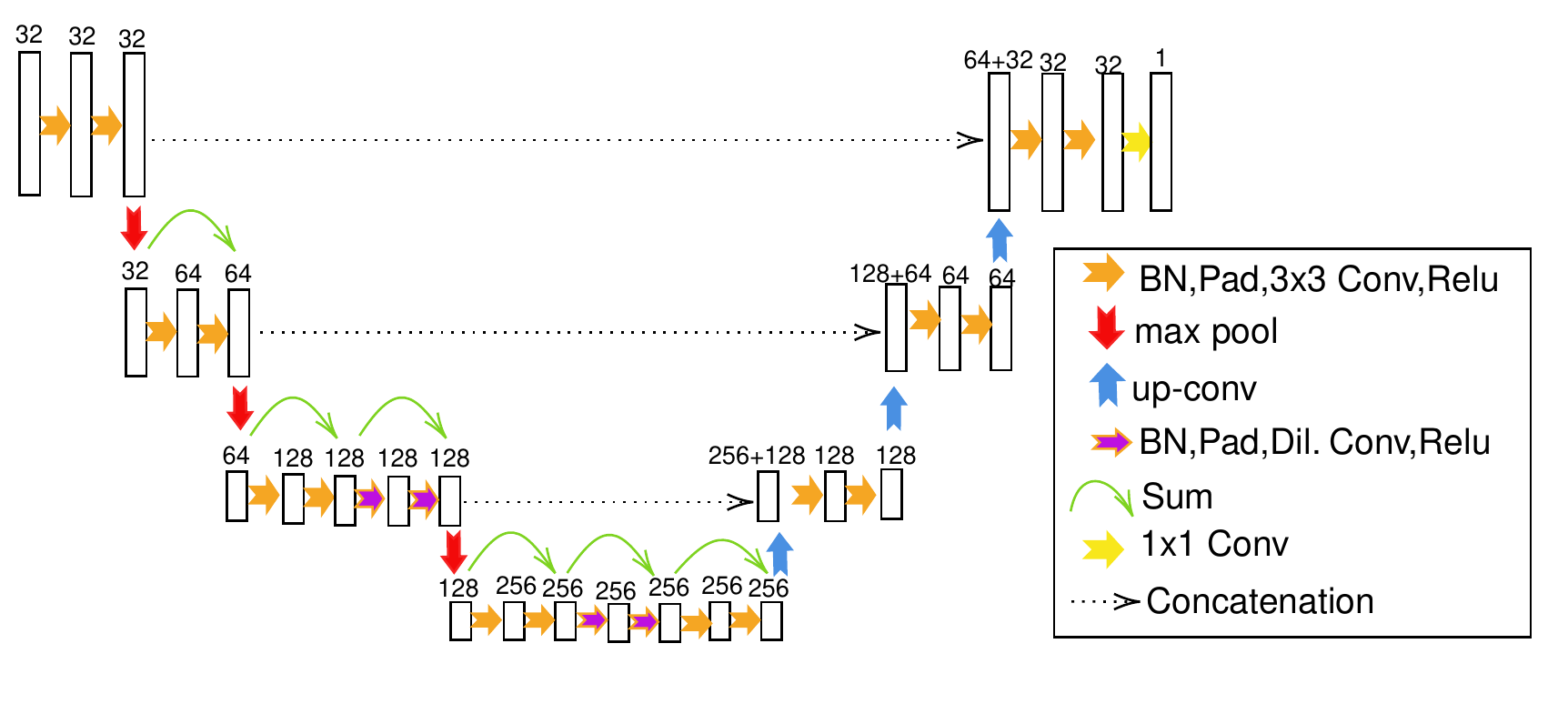}

\caption{U-net architecture used for the initial segmentation task. Each $3\times3\times3$ convolutional layer ($3\times3$ Conv) and each dilated convolutional layer (Dil. Conv)  is preceded by a batch normalization layer (BN) and a symmetric padding layer (Pad) of suitable width. }
\label{Fig:u-net}      
\end{figure}
\fi

\subsubsection{Building the ground truth: Annotation of 2D voxel slices}
\label{sec.gro.tru}
 As stated in \cite{3Dunet}, an advantage of using a 3D U-net is that only annotated 2D slices are required for training.  In this context, a 2D slice  is a subset of  voxels $\window_z\subset \window$ given by $\window_z=\{(\voxel_1,\voxel_2,\voxel_3)\in\window: \voxel_3=z\}$ for some integer  $z\in \mathbb{Z}$.  Then, an annotated 2D slice  is a mapping $L^{\mathrm{gt}}:\window_z\rightarrow\{0,1\},$ where
 the connected components of the set $\{\voxel\in\window_z: L^{\mathrm{gt}}(\voxel)=1\}$  correspond to  the voxel positions of the individual 3D particles intersected with the slice  $\window_z.$ 
 Meaning that  an annotated slice is a segmentation map, in this context referred to as ground truth, of a planar 2D section of 3D image data. 
 To avoid oversegmentation of non-convex particles, which can appear as disconnected sets in a given 2D slice, visual 3D information is used for the annotation of the ground truth labels.

 Once  suitable 2D slices $\window_{z_1},\ldots,\window_{z_n}$
 are chosen for some integers $z_1,\ldots,z_n\in\mathbb{Z}$, the  labels $L^{\mathrm{gt}}(\voxel)$ are determined for all $\voxel\in\window^\prime=\window_{z_1}\cup\ldots\cup\window_{z_n}$ by thresholding the image data first, followed by manual correction of particle boundaries to separate neighboring particles in the ground truth image data. Additionally,  thin         particles are manually enlarged to avoid over-segmentation. A cutout of the resulting ground truth labels for the slice $\window_z$ with slice number $z=387$ is shown in Fig.~\ref{Fig:2D_gt}. The amount of required hand-labeling can vary depending on the nature of the data. For reference, we annotated five $800\times800$ voxel-sized cutouts of the entire  $4086 \times 4086 \times 1498$ voxel-sized image data.  The annotated slices should be representative, meaning that they should include common features such as interfacing particles and imaging artifacts, like those shown in Fig.~~\ref{Fig:2D_gt}. 
 
 The goal of this section is to explain how this 2D ground truth can be used to create a 3D training data set for the 3D U-net.  Specifically, the remaining voxels  $\voxel\in \window\setminus\window^\prime$ of other slices are not annotated, but  placeholder labels for those voxels are still required  for the training process described in Section~\ref{sec:training}. Therefore, for a given ground truth labeling $L^{\mathrm{gt}}$ on $\window^\prime$, we  introduce a 3D voxelwise labeling $L:\window\rightarrow\{0,1\},$ given by 

 \begin{equation*}
        L(\voxel) = \begin{cases}
                         L^{\mathrm{gt}}(\voxel),\;\qquad \text{ if } \voxel \in \window^\prime, \\
                         0,\;\;\;\;\;\;\;\;\;\qquad \text{otherwise. } 
                    \end{cases}
\end{equation*}

\if\submission0
  \begin{figure}[ht]
   
  \includegraphics[width=1\linewidth]{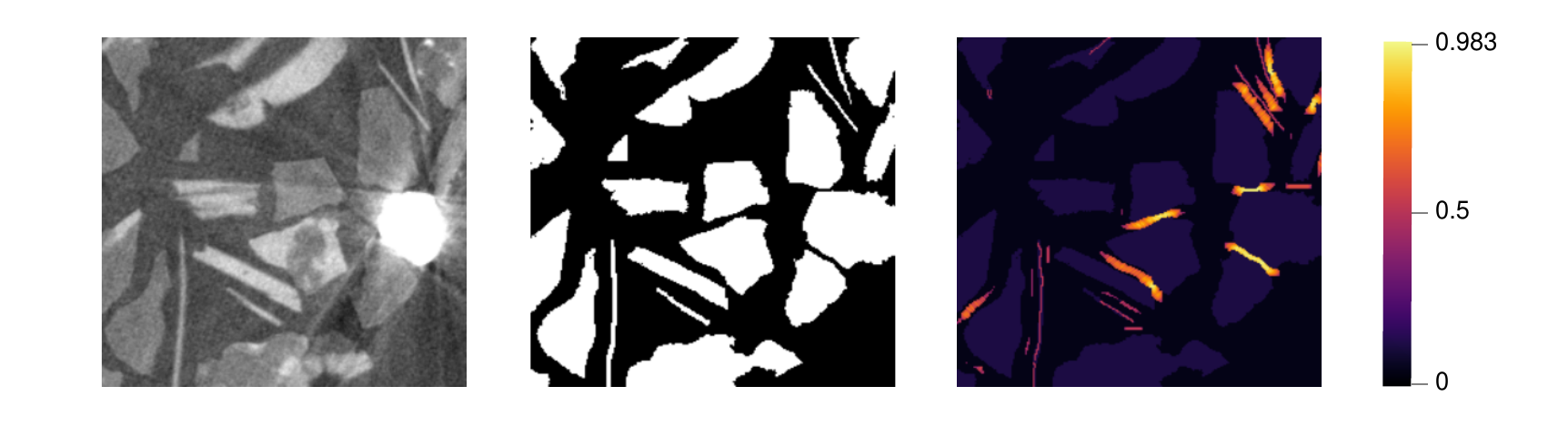}

  \caption{ (a) 2D cutout of raw CT data showing  a highly reflective particle, (b)  the corresponding ground truth labels  of particles (white) and background (black), and (c) the corresponding weight map, where brighter areas indicate higher weights.
Note that the weights of voxels corresponding to platelike particles are increased to avoid oversegmention of such particles.
 \label{Fig:2D_gt}
}
 \end{figure}
 
\fi

During network training, described below in  Section~\ref{sec:training}, voxels for which the ground truth is not known, i.e., for voxels $\voxel \in \window\setminus\window^\prime$, the eventually occurring discrepancy between the voxelwise labeling $L(\voxel)$ and the network's output $\widehat{L}(\voxel)$ should  be ignored in the loss function. This is achieved by means of a so-called voxel weight function. The weight function determines to which degree the network output at each voxel position contributes to the loss function during training. By setting the weight of a voxel equal to zero, the classification of that voxel does not impact the training loss. Vice versa, the voxels within regions between interfacing particles are most important for the quality of the obtained segmentation, and, thus, their weights should be increased. In general, any function $w: W\rightarrow [0,\infty)$ is a valid voxel weight function. However, in the present paper, we consider the following weight function  $w: W\rightarrow [0,\infty)$ 
with

\begin{equation}\label{Eq:weight}
    w(\voxel) =
        \left\{
            \begin{array}{ll}
                0.04+\exp\!\left(-\frac{\left(d_1^2(\voxel)+d_2^2(\voxel)\right)}{36}\right), &\qquad \text{ if } \voxel \in \window^\prime \text{ and } L(\voxel)=0, \\
                c_\mathrm{f}, &\qquad \text{ if } \voxel \in \window^\prime\text{ and } L(\voxel)=1,\\
                0, & \qquad\text{ otherwise,}
            \end{array}
        \right.
\end{equation}
where the constant $c_\mathrm{f} \ge 0$ is chosen such that  $\smashoperator[r]{\sum\limits_{\voxel\in \window^\prime \colon L(\voxel)=1}} w(\voxel)\quad\,= \smashoperator[r]{\sum\limits_{\voxel \in \window^\prime \colon L(\voxel)=0}} w(\voxel),$ and  $d_1(\voxel)$ and $d_2(\voxel)$ denote the distance of voxel $\voxel\in W^\prime$ to the closest and second closest particle  within the labeled slice to which  $\voxel$ belongs, respectively.  The weight function given in Eq.~(\ref{Eq:weight}) is similar to the weight function proposed in~\cite{unet}, with the difference  that we increase the weights for  voxels which are closer to the center of the gap  between two particles. Moreover, to avoid emphasizing gaps that are sufficiently large, we replace the distance functions $d_1$ and $d_2$ considered in Eq.~(\ref{Eq:weight}) by the truncated versions $\widetilde{d}_1:W^\prime\to[0,\infty)$ and $\widetilde{d}_2:W^\prime\to[0,\infty)$, where 
\begin{equation*}
        \widetilde{d}_i(\voxel) = \begin{cases}
                        d_i(\voxel), \qquad\text{ if } d_i(\voxel) \leq \widehat{d}, \\
                        \infty, \qquad \quad \; \text{ otherwise},
                    \end{cases} 
\end{equation*}
for $i=1,2$, for all $\voxel\in W^\prime$ and for some upper bound $\widehat d\ge 0$.  Putting
  $\widehat{d}=5$, the resulting weight map for a cutout of slice $\window_z$ with slice number $z=337$ is shown in Fig.~\ref{Fig:2D_gt}.

\subsubsection{Network training}\label{sec:training}

A convolutional neural network with an architecture as depicted in Fig.~\ref{Fig:u-net} can be described by the parameters  of its layers. In the following, we explain how the available image data as well as labels and voxel weights are used to train the network, i.e., to estimate optimum network parameters for the computation of segmentation maps from CT data.

 The training can be decomposed into different training steps, in each of which  the network's output $\widehat{L}:W\to[0,1]$ is compared to the voxelwise labeling $L:W\to\{0,1\}$, where the network parameters are updated to reduce the discrepancy between $\widehat L$ and $L$.
 For that purpose, a stochastic gradient descent algorithm is used to minimize the  weighted binary cross-entropy loss  $ H(L,\widehat{L},w)$, which is given by
\begin{equation}\label{Eq:loss}
    H(L,\widehat{L},w)=-
\sum_{\voxel \in \window} w(\voxel)\; \ell(L(\voxel),\widehat{L}(\voxel)),\end{equation} 
where $\ell(a,b)=a\,\text{log}\,b+\left(1-a\right)\text{log}\left(1-b\right)$  for any label $a\in\{0,1\}$ and predicted label $b\in(0,1),$ and $w:W\to[0,\infty)$ is  the weight function given in Eq.~(\ref{Eq:weight}).

Note that the output $\widehat{L}$ depends on the network's trainable parameters. Thus, by utilizing the gradient of the loss function in Eq.~(\ref{Eq:loss}) with respect to the neural network's parameters, the latter ones are updated according to the ADAM algorithm with a stepsize of $\alpha=10^{-4}$, see \cite{adam}. To accelerate learning in the early training steps, the  network parameters are initialized according to the distribution described in~\cite{he_normal}. The result of the   training process described above largely depends on which training images are used during the individual training steps.  Furthermore, the use of so-called data augmentation has been identified as a useful tool for training convolutional neural networks when an abundance of training data is not available, see~\cite{augmentation}. Therefore, the training images in each training step are chosen as follows:
   \begin{itemize}
       \item[(i)] An  $80\times80\times80$ sized cutout box is taken at random   from the  grayscale image $I:W\to\mathbb{R}$.
       \item[(ii)] The grayscale values of voxels belonging to the cutout box are normalized. 
       \item[(iii)] A random elastic transformation is applied on the cutout box.
       \item[(iv)] The cutout box is rotated by a random degree around each axis.
   \end{itemize}
 The same operations, with the exception of step (ii), are  performed on the  weight function $w:W\to[0,\infty)$ and the voxelwise labeling $L:W\to\{0,1\}$ in such a way that they correspond to the augmented training image. The cutout size in step (i) was chosen in accordance with GPU memory limitations.
 Furthermore, the cutout boxes are chosen such that their voxels intersect with at least one of the annotated slices. In step (ii), the grayscale values of the cutout boxes are normalized such that they have a mean value of 0 and a standard deviation of 1. The elastic transformations applied in  step (iii) are described in~\cite{elastic}.

  The training process itself can be decomposed into subsequent periods each of which consists of 2000 individual training steps. At the end of each period, the network is evaluated with the cost function given in Eq.~(\ref{Eq:loss}) on a separate validation set. If the value of the cost function on the validation set (validation score) does not decrease for four subsequent periods, then the network parameters are restored to the parameters of the most performant network. The training process concludes when the validation score does not improve for 40 subsequent periods (or after 200 periods all in all, to ensure that the training process terminates sufficiently fast).

  \subsubsection{Segmentation procedure}\label{sec:segmentation-procedure}
  After training the network is applied on the entire CT image data.
 While the network is trained on $80\times80\times80$ sized cutout boxes, its fully convolutional architecture allows for the application of the network to 3D input data of any size, as long as the size is divisible by 8 for each direction to accommodate the max-pooling layers. However, it turned out that due to memory limitations the application of the trained network on the entire 3D image data was not possible. Therefore, we used an overlap-tile strategy to remedy this issue, see \cite{unet}.  

  Then, the trained network's output is an image of the same size as the CT image data, which takes arbitrary values in the interval $[0,1]$, i.e., the output is not a segmentation of the CT image data yet. Therefore, we first compute an initial segmentation by binarizing  the network's output   with respect to the threshold of 0.5. Then, connected components within this binarization consisting of more than 50 voxels are identified as individual regions of the initial segmentation. However, certain features, such as gaps between interfacing particles, are overemphasized in the initial segmentation. In order to capture the particle shape more accurately, the final segmentation is computed by applying a marker-based watershed algorithm to the binarized CT image data, using the centroids of the connected components of the initial segmentation as markers, see \cite{soille2003}. Each connected (foreground) component
  $P \subset W$ of the final segmentation is then interpreted as a particle,  i.e., the segmented particles are considered to be sets of voxels.

  In the following  we will use the segmented image data in order to investigate various descriptors of the extracted particles  $P \subset W$ 
  describing their size, shape and texture---which will then be correlated with the  \composition{} of the particles determined by considering SEM-EDS data.

\subsection{Computation of  particle descriptors}\label{sec:stochastic-modeling-of-multidimensional-vectors-of-particle-characteristics}

We now explain how  particle descriptors characterizing the    size, shape and texture of particles can be computed, using the particle-wise segmented CT image data which has been obtained with methods of machine learning as stated in Section~\ref{Sec:Seg}. However, also the \composition{} of particles and the modeling thereof is  of great interest. In order to obtain such information which is not provided in the CT data considered in   Section~\ref{Sec:Seg},  phase-wise segmented SEM-EDS data will be utilized, which is available for three slices of the CT image data, see Fig.~\ref{fig:modeling-scheme} (first row; right).

 Later on, in Section~\ref{Sec:Vine}, the computed vectors of particle descriptors will be  stochastically modeled, which leads to  multivariate probability distributions of particle descriptors for the particle system described in Section~\ref{sec.des.mat}, see Fig.~\ref{fig:modeling-scheme} (second row; right).

\subsubsection{Size, shape and texture descriptors}\label{sec:characteristics}

Visual inspection of the available image data  indicates that particles consisting of different minerals differ significantly in size and shape.
In order to quantitatively distinguish irregularly shaped particles from each other, we consider several size and shape descriptors.
In particular,
the size descriptors of a particle $P\subset W$ considered in this paper are its volume $M_{\text{vol}}(P)$, given by the number of voxels associated with $P$,  and its surface area, denoted by $M_{\text{area}}(P)$, which is computed using an algorithm described in~\cite{schladitz2006}.

Regarding the description of particle shape, we  compute two so-called aspect ratios. More precisely, for a particle $P$ the elongation $M_{\text{elo}}(P)$ and the flatness $M_{\text{flat}}(P)$ are given by
\begin{equation}\label{flatness}
    M_{\text{elo}}(P)=\frac{a_2(P)}{a_1(P)}\quad\quad \text{and} \quad\quad  M_{\text{flat}}(P)=\frac{a_3(P)}{a_2(P)},
\end{equation}  
where $a_1(P),$ $a_2(P)$ and $a_3(P)$   denote the length of the longest, second longest and third longest axis of the (arbitrarily oriented) minimum-volume bounding box of $P,$ respectively, see~\cite{boundingbox}. Note that due to the definitions given in Eq.~(\ref{flatness}), a platelike particle would feature a smaller value of $M_{\text{flat}}$, whereas a more spherical particle would feature a larger value of $M_{\text{flat}}$. Furthermore, we compute the sphericity of particle $P$, denoted by $M_{\text{sphe}}(P)$. It quantifies how closely the shape of $P$ resembles that of a sphere, where the ratio of $M_{\text{vol}}(P)$ to $M_{\text{area}}(P)$ is compared to the ratio of volume to surface area of a sphere with the same volume as $P$. Thus, the sphericity $M_{\text{sphe}}(P)$ is given by
\begin{equation}\label{Eq:sphe}
    M_{\text{sphe}}(P)=\frac{M_{\text{vol}}(P)}{M_{\text{area}}(P)} \frac{S_{\text{area}}(r(P))}{S_{\text{vol}}(r(P))},
\end{equation}
where $S_{\text{vol}}(x)$ and $S_{\text{area}}(x)$ denote the volume and surface area of a sphere with radius $x>0$, respectively, and $r(P)$ is the radius of a sphere with the same volume as $P$. Alternatively,  instead of using Eq.~(\ref{Eq:sphe}),  the sphericity $M_{\text{sphe}}(P)$  can be computed using the following representation formula:
\begin{equation}
    M_{\text{sphe}}(P)=\frac{(36\pi M_{\text{vol}}(P)^2)^{\frac{1}{3}}}{M_{\text{area}}(P)}.
\end{equation}

Recall that the grayscale value $I(\voxel)\in\mathbb{R}$ of a voxel $\voxel\in W$ in CT image data  corresponds to  the (local) X-ray attenuation coefficient of the material at voxel $\voxel$ which depends on the material's mass density at this location, see e.g. \cite{GRODZINS1983541}. Thus, it seems to be plausible that the overall brightness of the set of voxels representing a particle can be correlated with its \composition{}. However,
in the present CT data, imaging artifacts exist which  could distort mean grayscale values of voxels associated to individual particles. Hence, instead of the mean grayscale value,  the median of grayscale values of voxels associated to a particle $P$ is considered,  which is denoted by $M_\text{med}(P)$. Similarly,  a  larger variability of grayscale values within a particle could indicate the presence of different mineralogical components. Therefore, besides the median $M_\text{med}(P)$, we consider the inter-quartile range $  M_{\text{IQR}}(P)$ of particle-wise grayscale values, which is given by
\begin{equation*}
    M_{\text{IQR}}(P)=Q_3(P)-Q_1(P), 
\end{equation*}
 where  $Q_1(P)$ and $Q_3(P)$  denote the 25-th and 75-th percentile of grayscale values associated to $P,$ respectively. The interquartile range $  M_{\text{IQR}}(P)$ is a robust measure of variability, frequently used instead of the standard deviation, see e.g.~\cite{KOLACZ2016930}.

\subsubsection{Mineralogical composition}\label{sec:mineralogical-composition-of-feed-particles}

 To correlate the morphological and textural 3D characterization of particles with their mineralogical composition, we use the SEM-EDS data described in Section~\ref{sec.des.mat} in order to determine a particle descriptor for quantifying the presence of valuable minerals, namely the \composition{}. 
 Since the distinction between valuable minerals (zinnwaldite) and non-valuable minerals (quartz, topaz, muscovite and others) is of interest, we group the detected minerals accordingly. The phase-wise segmented SEM-EDS data within a slice $\window_z\subset W$ for some $z\in\mathbb{Z}$ can then be described by a map $\phaseSegmentation: \window_z\rightarrow\{0,1,2\},$ where  $\phaseSegmentation(\voxel)=0$ indicates the presence of no particle at voxel $\voxel\in\window_z$. Moreover,  $\phaseSegmentation(\voxel)=1$ and $\phaseSegmentation(\voxel)=2$ indicate that a valuable mineral (zinnwaldite) and a non-valuable mineral (quartz, topaz, muscovite and others) was detected at $\voxel\in\window_z$, respectively.

The SEM-EDS data is available for three slices\footnote{The subsets of voxels, where SEM-EDS data are available, are not necessarily parallel to any of the axes, and, thus, they do not fit the usual definition of a 2D slice given in Section~\ref{sec.gro.tru}. However, for notational simplicity, we still refer to them as slices.} $\window_{z_1},\window_{z_2},\window_{z_3}$, i.e., the maps $\phaseSegmentation_j \colon \window_{z_j} \to \{0,1,2\}$ of the mineralogical composition are given for each  $j=1,2,3$.
Then, for each particle $P\subset \window$ such that
$P\cap \left(\window_{z_1} \cup \window_{z_2} \cup \window_{z_3} \right)\not=\emptyset$ the area fraction $M_{\mathrm{rat}}(P)$ of valuable minerals observed in voxels associated with the intersection $P\cap \left(\window_{z_1} \cup \window_{z_2} \cup \window_{z_3} \right)\not=\emptyset$ is given by
\begin{equation}
    M_\mathrm{rat}(P)= 
    \frac{\# \bigcup_{j=1}^3 \{x\in P\cap \window_{z_j} \colon \phaseSegmentation_j(x)=1\}}
    {\# \bigcup_{j=1}^3 \{x\in P\cap \window_{z_j} \colon \phaseSegmentation_j(x)>0\}},
\end{equation}
where $\#$ denotes the cardinality of a set. For this, to obtain a useful descriptor, we assume that the mineralogical composition of a particle $P$ observed in a SEM-EDS slice is representative for the entire 3D particle. Then, $M_\mathrm{rat}(P)$ can be interpreted as the \composition{} within $P$.

\subsection{Multivariate probabilistic modeling of particle descriptors}\label{Sec:Vine}

By computing the descriptors introduced in Section~\ref{sec:stochastic-modeling-of-multidimensional-vectors-of-particle-characteristics} 
for each segmented particle $P_1,\dots, P_n$ within the CT image data for which SEM-EDS information is available, we obtain a sample of  vector data
\begin{equation}
x^{(\ell)} =\bigl( M_{\text{med}}(P_\ell), M_{\text{IQR}}(P_\ell), M_{\text{vol}}(P_\ell), M_{\text{elo}}(P_\ell), M_{\text{flat}}(P_\ell), M_{\text{sphe}}(P_\ell), M_{\text{rat}}(P_\ell)\bigr)	
\label{eq:vector-of-characteristics}
\end{equation}
 for $\ell=1,\dots,n$.   We use the data set $D=\{x^{(\ell)}: \ell=1,\dots,n\}$ in order to fit a multivariate probability distribution of particle descriptors for the  particle system considered in this paper. 

Therefore, we interpret the data given in Eq.~(\ref{eq:vector-of-characteristics}) as   realizations of a certain random vector $X=(X_1,\dots,X_d)$ with $d=7$. Note that
the distribution of $X$ is uniquely determined by its   cumulative distribution function (CDF) $F_{1,\ldots,d}:\mathbb{R}^d\rightarrow [0,1]$, where $F_{1,\ldots,d}(x)=\mathbb{P}(X_1\leq x_1,\dots, X_d\leq x_d)$ for each $x=(x_1,\dots,x_d)\in\mathbb{R}^d$.
Thus, fitting a model for $F_{1,\ldots,d}$, or for the corresponding probability density $f_{1,\ldots,d}:\mathbb{R}^d\to[0,\infty)$, to the data set $D=\{x^{(1)},\ldots,x^{(n)}\}$
can provide valuable insight regarding the size, shape, texture and \composition{} of individual particles of  the particle system under consideration.

In~\cite{furat2019}, $d$-dimensional Archimedean copulas have been utilized to model the joint distribution of particle-wise descriptors determined in a similar type from sample data. In the present paper, we  extend this approach through the use of so-called R-vine copulas, which allow for a more flexible modeling of multivariate probability distributions. Therefore, to begin with, we briefly recall the definition of a copula.

\subsubsection{Copulas: Definition and Sklar's representation formula} 

A function $C : [0,1]^d \xrightarrow{}[0,1]$ with $d\geq 2$ is called a $d$-dimensional copula if $C$ is the  cumulative distribution function of a $d$-dimensional random vector with standard uniformly distributed marginals, i.e.,  $C : [0,1]^d \xrightarrow{}[0,1]$ is a component-wise non-decreasing function such that $C(1,\ldots,1,x_i,1,\ldots,1)=x_i$ for all $i=1,\ldots,d$ and $x_i\in[0,1]$.

Copulas are a powerful tool to parametrically model multivariate probability distributions. The reason for this is the fundamental representation formula discovered by A.~Sklar, see e.g. ~\cite{Nel06}, which states that the  multivariate CDF $F_{1,\ldots,d}:\mathbb{R}^d\rightarrow [0,1]$ of any random vector $X=(X_1,\ldots,X_d)$ can be expressed by its marginal CDFs $F_i:\mathbb{R}\rightarrow [0,1]$, $i=1,\ldots,d$, where $F_i(x_i)=\mathbb{P}(X_i\le x_i)$ for each $x_i\in\mathbb{R}$ and $i=1,\ldots,d$,
and a certain copula $C : [0,1]^d \xrightarrow{}[0,1]$, i.e., it holds that
\begin{equation}
     F_{1,\ldots,d}(x_1,\dots,x_d)=C(F_1(x_1),\dots,F_d(x_d)) \qquad \text{ for all } (x_1,\ldots,x_d)\in\mathbb{R}^d.
     \label{Eq:Sklar}
\end{equation}
Furthermore, if the CDFs $F_{1,\ldots,d}$ and $C$  are differentiable, then Eq.~(\ref{Eq:Sklar}) implies that 
 \begin{equation}
     f_{1,\ldots,d}(x_1,\dots,x_d)=c(F_1(x_1),\dots,F_d(x_d)) \prod_{i=1}^d f_i(x_i) \qquad \text{ for all } (x_1,\ldots,x_d)\in\mathbb{R}^d,
     \label{Eq:Sklar_density}
 \end{equation}  
where $f_{1,\ldots,d}:\mathbb{R}^d\to[0,\infty)$, $c:[0,1]^d\to[0,\infty)$ and $f_i:\mathbb{R}\to[0,\infty)$ denote the probability densities corresponding to the CDFs $F_{1,\ldots,d}$, $C$ and $F_i$, respectively.

On the other hand, copulas can be used for the construction of multivariate probability distributions. More precisely, by inserting any $d$-dimensional copula density $c$  and any combination of univariate CDFs $F_1,\dots, F_d$ with probability densities $f_1,\dots, f_d$ into  Eq.~(\ref{Eq:Sklar_density}), the function $f_{1,\ldots,d}$ given by  the right-hand side of Eq.~(\ref{Eq:Sklar_density}) is a multivariate probability density with marginal probability densities $f_1,\dots, f_d$.  Moreover,  by means of vine copulas, this procedure for the construction of multivariate distributions can be simplified by applying the so-called pair-copula construction method.  Then, for modeling the probability density of a $d$-dimensional random vector $X$  with $d>2$, two-dimensional copula densities are used to approximate conditional bivariate probability densities, see Section~\ref{sec.pai.cop}, where the three-dimensional case is considered for illustration.

\subsubsection{The pair-copula construction method}\label{sec.pai.cop}

Before explaining the pair-copula construction method in Section~\ref{sec.reg.vin},
we first illustrate the idea of this method  for the three-dimensional case. Let $X=(X_1,X_2,X_3)$ be  a random vector such that the CDFs $F_{1,2,3}:\mathbb{R}^3\to[0,1]$ and $C:[0,1]^3\to[0,1]$ appearing in  Eq.~(\ref{Eq:Sklar}) are continuously differentiable. Then, the density $f_{1,2,3}:\mathbb{R}^3\to[0,\infty)$ of $X$ can be written in the following form: For each  $x=(x_1,x_2,x_3)\in\mathbb{R}^3$ with $f_{1,2,3}(x)>0$, let us consider the identity
\begin{equation}
f_{1,2,3}(x)=f_{1,3\mid X_2=x_2}(x_1,x_3)\, f_{2}(x_2),
\label{eq.off.iks}
\end{equation}
where $f_{1,3\mid X_2=x_2}:\mathbb{R}^2\to[0,\infty)$ with $f_{1,3\mid X_2=x_2}(x_1,x_3)=f_{1,2,3}(x_1,x_2,x_3)/f_2(x_2)$ is the conditional density of the random vector $(X_1,X_3)$ given that $X_2=x_2$. Applying 
Eq.~(\ref{Eq:Sklar_density}) to the first factor on the right-hand side of Eq.~(\ref{eq.off.iks}), we get that
\begin{equation}
 f_{1,2,3}(x)=c_{1,3\mid X_2=x_2}\bigl(F_{1\mid X_2=x_2}(x_1),F_{3\mid X_2=x_2}(x_3)\bigr) f_{1\mid X_2=x_2}(x_1) \, f_{3\mid X_2=x_2}(x_3) \, f_{2}(x_2),
\label{eq.eff.mid}
\end{equation}
where $c_{1,3\mid X_2=x_2}:[0,1]^2\to[0,\infty)$ denotes the copula density corresponding to  $f_{1,3\mid X_2=x_2}$ and $F_{i\mid X_2=x_2}$ is the conditional CDF  of $X_i$ given that $X_2=x_2$ with density  $f_{i\mid X_2=x_2}$, $i=1,3$. Now, observe that similar to Eq.~(\ref{eq.off.iks}),
the identity
$f_{i\mid X_2=x_2}( x_i)=f_{i,2}(x_i,x_2)/f_2(x_2)$
holds, where $f_{i,2}:\mathbb{R}^2\to[0,\infty)$ denotes the (unconditional) probability density of $(X_i,X_2)$; $i=1,3$, Then, applying   
Eq.~(\ref{Eq:Sklar_density}) to the  two-dimensional probability densities 
$f_{i,2}$, $i=1,3$, we get from Eq.~(\ref{eq.eff.mid}) that  
  \begin{equation}
 f_{1,2,3}(x) 
=c_{1,3\mid X_2=x_2}(F_{1\mid X_2=x_2}(x_1),F_{3\mid X_2=x_2}(x_3)) \, c_{1,2}(F_1(x_1),F_2(x_2)) \, c_{3,2}(F_3(x_3),F_2(x_2))\prod_{i=1}^3 f_i(x_i) 
\label{Eq:example1}
\end{equation}
for each  $x=(x_1,x_2,x_3)\in\mathbb{R}^3$ with $f_{1,2,3}(x)>0$.
Thus, the (trivariate) probability density $f_{1,2,3}$ can be expressed as a product of bivariate  copula densities and the (univariate) densities $f_1,f_2,f_3$.  

Furthermore, instead of the identity given in Eq.~(\ref{eq.off.iks}), we can consider an alternative decomposition of $f_{1,2,3}$, using the fact that the identity $f_{1,2,3}(x)=f_{1,2|X_3=x_3}(x_1,x_2)\, f_{3}(x_3)$ holds for each  $x=(x_1,x_2,x_3)\in\mathbb{R}^3$ with $f_{1,2,3}(x)>0$. This leads to a different pair-copula construction given by
\begin{align*}
 f_{1,2,3}(x)&=c_{1,2|X_3=x_3}\bigl(F_{1|X_3=x_3}(x_1),F_{2|X_3=x_3}(x_2)  \bigr) \, c_{1,3}\bigl(F(x_1),F(x_3)\bigr) \, c_{2,3}(F_2(x_2), F_3(x_3))\prod_{i=1}^3 f_i(x_i)
\end{align*} 
for each  $x=(x_1,x_2,x_3)\in\mathbb{R}^3$ with $f_{1,2,3}(x)>0$. Moreover, a third pair-copula construction is obtained starting with the identity
 $f_{1,2,3}(x)=f_{2,3|X_1=x_1}(x_2.x_3)\, f_{1}(x_1)$ which holds for each  $x=(x_1,x_2,x_3)\in\mathbb{R}^3$ with $f_{1,2,3}(x)>0$.

Thus, in the three-dimensional case there are three different pair-copula constructions. Note that the number of possible pair-copula constructions grows exponentially with the dimension $d>3$, see \cite{numberofvines}. For that reason, so-called regular vines are used to describe  how different pair-copula constructions can be obtained in the general $d$-dimensional case, see Sections~\ref{sec.reg.vin} and \ref{sec.seq.fit} below.

\subsubsection{Regular vine copulas}\label{sec.reg.vin}

In the higher-dimensional case, the representation formula given in Eq.~(\ref{Eq:Sklar_density}) cannot be directly used for the fitting of multivariate probability densities to vector-valued data, as it would  require the fitting of a higher-dimensional copula density which can be difficult. Instead, we present an alternative representation formula which, under some simplifying assumptions (see Section~\ref{sec.seq.fit}), allows for the modeling of higher-dimensional probability densities using only bivariate copulas and marginal (univariate) densities. 

The vector $V = (T_1,\dots, T_{d-1})$ is called a \emph{regular vine} (or, briefly, R-vine) on a set of $d>1$ elements, identified with the set of integers $\{1,\ldots,d\}$, if the following three conditions hold: \begin{itemize}
  \item[(i)] $T_1$ is an undirected  tree with the set of nodes $N_1 = \{1,\dots, d\}$ and some set of edges denoted by $E_1$.
 \item[(ii)]  For $i = 2,\dots, d-1$, $T_i$ is an undirected tree with set of nodes $N_i = E_{i-1}$ and some set of edges $E_i$.
 \item[(iii)]  For $i = 2,\dots, d-1$ and $\{a, b\} \in E_i$  it holds that $|a \cap b| = 1$, where $|a \cap b|$ denotes the cardinality of the set $a \cap b$.
\end{itemize}
The edges  of a tree $T_i$ in a regular vine specify  conditioning and conditioned sets. For each edge $e=\{e_1,e_2\}\in E_1$, the conditioned set $O(e)$ is defined as  $O(e)=\{e_1,e_2\}$ and for the conditioning set we put $S(e)=\emptyset$, i.e., $S(e)$ is the empty set. Now, let $e=\{e_1,e_2\}\in E_2\cup\dots\cup E_{d-1}$ be an edge of any of the subsequent trees. Then, we recursively define 
\begin{equation*}
    S(e)=S(e_1)\cup S(e_2)\cup(O(e_1)\cap O(e_2)) \quad\text{ and }\quad O(e)=(O(e_1)\cup O(e_2))\setminus S(e).
\end{equation*} 
By construction, the sets $S(e)$ and $O(e)$ defined in this way are subsets of  $\{1,\dots,d\}$ for any edge $e\in E(V)$, where $E(V)=E_1\cup\dots\cup E_{d-1}$ denotes the set of all edges in  $V$. 
According to~\cite{kurowicka2010},
 the conditioned set $O(e)$ consists of two elements for every edge $e \in E(V)$.
 Moreover, for each pair of indices $\{i,j\}\in \{1,\dots,d\}\times \{1,\dots,d\}$ with $i<j$, there is exactly one edge $e\in E(V)$ with $\{i,j\}=O(e)$. Thus, it holds that $\{e_{t_1},e_{t_2}\}=O(e)$ for some indices $e_{t_1}<e_{t_2}$, where for notational simplicity we will shortly write $e_1$ and $e_2$ instead of $e_{t_1}$ and $e_{t_2}$, respectively.
 
Let $X=(X_1,\dots,X_d)$ be a random vector with continuous probability density $f_{1,\dots,d}:\mathbb{R}^d\rightarrow[0,\infty]$ and marginal densities $f_1,\dots,f_d:\mathbb{R}\rightarrow[0,\infty].$ Moreover, let $V=(T_1,\dots,T_{d-1})$ be a $d$-dimensional R-vine as described above. Then, for any $x=(x_1,\dots,x_d)\in\mathbb{R}^d$ with $f_{1,\ldots,d}(x)>0,$ the following representation holds
(see~\cite{Czado2019})
\begin{equation}\label{Eq:vinecopula}
f_{1,\ldots,d}(x)=
\prod_{e=\{e_1,e_2\}\in E(V)}c_{{e_1},{e_2}\mid X_{S(e)}=x_{S(e)}}(F_{{e_1}\mid X_{S(e)}=x_{S(e)}}(x_{{e_1}}),F_{{e_2} \mid X_{S(e)}=x_{S(e)}}(x_{{e_2}}))\prod_{i=1}^{d}f_j(x_i),
\end{equation}
where $X_{S(e)}$ denotes the  random vector consisting of those components of $X$ whose indices belong to the set $S(e),$ and $x_{S(e)}$ is the corresponding subvector of $x$. Moreover, $c_{e_1,e_2\mid X_{S(e)}=x_{S(e)}}:\mathbb{R}^2\rightarrow[0,\infty]$ denotes the bivariate copula density of the conditional probability distribution of the random vector $(X_{e_1},X_{e_2})$ under the condition that $X_{S(e)}=x_{S(e)},$ and $F_{e_i\mid X_{S(e)}=x_{S(e)}}$   denotes the conditional CDF of $X_{e_i}$ under the condition that $X_{S(e)}=x_{S(e)}$, with $i\in\{1,2\}$, which can be determined by using the recursion formula (see~\cite{diss})

\begin{equation}
    F_{
        e_i|X_{S(e)\cup e_{j}}=x_{S(e)\cup e_{j}}
        }(x_{e_i})=
    \frac{\frac{\text{d}}{\text{d}x_{e_j}}
        C_{e_1,e_2|X_{S(e)}=x_{S(e)}} 
        \left(
        F_{e_1|X_{S(e)}=x_{S(e)}}(x_{e_1}),
        F_{e_2|X_{S(e)}=x_{S(e)}}(x_{e_2})
        \right)}{
    \frac{\text{d}}{\text{d}x_{e_j}}
    F_{e_j|X_{S(e)}=x_{S(e)}}(x_{e_j})
    },
    \label{eq.rec.con}
\end{equation}

where  $j\in\{1,2\}\setminus \{i\}$.

The representation formula given in Eq.~(\ref{Eq:vinecopula}), called vine copula representation  of the multivariate probability density  $f_{1,\ldots,d} \colon \R^d \to [0,\infty)$, is a central result of copula theory, see e.g.~\cite{Czado2019} where further details can be found.

\subsubsection{Sequential fitting procedure}\label{sec.seq.fit}

The goal is now to fit the probability density  $f_{1,\ldots,d} \colon \R^d \to [0,\infty)$ to empirical data of particle descriptors, using the  formulas given in Eqs.~(\ref{Eq:vinecopula}) and (\ref{eq.rec.con}). However, estimating the  density $f_{1,\ldots,d}$ by directly utilizing  Eqs.~(\ref{Eq:vinecopula}) and (\ref{eq.rec.con})
would require a separate fitting of the conditional bivariate copula densities $c_{e_1,e_2\mid X_{S(e)}=x_{S(e)}}$ for each realization $x_{S(e)}\in\mathbb{R}^{|S(e)|}$ of $X_{S(e)}$. Instead, it is common to suppose that the so-called \emph{simplifying assumption} is true, i.e., the bivariate copula densities $c_{e_1,e_2\mid X_{S(e)}=x_{S(e)}}$ do not depend on the  realizations $x_{S(e)}$ of $X_{S(e)}$, but just on the conditioning set $S(e)$, see~\cite{killiches2016examination}. Note that the  use of copulas generally allows for the subsequent estimation of marginal distributions and, in a second step, interdependencies.  Therefore, after fitting suitable (univariate) probability densities $\widehat{f}_1, \dots, \widehat{f}_d:\mathbb{R}\to[0,\infty)$ 
to a given sample 
$D=\{x^{(k)}:k\in\{1,\dots,n\}\}$

of $n$ realizations $x^{(1)}=(x^{(1)}_1,\ldots,x_d^{(1)}), \ldots, x^{(n)}=(x^{(n)}_1,\ldots,x_d^{(n)}) \in \R^d$ of the random vector $X=(X_1,\ldots,X_d)$, 
the task of estimating the  density $f_{1,\dots,d}$ is reduced to selecting a vine structure $\widehat{V}=(\widehat{T}_1,\dots,\widehat{T}_{d-1})$ and a family of bivariate copula densities $\widehat{\cal C}=\{\widehat{c}_{e_1,e_2|S(e)}: e \in E(\widehat V)\}$.

We start by modeling the marginal probability densities ${f}_1, \dots, {f}_d$ 
of individual particle descriptors. Since the 
particles considered in this paper are often composites of different mineralogical components, mainly quartz and zinnwaldite, multimodal distributions are chosen as candidates for the modeling of the marginal probability densities. More precisely, we consider a mixture $f_\mathrm{mixed}:\mathbb{R}\to[0,\infty)$ of two probability densities $f^{(1)},f^{(2)}:\mathbb{R}\to[0,\infty)$ given by
\begin{equation}
f_\mathrm{mixed}(x) = \lambda \, f^{(1)}(x) + (1-\lambda) \, f^{(2)}(x)\qquad\mbox{for each $x\in\mathbb{R}$},
\label{eq:mixture}
\end{equation}
where  $\lambda\in [0,1]$ is some mixing ratio.
The parameters of $f_\mathrm{mixed}$, i.e., the parameters of the mixing components $f^{(1)}$, $f^{(2)}$ and the mixing ratio $\lambda$, are fitted using the expectation-maximization algorithm, see  \cite{kroese2019DSML}.
For example, we fit the (marginal) probability distributions
of the descriptors $M_\mathrm{med}$, $M_\mathrm{IQR}$ and $M_\mathrm{vol}$ introduced in Section~\ref{sec:stochastic-modeling-of-multidimensional-vectors-of-particle-characteristics}, using  mixtures of two gamma distributions, the probability density $f:\mathbb{R}\to[0,\infty)$ of which is given by
\begin{equation}
f(x)=\frac{
	x^{\alpha-1} \, \exp\bigl(-x/\beta\bigr)	
}{
\beta^\alpha \Gamma(\alpha)
}
\, \mathbbm{1}_{[0,\infty)}(x)
\qquad \text{for each } x \in \mathbb{R},
\end{equation}
where $\alpha,\beta>0$ are model parameters, $\Gamma:[0,\infty)\to(0,\infty)$ is the gamma function and $\mathbbm{1}_A$ denotes the indicator of the set $A\subset\mathbb{R}$, i.e., $\mathbbm{1}_A(x)=1$ if $x\in A$ and $\mathbbm{1}_A(x)=0$ if $x\not\in A$, see  \cite{johnson1995continuous_vol1}.
Analogously, the remaining descriptors $M_\mathrm{elo}$, $M_\mathrm{flat}$, $M_\mathrm{sphe}$ and $M_\mathrm{rat}$ introduced in Section~\ref{sec:stochastic-modeling-of-multidimensional-vectors-of-particle-characteristics}, which take values in the interval $[0,1]$ are modeled using mixtures of beta distributions, the probability density $f:\mathbb{R}\to[0,\infty)$ of which is given by
\begin{equation}
f(x)=
\frac{
	x^{p-1} (1-x)^{q-1}
}{
B(p,q)
}
\mathbbm{1}_{[0,1]}(x)
\qquad \text{for each } x \in \mathbb{R},
\end{equation}
where $p,q>0$ are model parameters  and
$B:[0,\infty)^2\to(0,\infty)$ 
is the beta function, see  \cite{johnson1995continuous}. By fitting such mixtures of distributions to the seven particle descriptors
$M_\mathrm{med}$, $M_\mathrm{IQR}$, $M_\mathrm{vol}$, $M_\mathrm{elo}$, $M_\mathrm{flat}$, $M_\mathrm{sphe}$ and $M_\mathrm{rat}$, we obtain the parametric densities $\widehat{f}_1, \dots, \widehat{f}_7$ and, by numerical integration, the corresponding CDFs $\widehat{F}_1, \dots, \widehat{F}_7$.

  To select a suitable vine structure $\widehat{V}$ and a family $\widehat{\cal C}$ of bivariate copula densities, the sequential algorithm described in \cite{diss} is used, which iteratively models the trees $T_i, i=1,\ldots,d-1$ and the corresponding pair copulas starting with  $T_1$. The motivation of this sequential estimation procedure is the supposition that the edges of the first trees are   modeling the  dependencies in the data more accurately than those considered in higher levels of the vine structure. 
  
  The strength of  dependency is quantified using the pairwise Kendall rank correlation coefficient $\tau$, which for two  vectors $y=(y_1,\dots,y_n),y^\prime=(y^\prime_1,\dots,y^\prime_n)\in\mathbb{R}^n$, is defined by
 \begin{equation*}
  \tau(y,y^\prime)=\frac{2}{n(n-1)}\sum_{i<j}\text{sgn}(y_i-y_j)\; \text{sgn}(y^\prime_i-y^\prime_j).
  \label{eq:kendall}
\end{equation*}
Starting with the first level  of the vine structure, the  configuration of $T_1$, which maximizes the sum (over all edges of $T_1$) of the absolute values of pairwise Kendall rank correlation coefficients, is selected and for each edge of the resulting tree $T_1$ a pair copula is chosen based on the maximum-likelihood criterion.  This process is repeated for each subsequent level of the vine structure, as detailed below in Algorithm 1, where  $\mathbb{T}(T_{i-1})$ denotes  the set of tree structures, which satisfy the requirements  of regular vines given that $T_{i-1}$ is the tree at the preceding level. The parametric families of the Archimedean copulas Frank, Joe, Clayton and Gumbel, and their  rotations by 90, 180 and 270 degrees  (see~\cite{kurowicka2010}), are chosen as candidates for each pair copula. The set of their densities is denoted by $H$.

\vspace{0.2cm}
\begin{algorithm}[H]
\textbf{Algorithm 1. Sequential estimation of vine copulas }
 \hrule
 \vspace{1mm}
 \KwData{ training data $D$, set $H$ of candidates for bivariate copula densities, marginal CDFs $\widehat{F}_1,\dots,\widehat{F}_d$}
 \KwResult{vine structure and its (conditional) bivariate copula densities}

 \For{$i\,\in \{1,\dots,d-1\}$}{
  $T_i=\underset{T\in \mathbb{T}(T_{i-1})}{\operatorname{argmax}} \left[\underset{e\in E(T)}{\sum }\left|\tau (\widehat{F}_{e_1\mid S(e)}(D_{e_1}),\widehat{F}_{e_2 \mid S(e)}(D_{e_2}))\right|\right]$
  
  \For{$e\in E(T_i)$}{
   \If{$|\tau(\widehat{F}_{e_1\mid S(e)},\widehat{F}_{e_2 \mid S(e)})|\sqrt{\frac{9n(n-1)}{2(2n+5)}}\leq 1.96$}{
  \vspace{3mm}
     $\widehat{c}_{e_1,e_2|S(e)}=1$ (bivariate independence copula density)

  } \Else{ $
    \widehat{c}_{e_1,e_2|S(e)}=\underset{h\in H}{\operatorname{argmax}}\left[ \overset{n}{\underset{l=1 }{\prod}} h\left(\widehat{F}_{e_1\mid S(e)=x_{l,S(e)}}(x_{l,e_1}),\widehat{F}_{e_2 \mid S(e)=x_{l,S(e)}}(x_{l,e_1})\right)\right]
    $}
   }

 }
 $\widehat{V}=(T_1,\dots,T_{d-1})$
 \hrule
\end{algorithm}
 \vspace{0.3cm}  

By means of the fitting approach described above, we can compute multivariate probability densities for the  size, shape, texture and \composition{} descriptors introduced in Section~\ref{sec:stochastic-modeling-of-multidimensional-vectors-of-particle-characteristics}, such that we achieve an efficient characterization of the considered particle system. In the next section we not only explain in detail how the descriptor vectors $x^{(1)},\dots,x^{(n)}$ determined in   Section~\ref{sec:stochastic-modeling-of-multidimensional-vectors-of-particle-characteristics} are leveraged to characterize the particle system using multivariate probability densities, but also how we use the resulting fits to derive prediction models that allow us to estimate the \composition{} of particles using only CT data.

\subsection{Quantitative prediction of mineralogical composition from CT}\label{pre.min.sub}

\subsubsection{Modification of the copula-based modeling approach}\label{sec:multiv ariateCharacterization}
In this section, we show how the copula-based modeling approach stated in Section~\ref{Sec:Vine} has to be modified in order to capture the particularities of the data set   $D=\{x^{(\ell)}: \ell=1,\dots,n\}$.   Since the nature of the data given in $D$ suggests that the  distribution of $M_{\mathrm{rat}}$ might have atoms at 0 and 1, a seven-variate probability density cannot be directly fitted to $D$ using the algorithm  stated in Section~\ref{sec.seq.fit}. The apparent occurrence of particles  with $M_{\mathrm{rat}}=0$ and $M_{\mathrm{rat}}=1$, respectively, might be caused by the limited resolution of image data and limitations of the stereological approach used for the computation of $M_\mathrm{rat}$. Thus, to construct a seven-variate probabilistic characterization of the particle system considered in this paper, the data set $D$ of descriptor vectors is represented by three disjoint sets: 
(i) the set $D_{\zin}\subset \R^6$ of CT-based descriptor vectors of particles consisting almost exclusively of valuable minerals, which is given by
\begin{equation*}
   D_{\zin}=\{x_{1,\dots,6}^{(\ell)}\in\R^6:  \ell=1,\dots,n, \text{ where } (x_{1,\dots,6}^{(\ell)}, x^{(\ell)}_7)\in D  \text{ with } x^{(\ell)}_7\geq 0.99\}; 
\end{equation*}
(ii) the set $D_{\nzin}\subset \R^6$ of CT-based descriptor vectors of  particles with almost no valuable minerals, which is given by
\begin{equation*}
   D_{\nzin}=\{x_{1,\dots,6}^{(\ell)}\in\R^6:  \ell=1,\dots,n, \text{ where } (x_{1,\dots,6}^{(\ell)}, x^{(\ell)}_7)\in D  \text{ with } x^{(\ell)}_7\leq 0.01\}; 
\end{equation*}

and (iii) the set $D_{\czin}\subset \R^7$ of descriptor vectors of composite particles, meaning those particles which contain significant fractions of both, valuable and non-valuable minerals, given by
\begin{equation*}
   D_{\czin}=\{(x_{1,\dots,6}^{(\ell)}, x^{(\ell)}_7)\in D\subset \R^7: \ell=1,\dots,n, \text{ where } 0.01<x^{(\ell)}_7< 0.99\}.
\end{equation*}

Now, using the vine copula approach stated in Section~\ref{Sec:Vine}, six-variate probability densities $\widehat{f}^{\zin}, \widehat{f}^{\nzin} \colon \R^6 \to [0,\infty)$ can be fitted to the data sets $D_{\zin}$ and  $D_{\nzin}$, respectively. Similarly, a seven-variate  probability density $\widehat{f}^{\czin}\colon \R^7\to [0,\infty)$ is fitted to the data set $D_{\czin}$ such that its univariate marginal density $\widehat{f}_7^{\czin}\colon\R\to[0,\infty)$ vanishes outside of the interval $(0.01,0.99)$ (e.g., by fitting a truncated mixed beta distribution).  

To describe the distribution of the seven-dimensional descriptor vectors considered in Eq.~(\ref{eq:vector-of-characteristics}) for the entire data set $D$, we make the following model assumption. Suppose
that the random variable $M_{\mathrm{rat}}$ describing the \composition{}  is conditionally independent of the remaining six (CT-based) particle descriptors  
and uniformly distributed on $[0,0.01]$ and $[0.99,1]$. Note that these two intervals can be chosen arbitrarily small around 0 and 1, respectively.
Then, a seven-variate probability density $ \widehat{f}\colon \R^7 \to [0,\infty)$ of the particle descriptors considered in 
Eq.~(\ref{eq:vector-of-characteristics})   can be constructed   as follows. For each  $x=(x_{1\ldots,6},x_7)\in \R^6\times \R$, we put
\begin{equation}\label{eq:multivaraiteCharacterizationDensity}
    \widehat{f}(x)=
    \left\{
    \begin{array}{ll}
        \frac{n_\nzin}{n}\,\frac{1}{0.01}\,  \widehat{f}^\nzin(x_{1,\dots,6}), & \text{if } 0\leq x_7\leq 0.01, \vspace{0.1cm} \\
        \frac{n_\czin}{n} \widehat{f}^\czin(x), & \text{if } 0.01< x_7\leq 0.99, \vspace{0.1cm} \\
        \frac{n_\zin}{n}\,\frac{1}{0.01}\,   \widehat{f}^\zin(x_{1,\dots,6}), & \text{if } 0.99 < x_7\leq 1.0, \vspace{0.1cm}\\
        0, & \text{otherwise,}
    \end{array}
    \right.
\end{equation}
 where $n_{\zin},n_{\nzin}, n_{\czin}$ and $n= n_{\zin}+n_{\nzin}+n_{\czin} $ denote the cardinalities of the data sets $D_{\zin},D_{\nzin}, D_{\czin}$ and $D$, respectively.

\subsubsection{Prediction model}\label{sec.pre.min}

Finally, we show how multivariate probability densities of particle descriptor vectors, as considered in Section~\ref{sec:multiv ariateCharacterization}, can be used for  predicting the \composition{} of a particle using  size, shape and texture descriptors computed solely from CT image data, see Fig.~\ref{fig:modeling-scheme} (second row; left). 
Recall that the first six entries $x_1^{(\ell)},\ldots, x_6^{(\ell)}\in \R$ of a descriptor vector $x^{(\ell)}=(x_1^{(\ell)},\ldots, x_7^{(\ell)})
\in D\subset \R^7$ are determined from CT image data, whereas its seventh entry $x_{7}^{(\ell)}$ which characterizes the \composition{} was determined from SEM-EDS  data, see Section~\ref{sec:stochastic-modeling-of-multidimensional-vectors-of-particle-characteristics}. Thus, we call  $x_{1,\dots,6}^{(\ell)}=(x_1^{(\ell)},\ldots, x_6^{(\ell)})\in\R^6$ the CT-based descriptor vector of $x^{(\ell)}$.

Using the  probability densities $\widehat{f}^{\zin},\widehat{f}^{\nzin},\widehat{f}^{\czin}$ introduced in Section~\ref{sec:multiv ariateCharacterization}, a  prediction model $\predictionModel \colon \R^6 \to [0,1]$ can be constructed as follows:
For any given CT-based descriptor vector $x\in\R^6$, the corresponding particle  is  classified as either a particle with only valuable minerals, a particle with no valuable minerals, or as a composite particle, based on 
which of the six-variate probability densities $\widehat{f}^{\zin},\widehat{f}^{\nzin},\widehat{f}^{\czin}_{1,\dots,6}$ has the largest likelihood at $x\in\R^6$, where the partially marginalized probability density $\widehat{f}^{\czin}_{1,\dots,6}: \R^6\to[0,\infty)$ is given by 
$\widehat{f}^{\czin}_{1,\dots,6}(x)=\int_0^1\widehat{f}^{\czin}(x,x_7)\,\text{d}x_7$ for each $x\in\R^6$. Such prediction models are referred to as Bayes classifiers \citep{hastie2009}. 
Thus, in case of a classification as particle with no valuable minerals (i.e., $M_{\mathrm{rat}}=0$) or as particle with only valuable minerals (i.e., $M_{\mathrm{rat}}=1$)  the output of $\predictionModel$ is set to $0$ or $1$, respectively. If the partially marginalized probability density $\widehat{f}^{\czin}_{1,\dots,6}$ of composite  particles has the largest likelihood at $x\in\R^6$, the output of $\predictionModel$ will be set to the median $\phi(x)$ of the conditional probability density $\widehat{f}^{\czin}_{7\mid x}$, where $\widehat{f}^{\czin}_{7\mid x}(x_7)=\widehat{f}^{\czin}(x,x_7)/\widehat{f}^{\czin}_{1,\ldots,6}(x)$
for each $x_7\in\R$. Thus, the  prediction model $\predictionModel \colon \R^6 \to [0,1]$ is given by

\begin{equation}
\label{eq.dec.two}
    \predictionModel(x)=\left\{
            \begin{array}{ll}
                1, &\qquad \text{ if }\frac{n_\zin}{n} \widehat{f}^{\zin}(x)\geq \max\!\left\{
                \frac{n_\czin}{n} \widehat{f}_{1,\dots,6}^\czin(x),\frac{n_\nzin}{n}\widehat{f}^{\nzin}(x)
                \right\}, \\
                0, &\qquad \text{ if } \frac{n_\nzin}{n} \widehat{f}^{\nzin}(x)> \max\!\left\{
                \frac{n_\czin}{n}\widehat{f}_{1,\dots,6}^\czin(x),\frac{n_\zin}{n}\widehat{f}^{\zin}(x)
                \right\},
                \\
                
                \phi(x),  &\qquad  \text{ otherwise, }
                
            \end{array}
    \right.
\end{equation}
for each CT-based descriptor vector $x\in \R^6$.
Note, however, that the prediction $\predictionModel(x)$ for a particle with CT-based descriptor vector $x\in\R^6$ is not necessarily equal to the ``true value'' 
of  its composition descriptor $M_{\mathrm{rat}}$. Thus, in Section~\ref{sec.sec.res}, we  will  evaluate model fits and the predictive power of the prediction model introduced above.

\section{Results}\label{sec.sec.res}

Within the set of 1341 particles observed in the CT data which intersect with one of the 2D voxel slices where SEM-EDS data is available, we identified $n_{\zin}=227$ particles which almost exclusively consist of valuable minerals and $n_{\nzin}=489$ particles which contain almost no valuable minerals.  Even though many particles in the considered data set are almost purely composed of either valuable or non-valuable minerals,  see Fig.~\ref{Fig:fractions} 
for the histogram of the  VFVM, there is also a non-negligible subset of $n_{\czin}=625$  particles which contain significant fractions of both, valuable and non-valuable minerals.  
We determine the probability densities $\widehat{f}^{\zin},\widehat{f}^{\nzin}$ and $\widehat{f}^{\czin}$ of descriptor vectors using the fitting procedure described in Section~\ref{sec.seq.fit}. For that purpose, the (univariate) marginal distributions of $\widehat{f}^{\zin},\widehat{f}^{\nzin}$ and $\widehat{f}^{\czin}$ are fitted using mixed beta and mixed gamma distributions, see Fig.~\ref{fig:histograms_all}.

\if\submission0
\begin{figure}[ht]

\begin{subfigure}{.7\textwidth}

  \centering
  \includegraphics[width=0.7\linewidth]{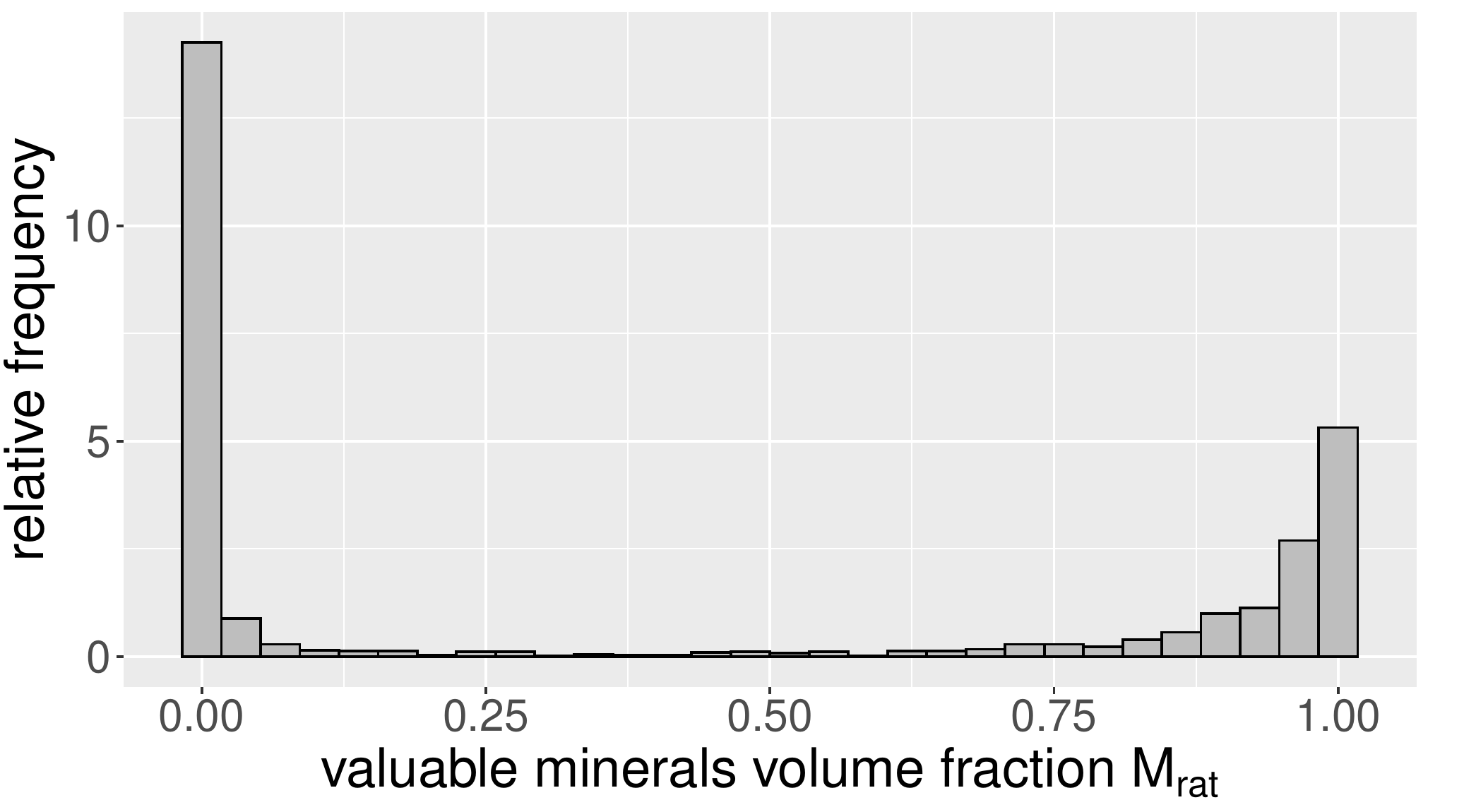}
\end{subfigure}

\caption{Histogram of the VFVM of particles intersecting with a SEM-EDS slice.  }
\label{Fig:fractions}
\end{figure}

\fi

\if\submission0

\newpage

\begin{figure}[H]

\newpage

\tikzset{every picture/.style={line width=0.75pt}} 

\begin{tikzpicture}[x=0.75pt,y=0.75pt]

\draw (113.08,77.73) node  {\includegraphics[width=139.53pt,height=70.58pt]{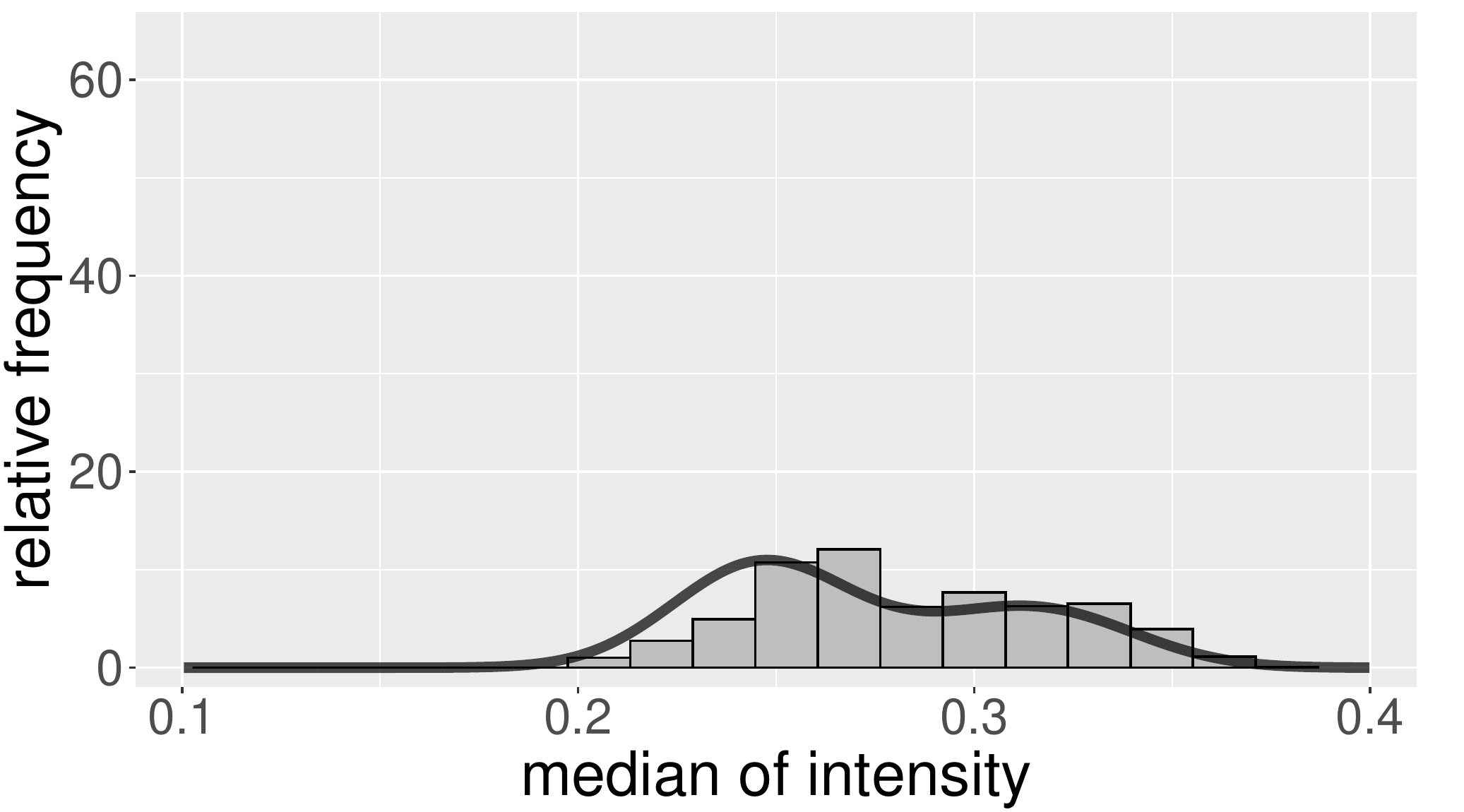}};

\draw (315.52,77.79) node  {\includegraphics[width=139.53pt,height=70.58pt]{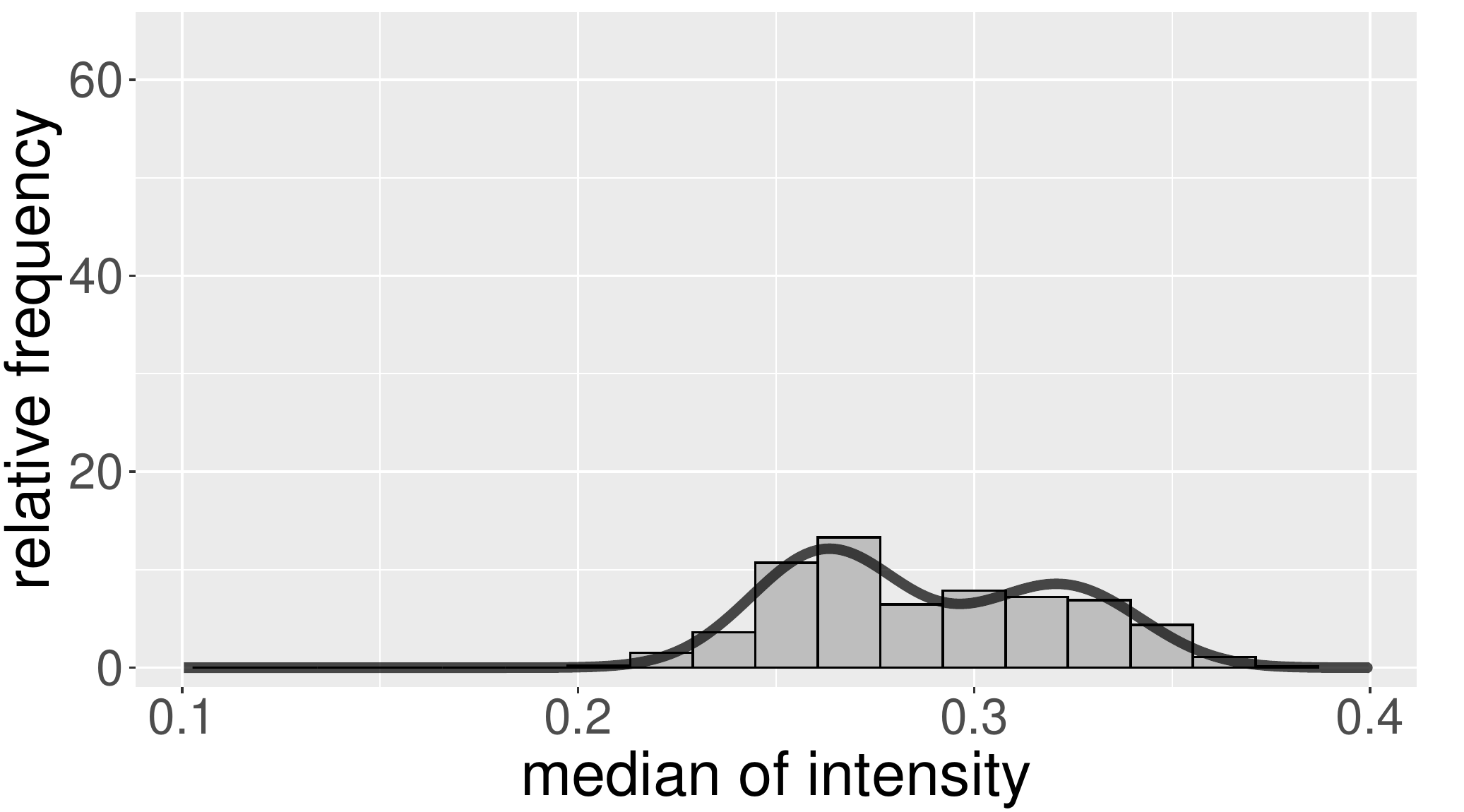}};

\draw (516.41,77.68) node  {\includegraphics[width=139.53pt,height=70.58pt]{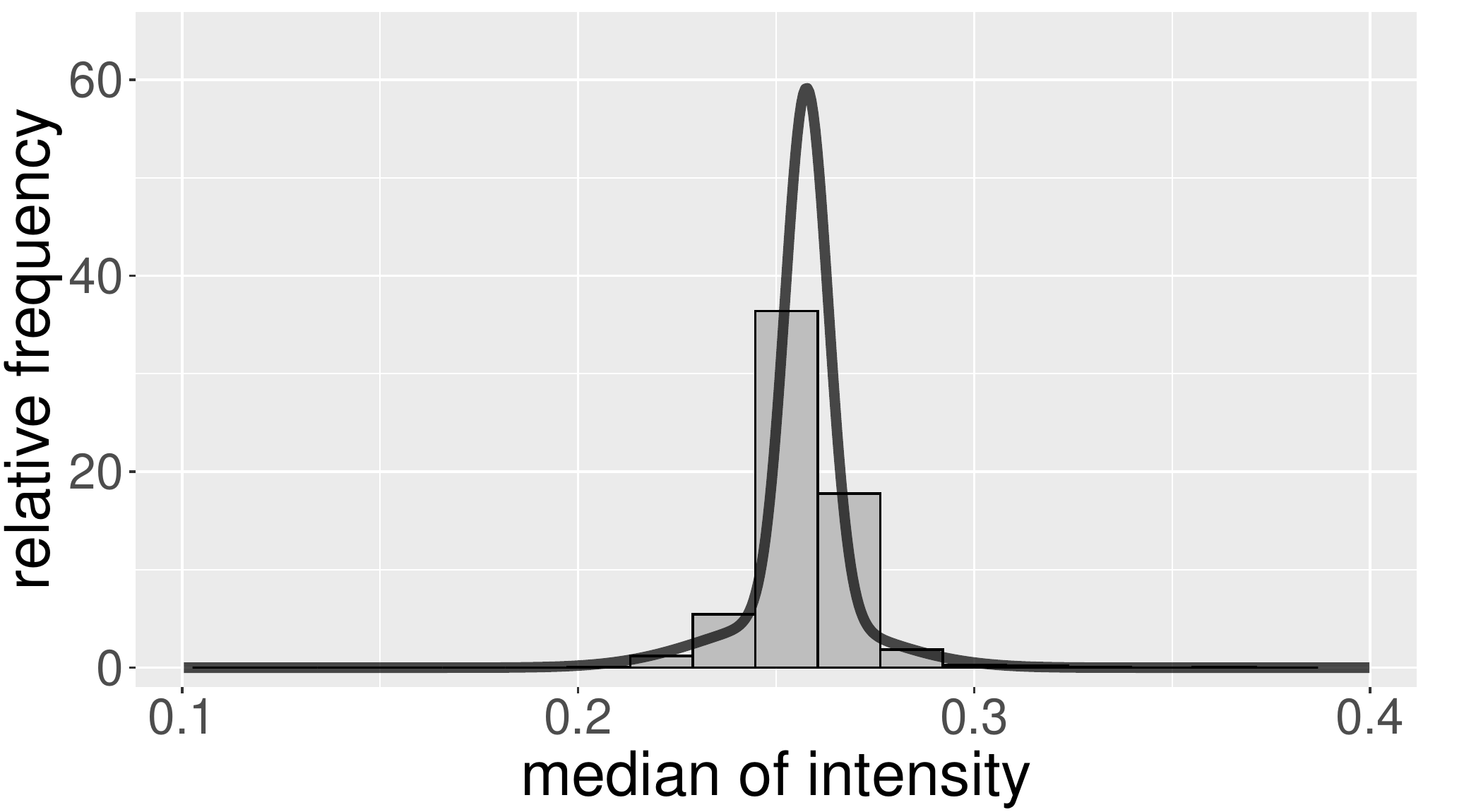}};
 
\draw (113.65,188.73) node  {\includegraphics[width=139.53pt,height=70.58pt]{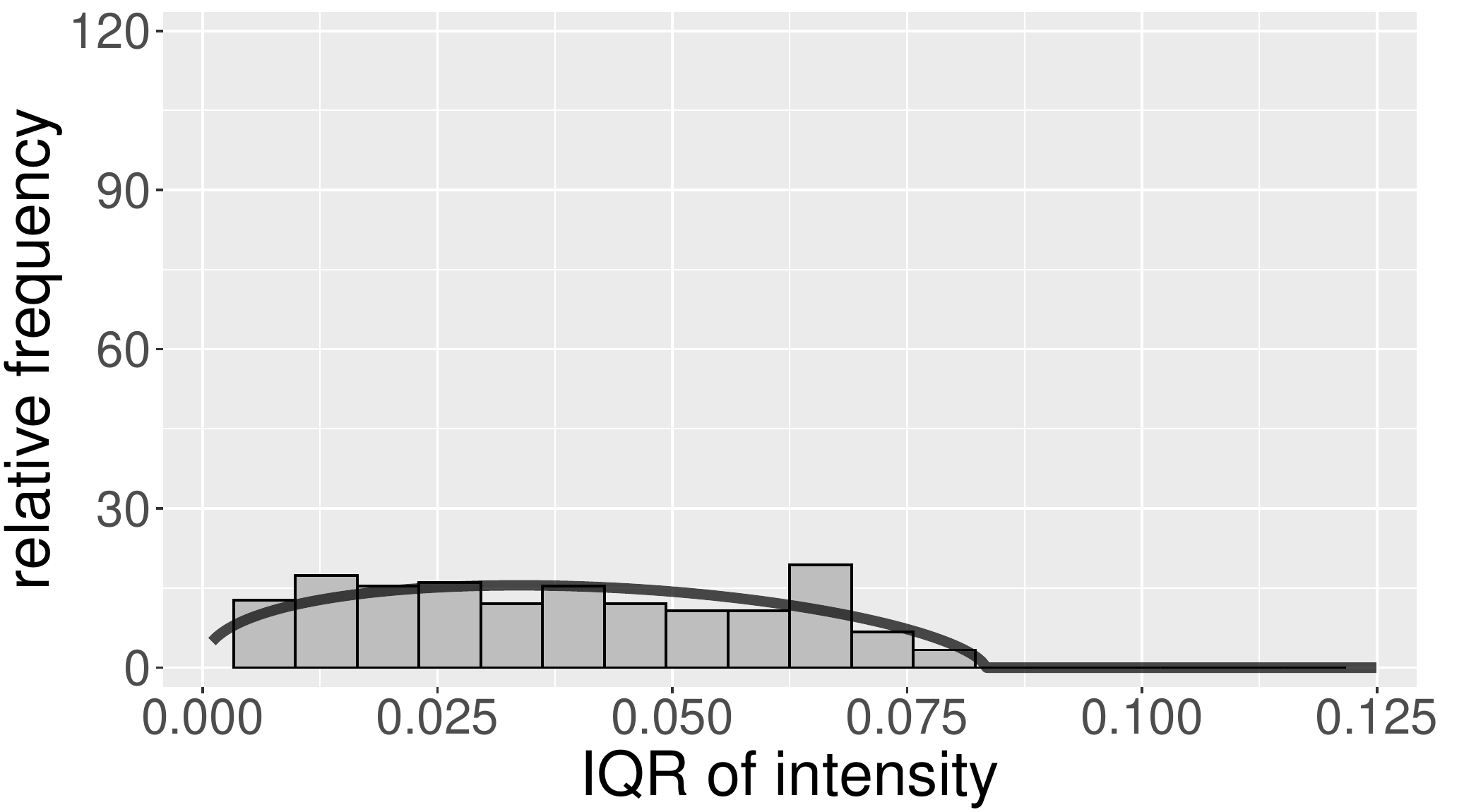}};

\draw (316.09,188.79) node  {\includegraphics[width=139.53pt,height=70.58pt]{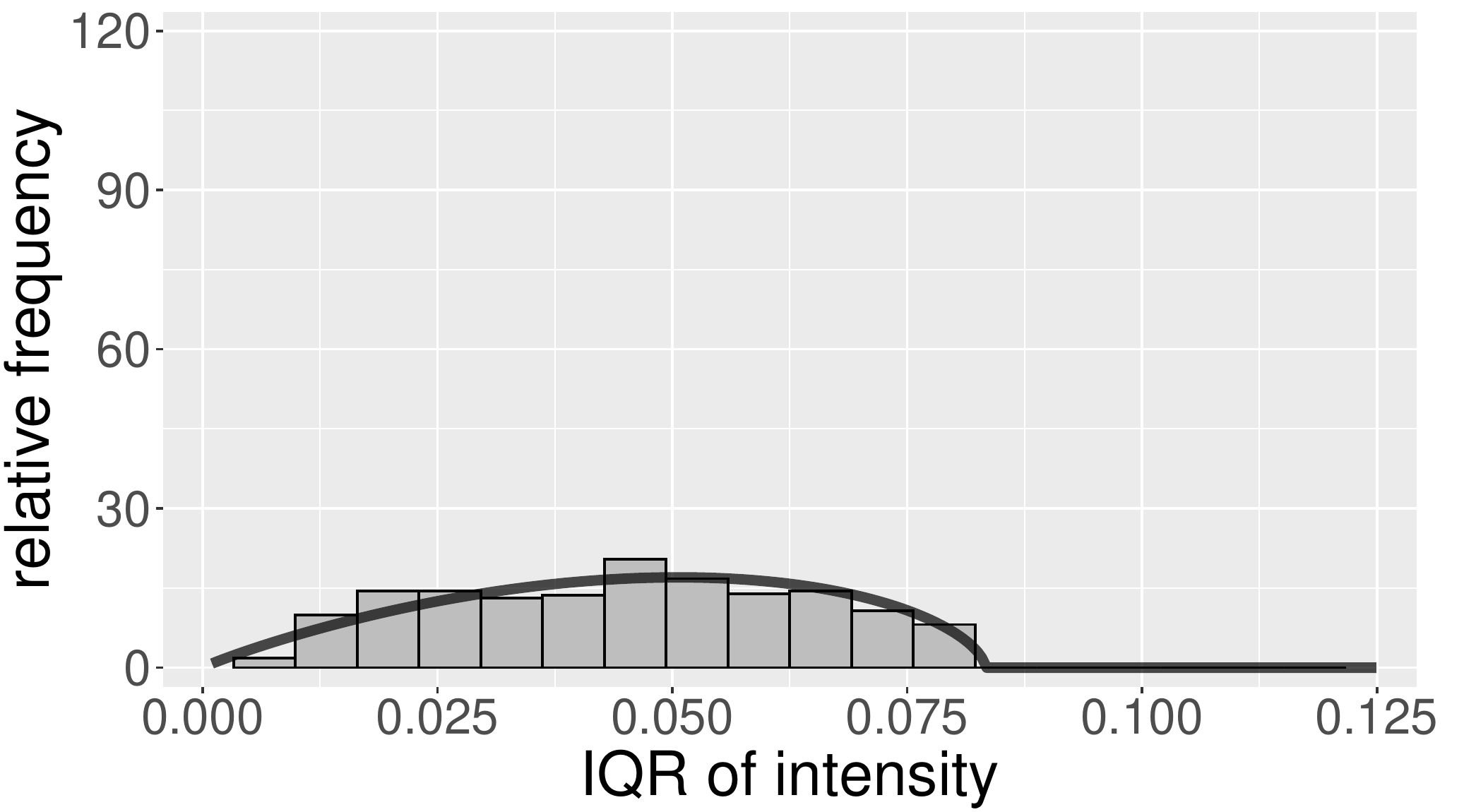}};

\draw (516.98,188.68) node  {\includegraphics[width=139.53pt,height=70.58pt]{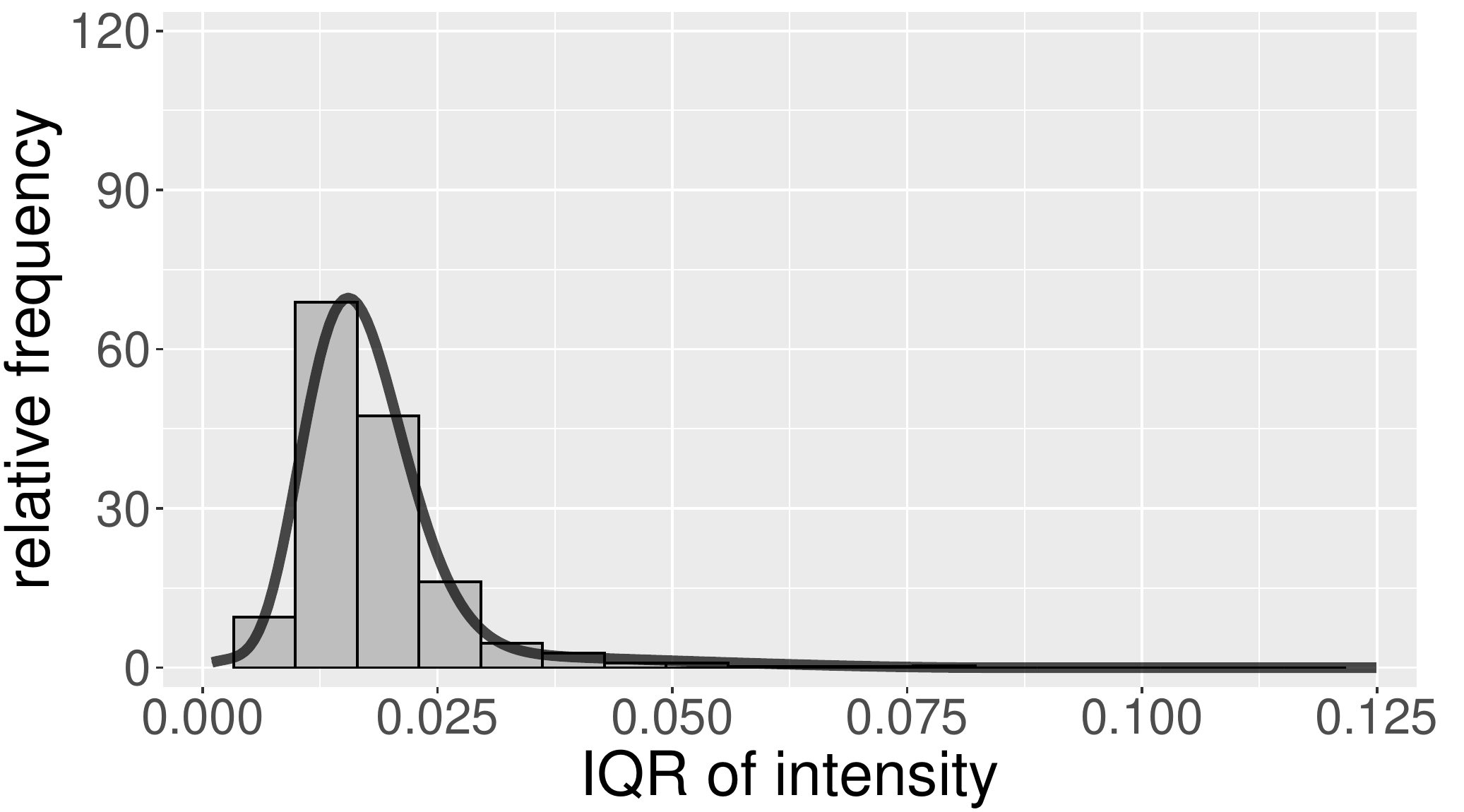}};

\draw (111.08,299.73) node  {\includegraphics[width=139.53pt,height=70.58pt]{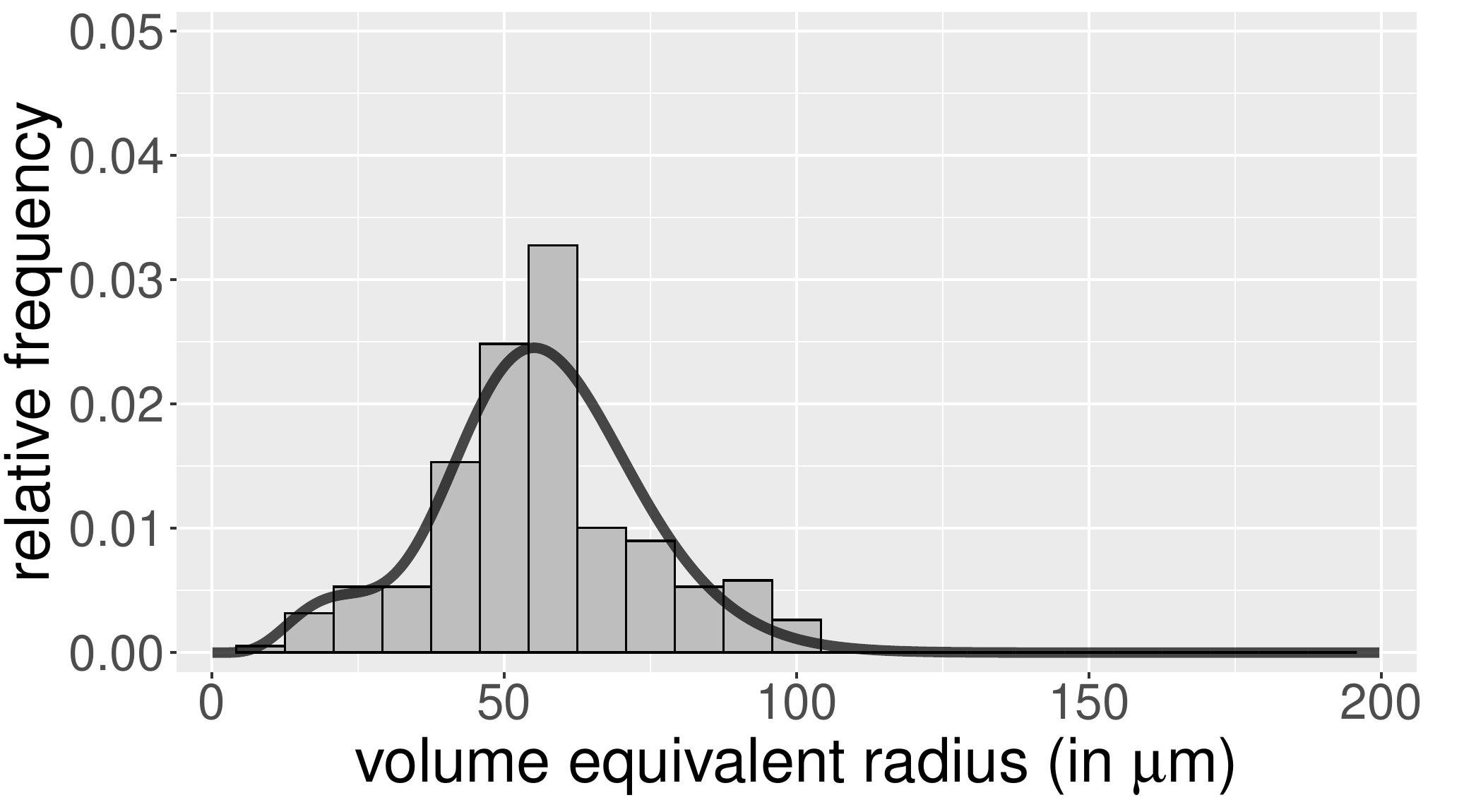}};

\draw (313.52,299.79) node  {\includegraphics[width=139.53pt,height=70.58pt]{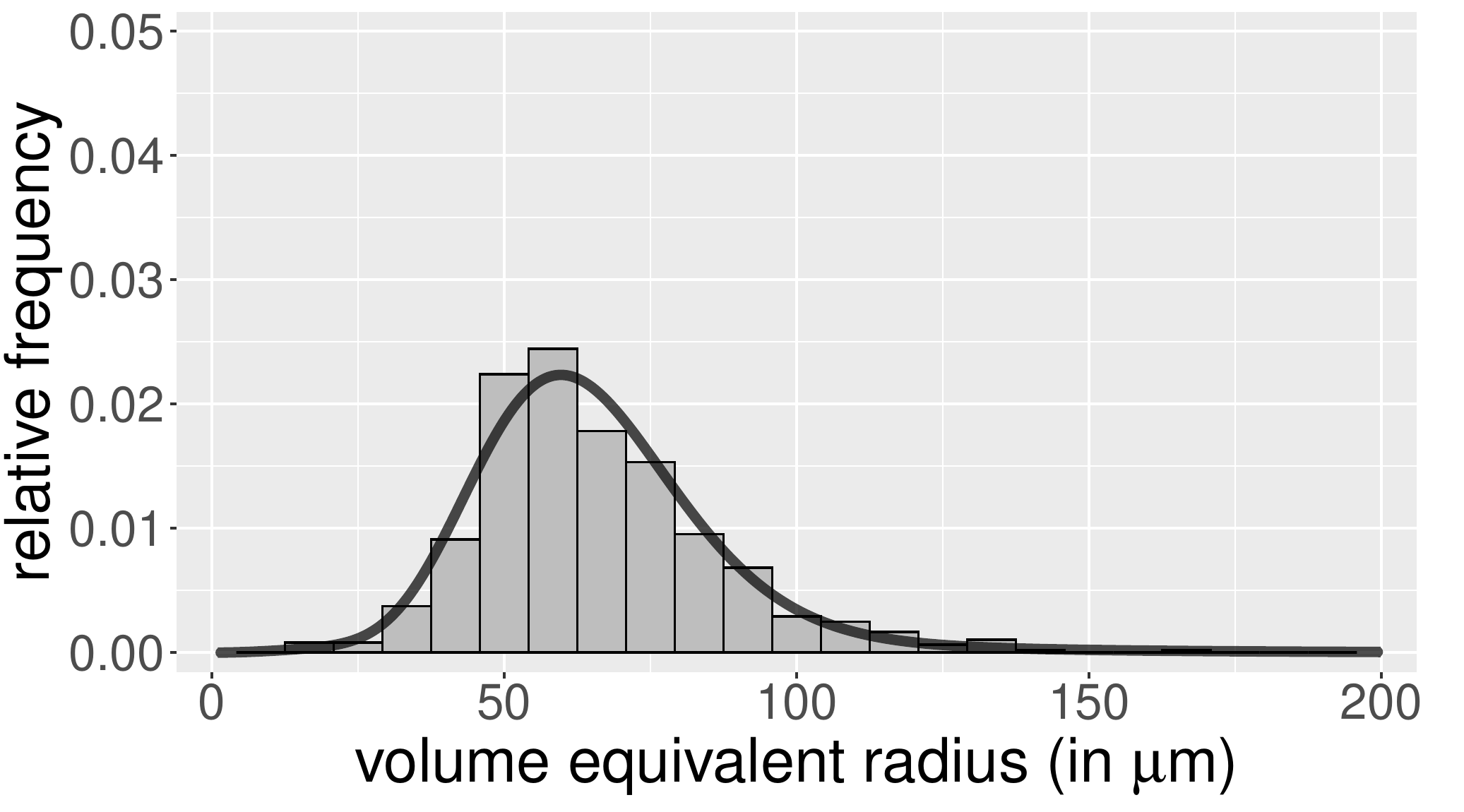}};

\draw (514.41,299.68) node  {\includegraphics[width=139.53pt,height=70.58pt]{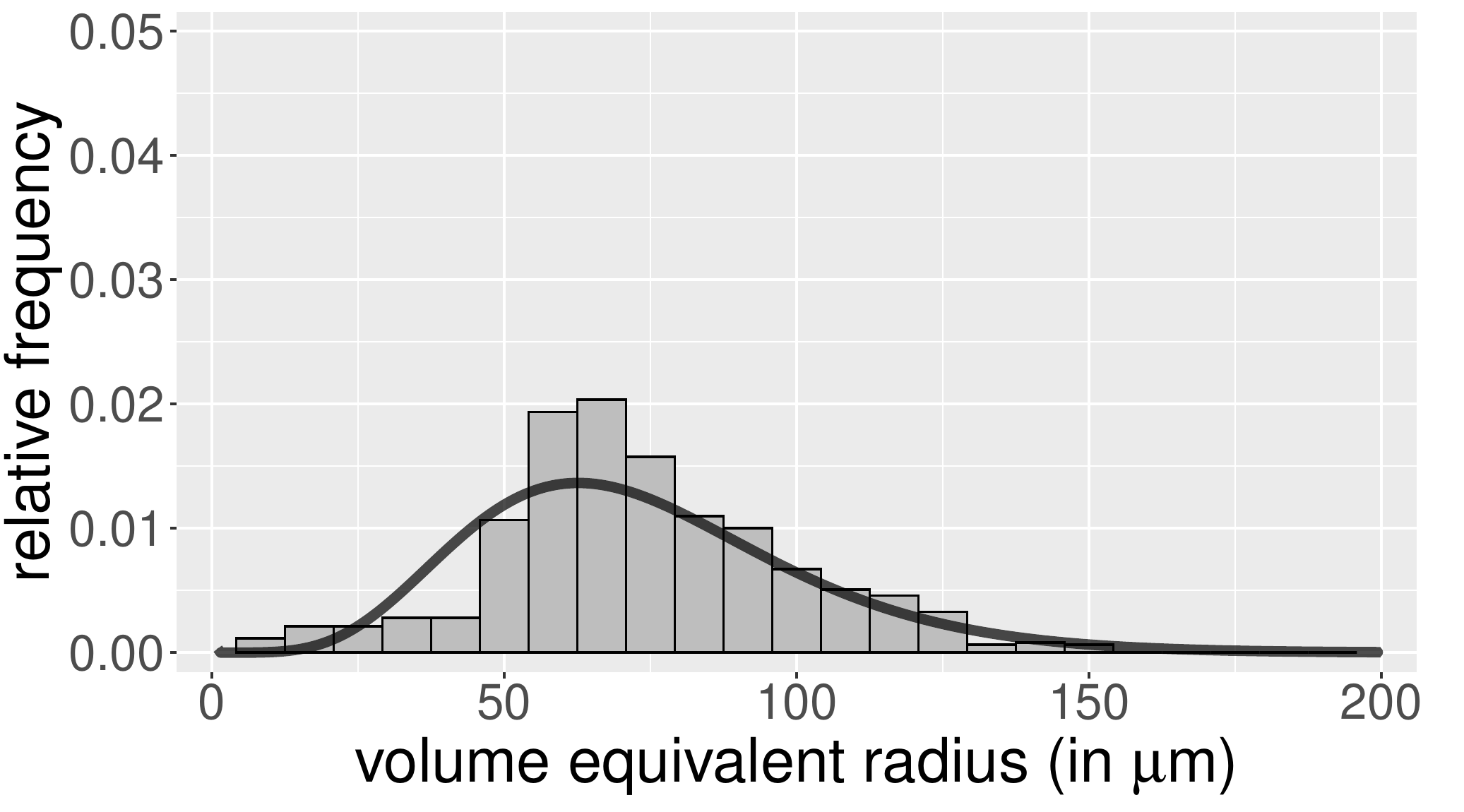}};

\draw (111.65,410.73) node  {\includegraphics[width=139.53pt,height=70.58pt]{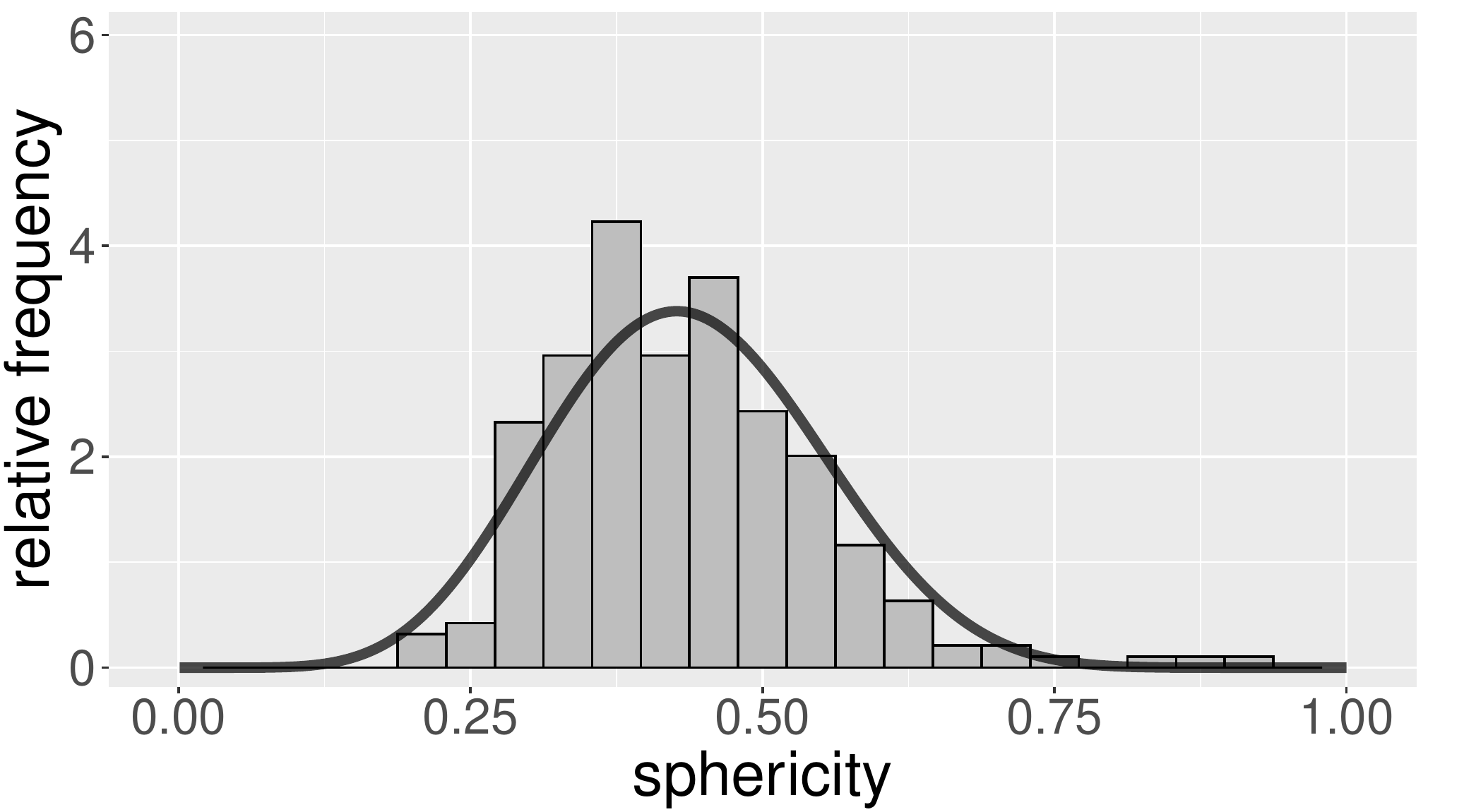}};
 
\draw (314.09,410.79) node  {\includegraphics[width=139.53pt,height=70.58pt]{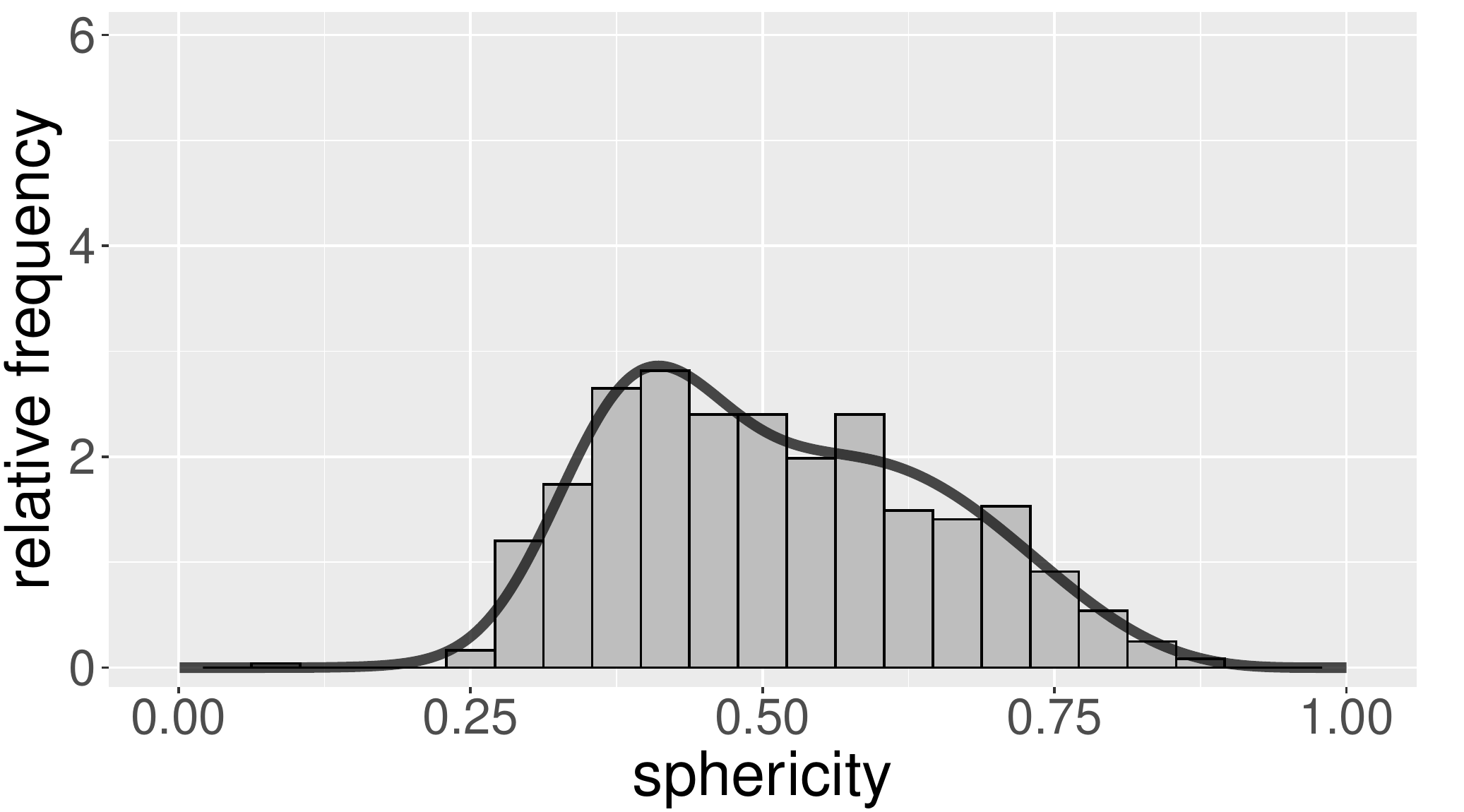}};
 
\draw (514.98,410.68) node  {\includegraphics[width=139.53pt,height=70.58pt]{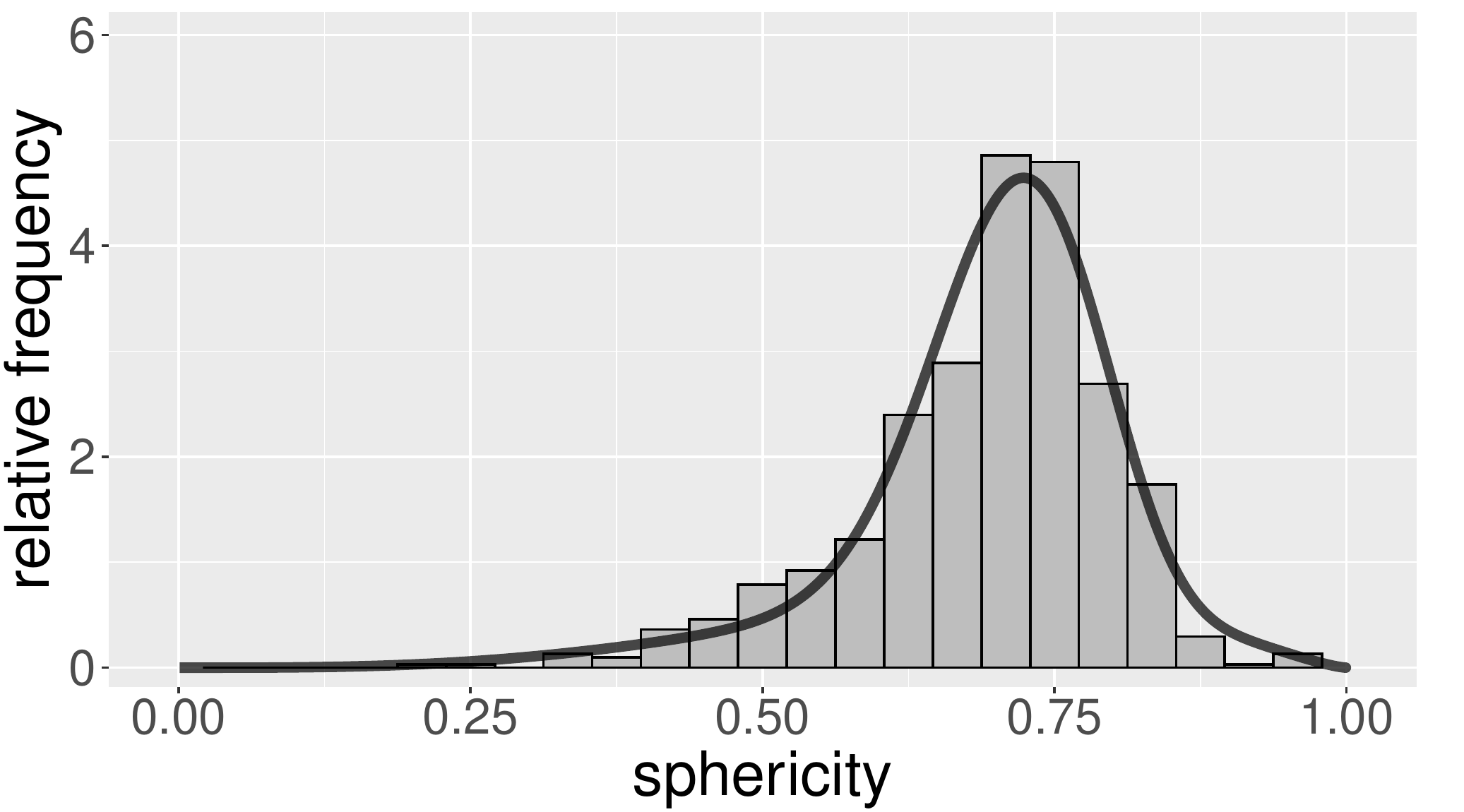}};
 
\draw (113.65,521.73) node  {\includegraphics[width=139.53pt,height=70.58pt]{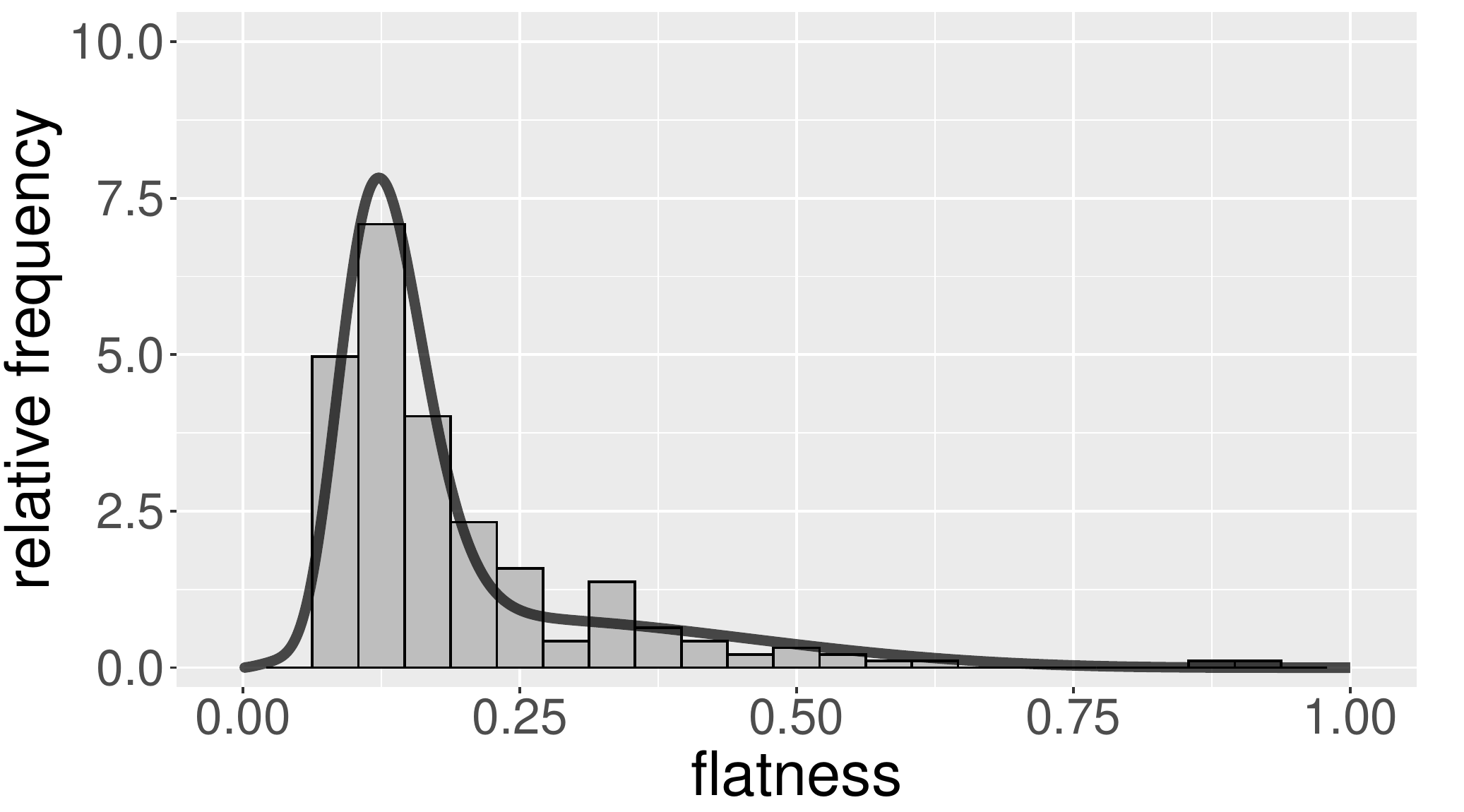}};
 
\draw (316.09,521.79) node  {\includegraphics[width=139.53pt,height=70.58pt]{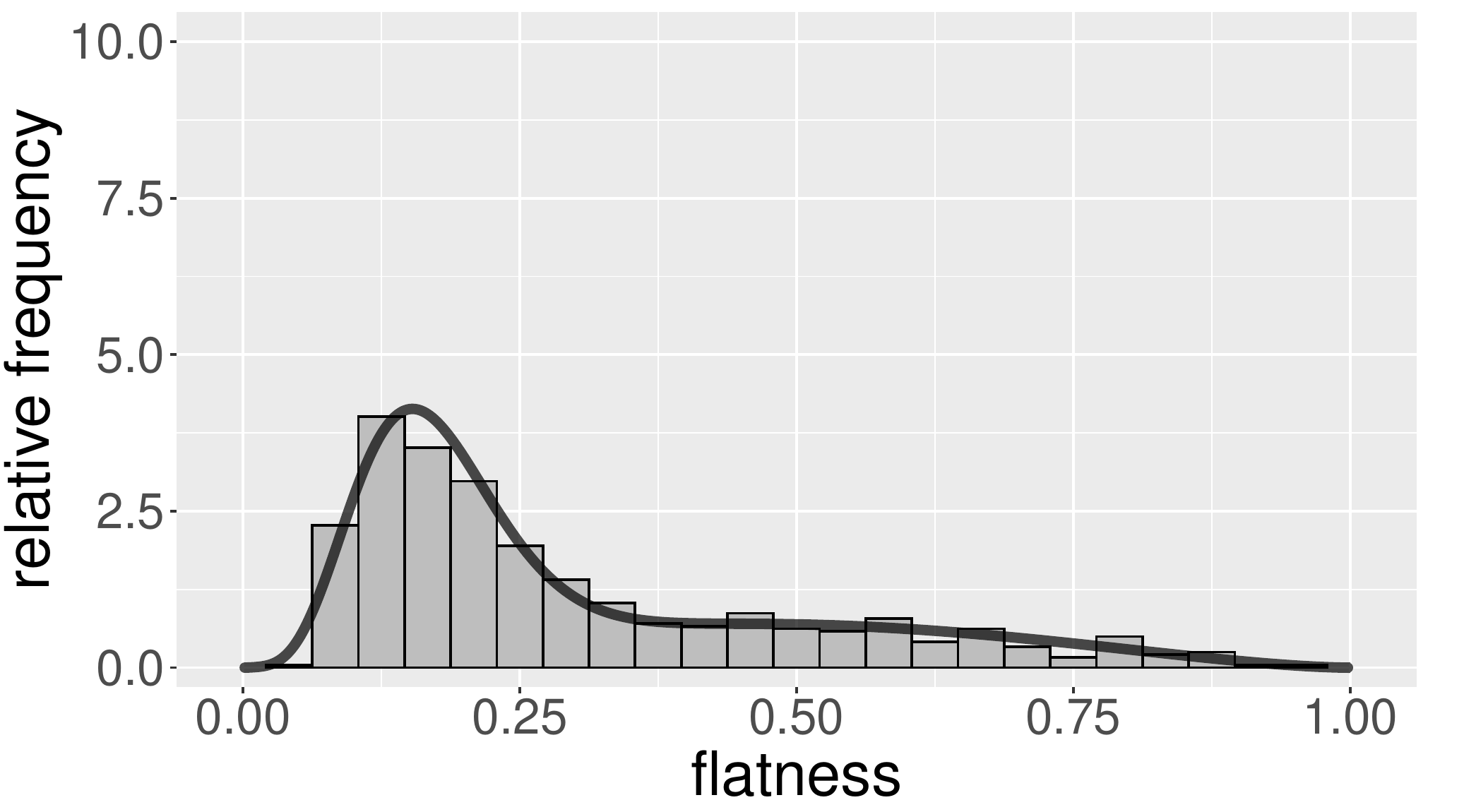}};
 
\draw (516.98,521.68) node  {\includegraphics[width=139.53pt,height=70.58pt]{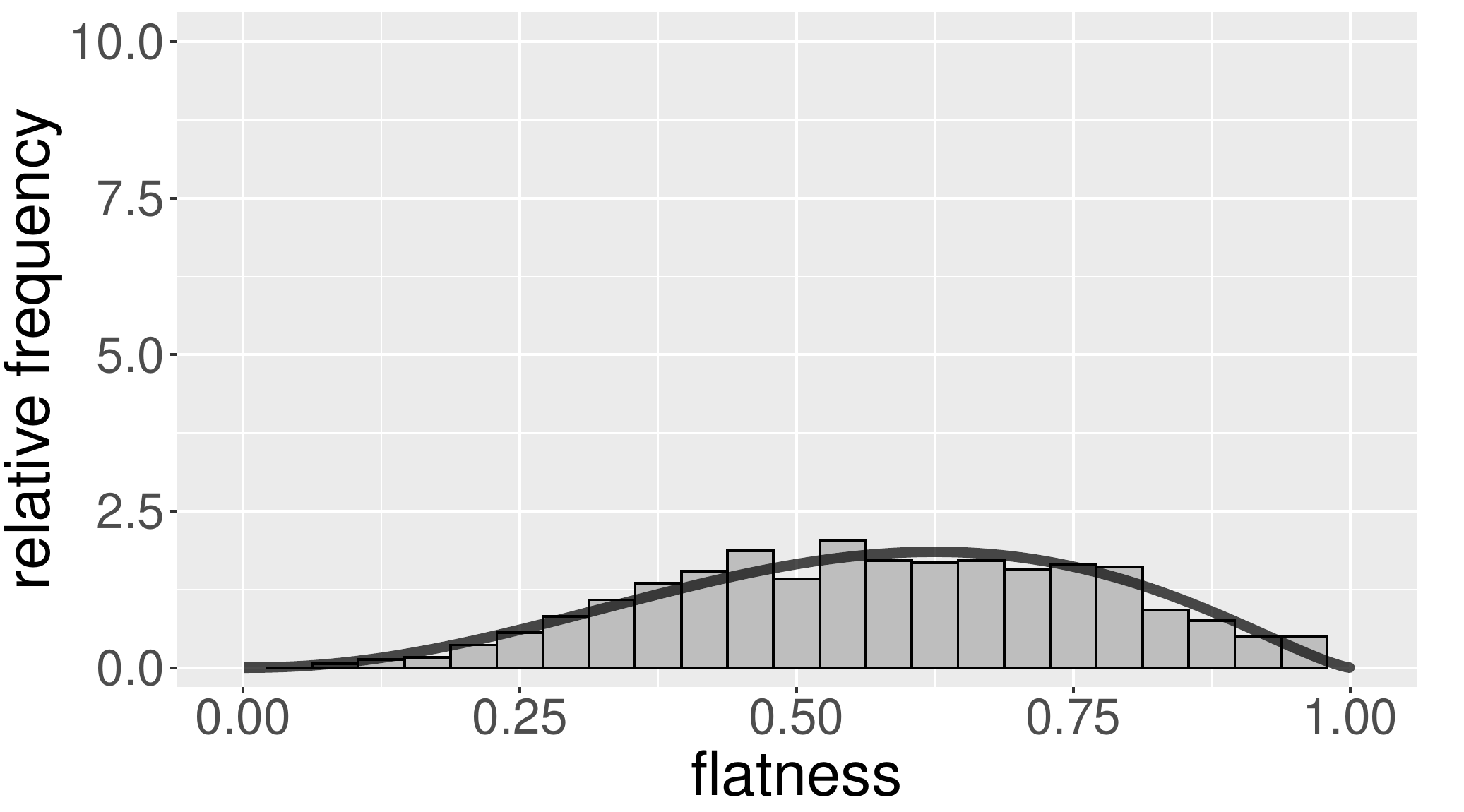}};
 
\draw (111.08,632.73) node  {\includegraphics[width=139.53pt,height=70.58pt]{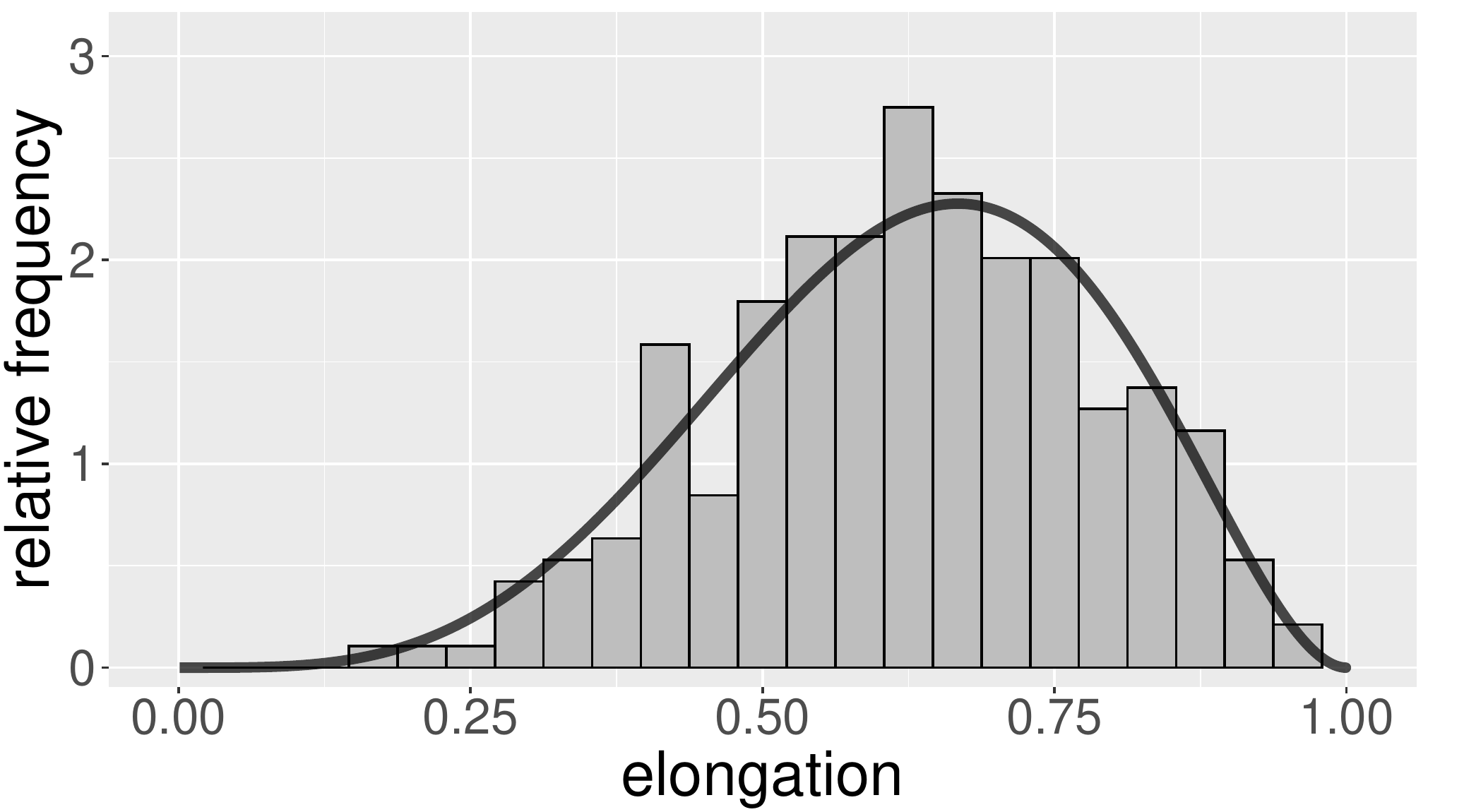}};
 
\draw (313.52,632.79) node  {\includegraphics[width=139.53pt,height=70.58pt]{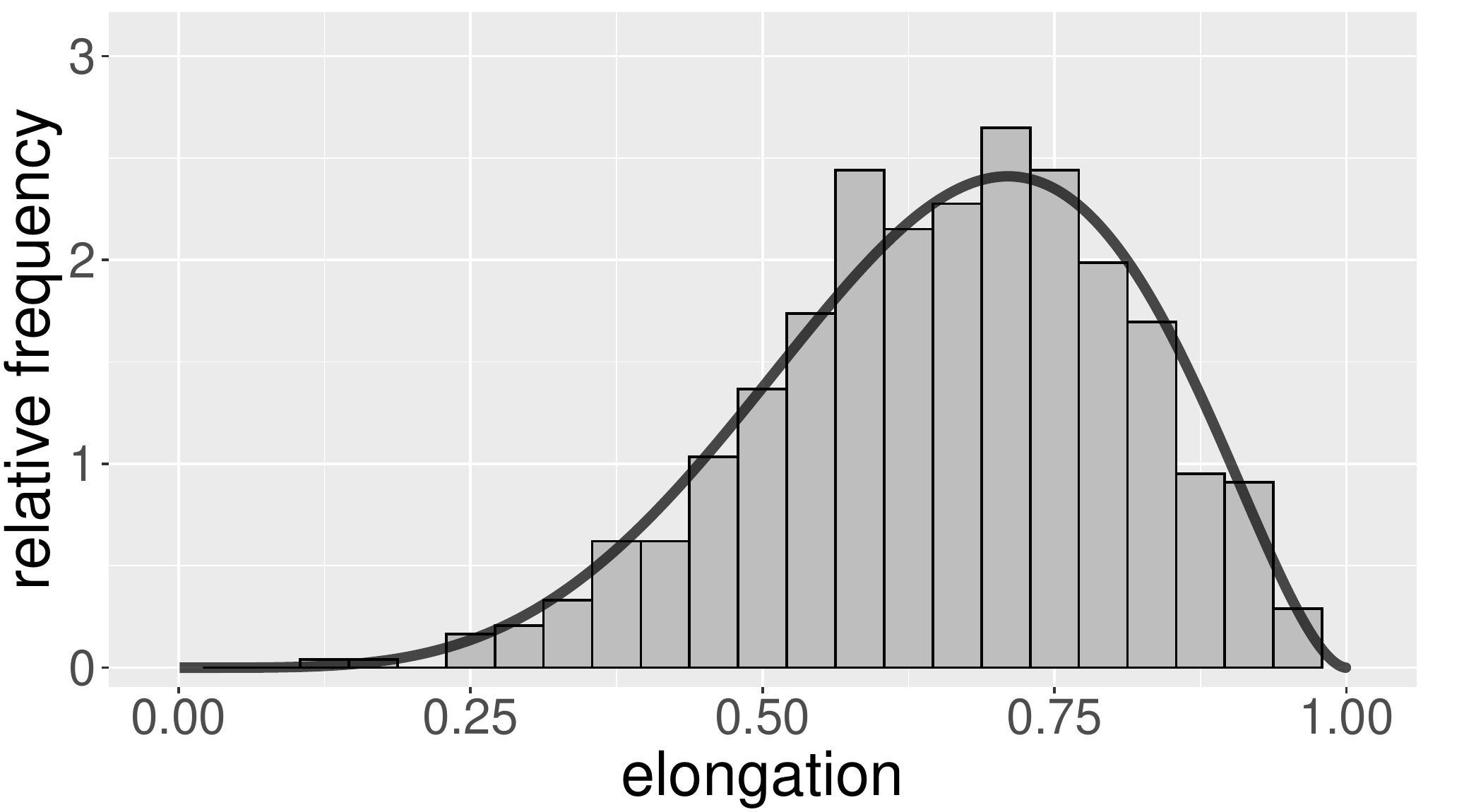}};
 
\draw (514.41,632.68) node  {\includegraphics[width=139.53pt,height=70.58pt]{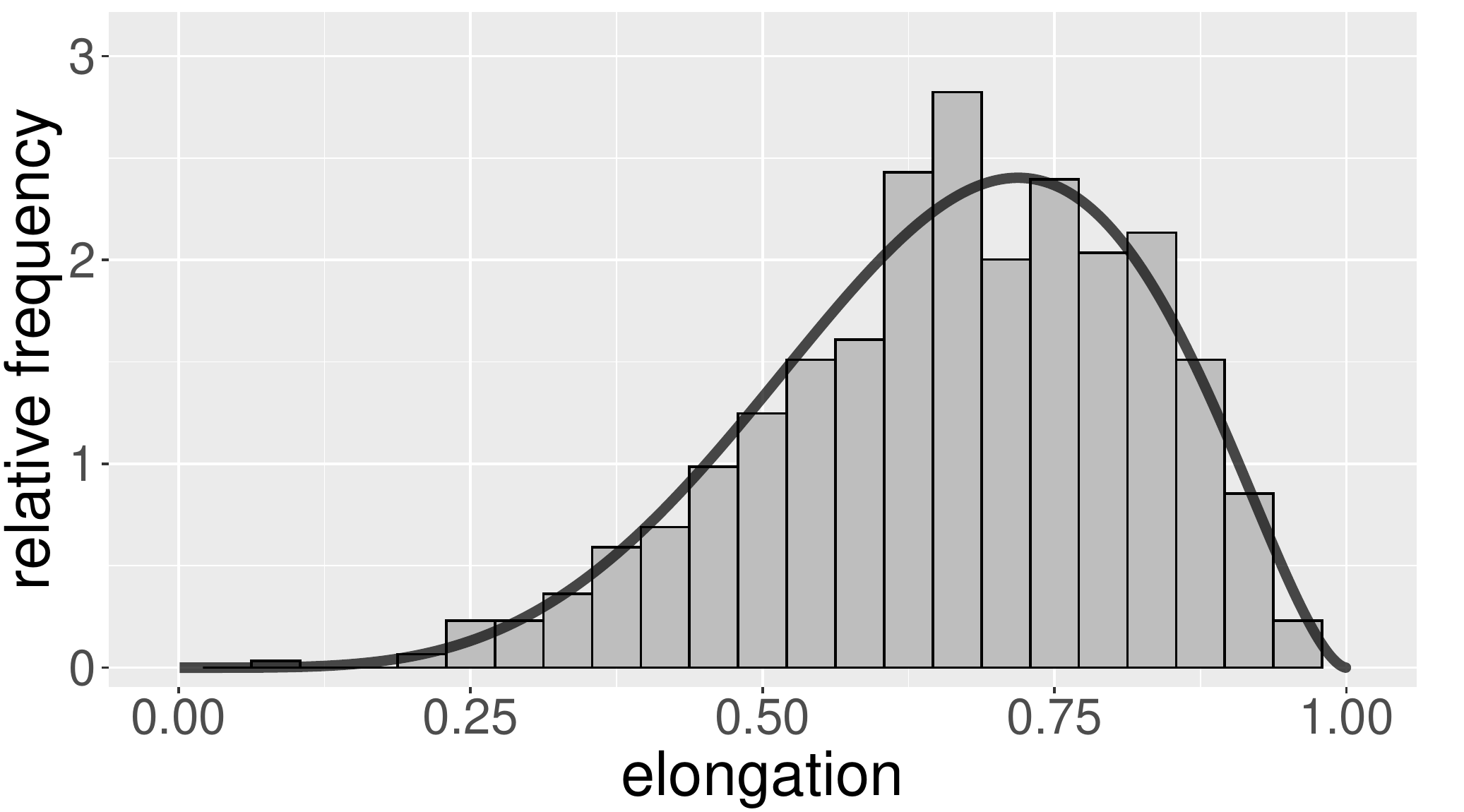}};

\end{tikzpicture}

\caption{Histograms and fitted marginal probability densities of shape, size and texture descriptors of particles  which almost exclusively consist of valuable minerals (left), particles which contain almost no valuable minerals (right) and  particles which contain significant fractions of both, valuable and non-valuable minerals (middle). }
\label{fig:histograms_all}
\end{figure}
\fi

 Furthermore, similar to the modeling approach considered in~\cite{furat2019}, we also use the (simpler) model of Archimedean copulas for the fitting of the multivariate probability densities $\widehat{f}^{\nzin},\widehat{f}^{\zin}$ and $\widehat{f}^{\czin}$, in addition to using R-vine copulas.
To evaluate the model fit, we investigate various scores in order to compare the accuracy of the fitted density $\widehat{f}$  achieved either by means of R-vine or  Archimedean copulas. First, we consider the log-likelihood $\mathbb{L} $ given by $ \mathbb{L} =\sum_{x\in D}\mathrm{ln}\bigl(\widehat{f}(x)\bigr)$.

Furthermore, we consider the  Akaike information criterion 
$\mathrm{AIC} =2k -2\mathbb{L}  $ and the Bayesian information criterion  $ \mathrm{BIC} =k\, \mathrm{ln}(n)-2\mathbb{L} $,
where $k $ denotes the number of model parameters of  $\widehat{f}$ and $n$ is the cardinality of $D$.

\if\submission0
     
\begin{figure}[ht]

\tikzset{every picture/.style={line width=0.75pt}} 
\begin{tikzpicture}[x=0.75pt,y=0.75pt,yscale=-1,xscale=1]

\draw (161.52,91.83) node  {\includegraphics[width=231.46pt,height=115.25pt]{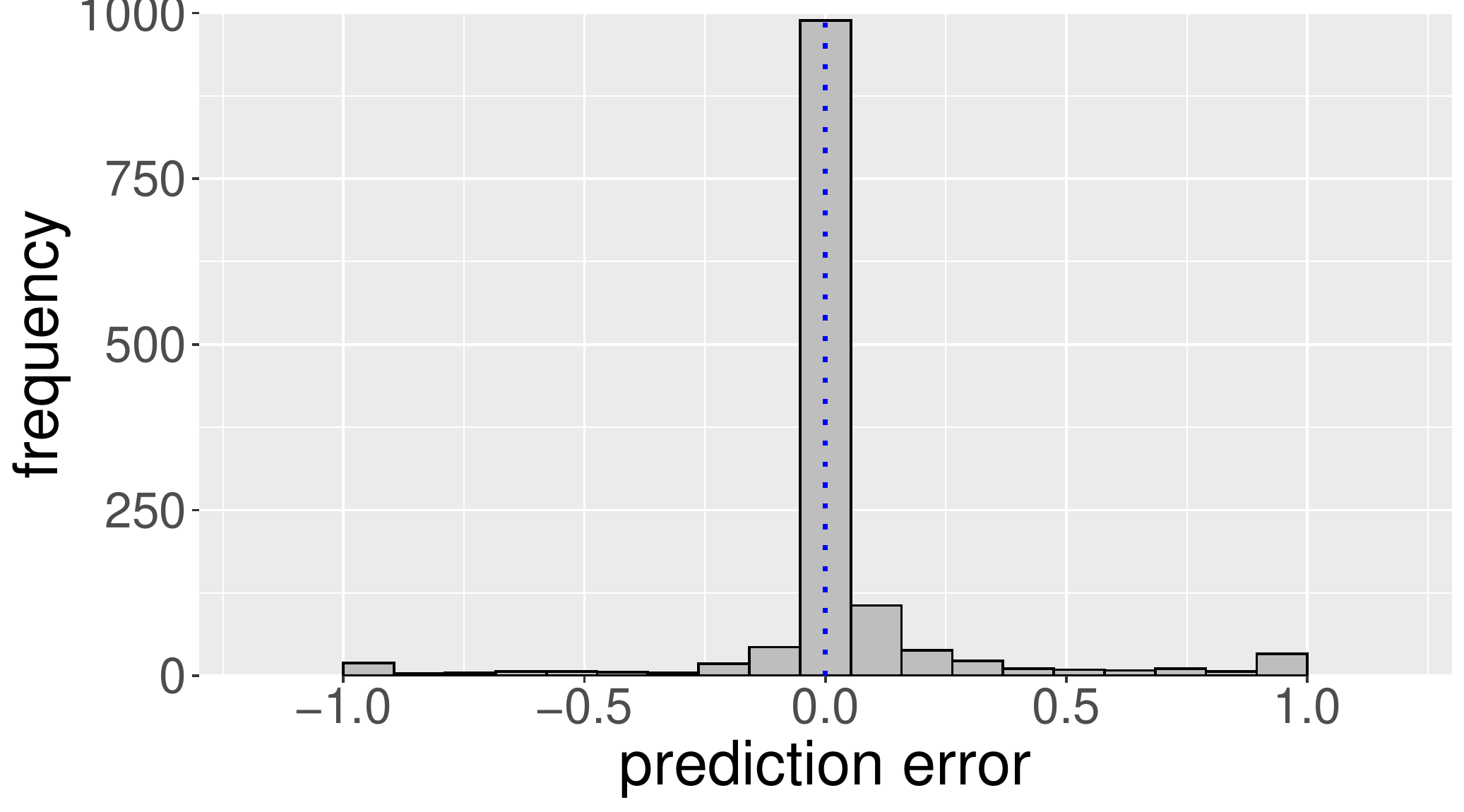}};

\draw (460.7,91.5) node  {\includegraphics[width=231.46pt,height=115.25pt]{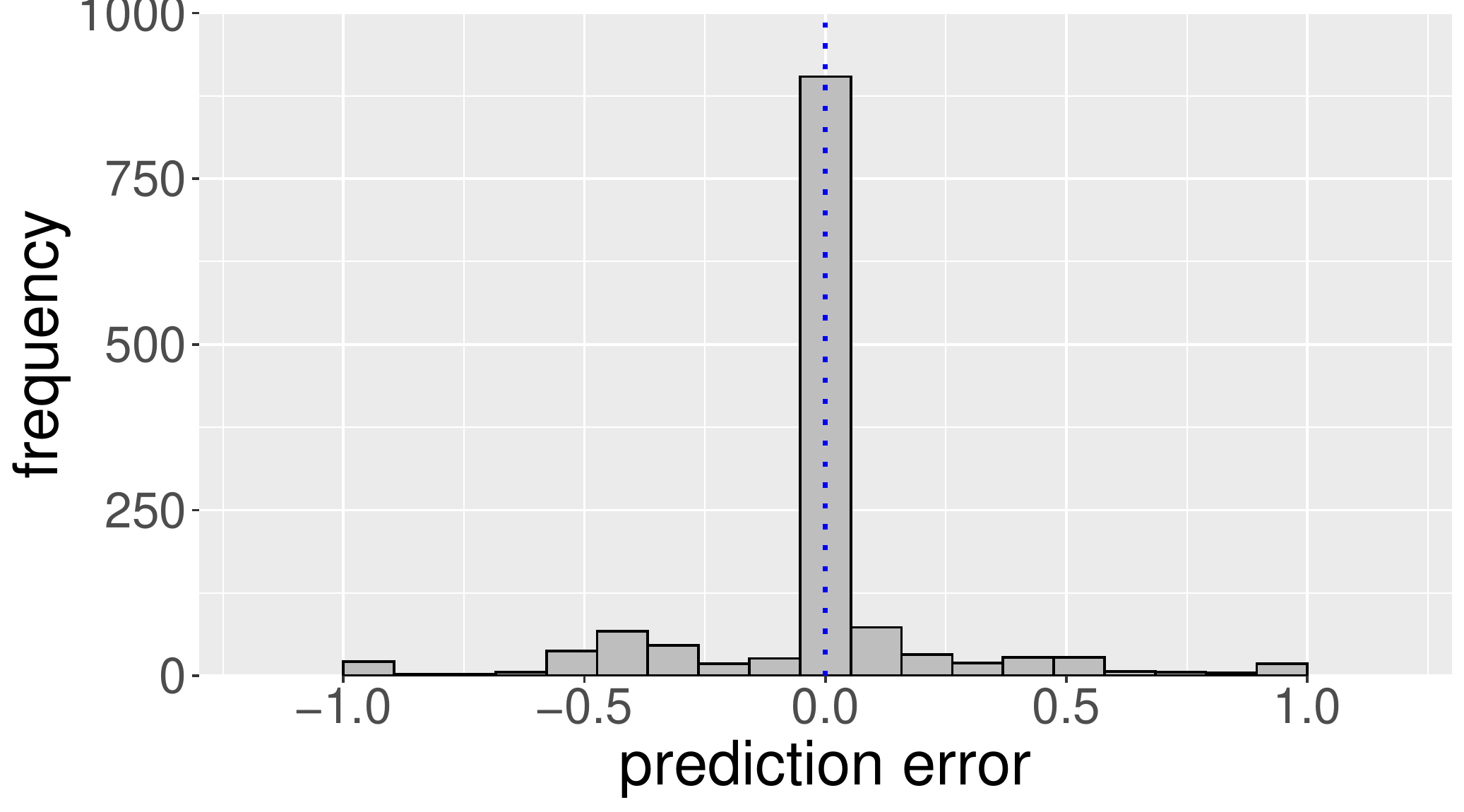}};

\end{tikzpicture}

 \caption{
 Histograms of prediction errors
  $g_{D_\czin\setminus\{x\}}(x_{1,\dots,6})-x_7$  resulting from the use of R-vine  (left) and Archimedean  (right) copulas, computed for each particle descriptor vector $x=(x_{1,\dots,6},x_7)\in D$.
 }
    \label{fig:pred_all}

\end{figure}

\fi

To evaluate the power of the  prediction model $\predictionModel$ we use leave-one-out cross-validation  
to obtain valid prediction scores. For each particle descriptor vector $x=(x_{1,\dots,6},x_7)\in D$,
the probability densities
 $\widehat{f}^{\nzin},\widehat{f}^{\zin}$ and $\widehat{f}^{\czin}$
required for the calibration of the prediction model $\predictionModel$ are fitted on the data set $D\setminus\{x\}$, and the resulting prediction for the CT-based descriptor vector $x_{1,\dots,6}$, denoted by $\predictionModel_{D\setminus\{x\}}(x_{1,\dots,6}),$ is compared to the actual \composition{} $x_7$. In Fig.~\ref{fig:pred_all}  histograms of the discrepancies $\predictionModel_{D\setminus\{x\}}(x_{1,\dots,6})-x_7$ are shown, where
the prediction model $\predictionModel$ fitted by means of R-vine copulas seems to lead to better results than the model obtained by Archimedean copulas.
Additionally, the predictive power is quantified by means  of the  mean absolute error MAE and the mean squared error MSE, given by

\begin{equation*}
    \mathrm{MAE} =
    \frac{1}{n}
    \sum_{x\in  D}\left|\predictionModel_{D\setminus\{x\}}(x_{1,\dots,6})-x_7\right|\quad\quad\text{and}\quad\quad
    \mathrm{MSE} =\frac{1}{n}\sum_{x\in  D}{\left(\predictionModel_{D\setminus\{x\}}(x_{1,\dots,6})-x_7\right)^2}.
\end{equation*}
 Again, the scores for the prediction model $\predictionModel$ fitted by means of R-vine copulas are better than those obtained by Archimedean copulas, see Table~\ref{tab:scores_all}.

 \if\submission0
 \begin{table}[ht]
        \centering

\begin{tabular}{p{0.24\textwidth}|p{0.24\textwidth}|p{0.3\textwidth}}

  & using R-vine copulas & using Archimedean copulas \\
\hline 
 log-likelihood $\mathbb{L}$ & \textcolor[rgb]{0,0,0}{\hspace{1.2mm}\bf{4963.41}} & \hspace{0.8mm} 4034.87 \\
\hline 
 AIC & \bf{-9526.82} & -7867.74 \\
\hline 
 BIC & \bf{-8486.58} & -7342.42 \\
 \hline 
  MAE & \hspace{1.05mm}\bf{0.0990} & \hspace{1.05mm}0.1304 \\
\hline 
  MSE & \hspace{0mm} \bf{0.0622} & \hspace{0mm} 0.0707\\
 
\end{tabular}

        \caption{
        Validation scores of the prediction model $\predictionModel$ calibrated by means of vine   (left) and  Archimedean (right) copulas, where the scores were computed on the data set $D$ of particle descriptor vectors.        }
    \label{tab:scores_all}
\end{table}

\fi

Finally, to evaluate model fit and prediction power   with respect to composite particles, we additionally consider scores which are computed on the set $D_{\czin}$ only. For that purpose, we determine  the log-likelihood $   \mathbb{L}_\czin$ given by
$   \mathbb{L}_\czin=\sum_{x\in D_\czin}\mathrm{ln}\bigl(\widehat{f}^{\czin}(x)\bigr)$ as well as
 the corresponding   Akaike and Bayesian information criteria 
$\mathrm{AIC}_{\czin}=2k_{\czin}-2\mathbb{L}_{\czin} $ and  $ \mathrm{BIC}_{\czin}=k_{\czin}\,\mathrm{ln}(n_\czin)-2\mathbb{L}_{\czin}$.
Moreover, for all $x=(x_{1,\dots,6},x_7)\in D_{\czin}$, the histograms of the prediction errors $\predictionModel_{D\setminus\{x\}}(x_{1,\dots,6})- x_7$ are determined, 
and the  mean (absolute/squared) errors MAE$_{\czin}$ and  $\mathrm{MSE}_{\czin}$ are computed, which are given by

\begin{equation*}
    \mathrm{MAE}_\czin=
    \frac{1}{n_\czin}
    \sum_{x\in  D_\czin}\left|\predictionModel_{D\setminus\{x\}}(x_{1,\dots,6})-x_7\right|\qquad\text{and}\qquad
    \mathrm{MSE}_{\czin}=\frac{1}{n_\czin}\sum_{x\in  D_\czin}{\left(\predictionModel_{D\setminus\{x\}}(x_{1,\dots,6})-x_7\right)^2}.
\end{equation*}

The obtained results are shown in Fig~\ref{fig:pred_mix} and
Table~\ref{tab:scores_mix}.

\if\submission0
\tikzset{every picture/.style={line width=0.75pt}} 
\begin{figure}[ht]

\begin{tikzpicture}[x=0.75pt,y=0.75pt,yscale=-1,xscale=1]

\draw (161.52,91.83) node  {\includegraphics[width=231.46pt,height=115.25pt]{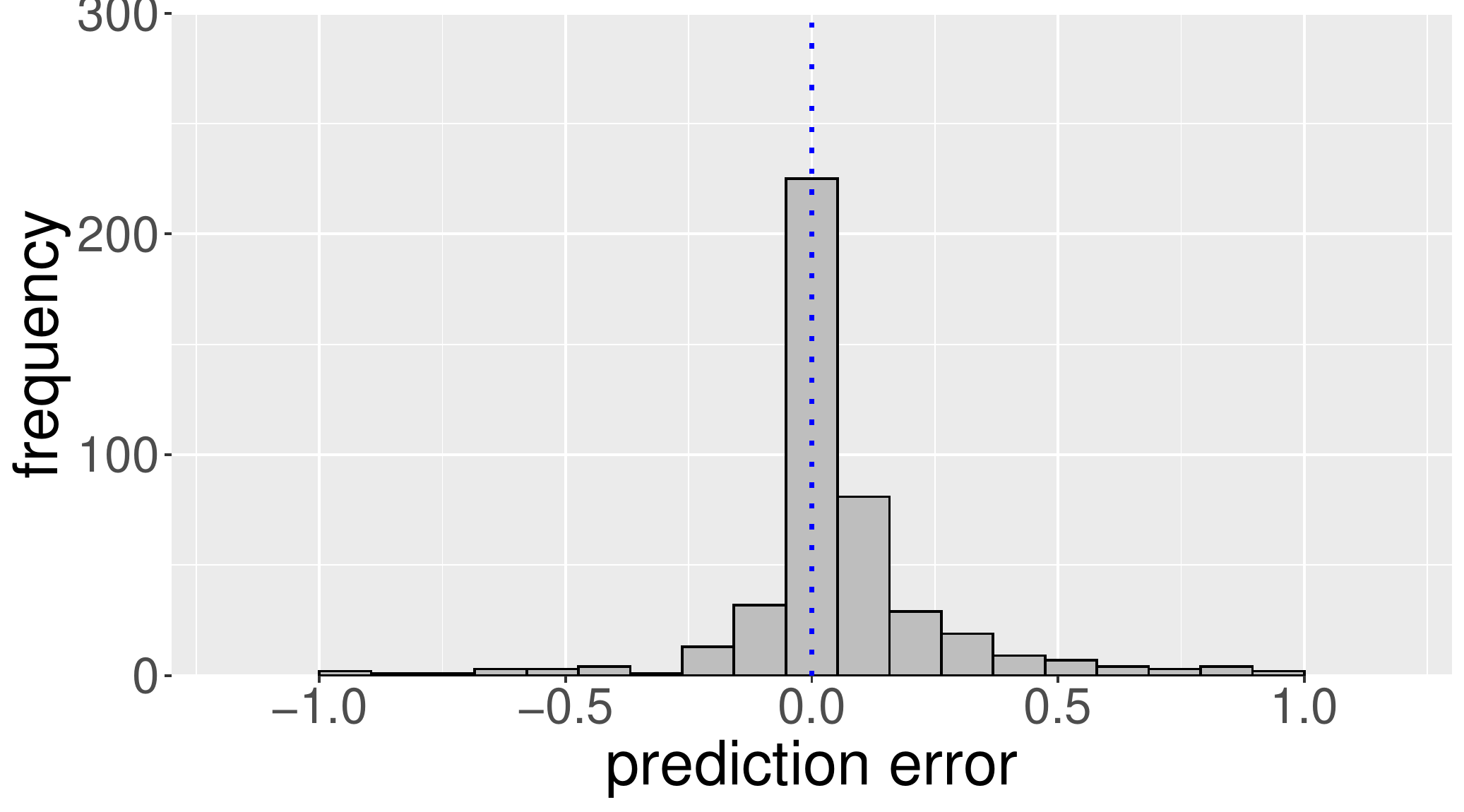}};

\draw (460.7,91.5) node  {\includegraphics[width=231.46pt,height=115.25pt]{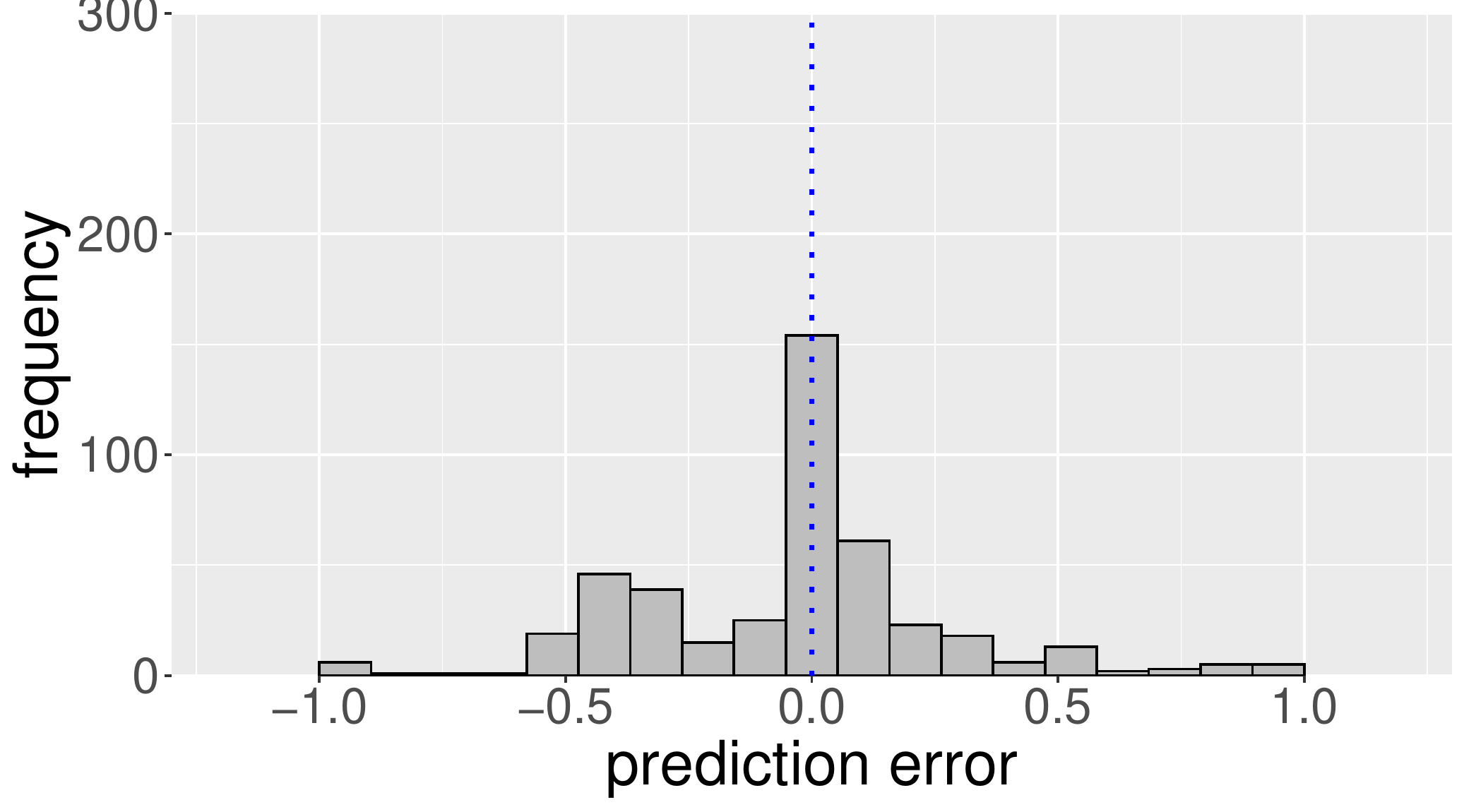}};
\end{tikzpicture}

 \caption{Histograms of prediction errors
  $\predictionModel_{D\setminus\{x\}}(x_{1,\dots,6})-x_7$  resulting from the use of R-vine  (left) and Archimedean  (right) copulas, computed for each particle descriptor vector $x=(x_{1,\dots,6},x_7)\in D_{\czin}$.   }
    \label{fig:pred_mix}

\end{figure}
\fi

Note that all scores  presented in
Table~\ref{tab:scores_mix}  for the density $\widehat{f}^{\czin}$ fitted on $D_{\czin}$  and for the corresponding prediction models  $\predictionModel_{D\setminus\{x\}}$  improve if  R-vine copulas (instead of  Archimedean copulas) are used for model fitting. In particular, 
the
scores for the log-likelihood $   \mathbb{L}_\czin$ and for the
 Akaike and Bayesian information criteria 
$\mathrm{AIC}_{\czin}$ and  $ \mathrm{BIC}_{\czin}$
obtained
by means of R-vine copulas are clearly better than those obtained by Archimedean copulas.

\if\submission0
 \begin{table}[ht]
        \centering
     
\begin{tabular}{p{0.24\textwidth}|p{0.24\textwidth}|p{0.3\textwidth}}

  & using R-vine copulas & using Archimedean copulas \\
\hline 
 log-likelihood  $\mathbb{L}_\czin$& \textcolor[rgb]{0,0,0}{\hspace{1.2mm}\bf{1315.06}} & 725.51 \hspace{0.8mm}  \\
\hline 
 $\mathrm{AIC}_{\czin}$ & \bf{-2454.12
} & -1377.02
 \\
\hline 
 $\mathrm{BIC}_{\czin}$ & \bf{-2093.88} & -1225.55 \\
 \hline 
   $\mathrm{MAE}_\czin$ & \hspace{1.05mm}\bf{0.1378} & \hspace{1.05mm}0.2153 \\
\hline 
 $\mathrm{MSE}_\czin$ & \hspace{0mm} \bf{0.0631} & \hspace{0mm} 0.0952 \\
 
\end{tabular}

        \caption{Validation scores for the prediction model $\predictionModel$ calibrated by means of vine   (left) and  Archimedean (right) copulas, where the scores were computed on the data set $D_\czin\subset D$ composed of descriptor vectors for composite particles. 
        }
    \label{tab:scores_mix}
\end{table}
\fi

\section{Discussion}\label{sec.dis}
The power of the prediction model $\predictionModel$, which was introduced in Section~\ref{sec.pre.min}
for estimating the \composition{} of particles  by means of CT-based descriptor vectors,
 was evaluated in Section~\ref{sec.sec.res}, by applying the model to a particle system consisting of zinnwaldite, quartz, topaz and muscovite composites. Additionally,  the results obtained for  the goodness of fit and the predictive power of  $\predictionModel$  were compared to those achieved by using the (simpler) model of Archimedean copulas instead of using R-vine copulas.

The results shown in Fig.~\ref{fig:pred_mix} and Table~\ref{tab:scores_mix}   indicate that the use of Archimedean copulas leads to significantly larger prediction errors, when we investigate the performance of the  prediction model $\predictionModel$ for composite particles. In particular, the mean absolute error of 0.2153 when using Archimedean copulas reduces to an error of 0.1378 achieved by means of R-vine copulas. 
Furthermore, Fig.~\ref{fig:pred_all} and Table~\ref{tab:scores_all}   indicate that the prediction model $\predictionModel$  also estimates the \composition{} of all considered particles (i.e., not only composite particles but also particles which consist of either almost exclusively of valuable or non-valuable minerals) reasonably well when R-vine copulas are used for model fitting.

Recall that the results stated in Section~\ref{sec.sec.res} for the  power of the prediction model $\predictionModel$ were achieved via cross-validation. Thus, it is expected that prediction results would be similarly accurate for particles not intersecting with the 2D voxel slices where SEM-EDS data is available, i.e., for particles for which only CT-based descriptor vectors are known. In particular, the prediction model $\predictionModel$ can be used to estimate the \composition{} of all particles observed in the CT data, allowing for a quantitative mineralogical characterization  of the entire particle system reconstructed by CT imaging. 

The capability to predict the \composition{} of particles quantitatively is only one possible application of the multivariate probabilistic characterization of particle descriptor vectors. More precisely, the multivariate probability density $\widehat{f}$ 
considered in Eq.~(\ref{eq:multivaraiteCharacterizationDensity})
can be used as a basis for an in-depth analysis of the quality of separation processes. This will be discussed in detail in a
forthcoming paper.

Moreover, even though the prediction model $\predictionModel$ was calibrated to predict the mineralogical composition with respect to the \composition{} of particles based on six CT-based descriptors, it can be adapted to predict the prevalence of $K\geq2$ different minerals.
In such a scenario, the mineralogical composition of a particle can be described by a vector $M_\mathrm{rat}=(m_1,\dots,m_{K}) \in [0,1]^{K}$ with $\|M_\mathrm{rat}\|=1$, where  $\|M_\mathrm{rat}\|=\sum_{i=1}^{K} |m_i|$ and $m_i$ describes the fraction of the $i$-th  mineral within the particle for $i=1,\dots,K$. 
Note that similar to the prediction model given in (\ref{eq.dec.two}) for the case $K=2$, it suffices to consider the first $K-1$ entries of $M_\mathrm{rat}$, since $m_K=1-\sum_{k=1}^{K-1}m_k$.
Thus, for some $d\ge 2$,  the joint distribution of $d$  CT-based particle descriptors and the mineralogical volume fractions of the first $K-1$ minerals of composite particles can be modeled by some $(d+K-1)$-variate probability density $\widehat{f}^{\mathrm{c}}$. 
Note that such a model for the joint distribution of the mineralogical volume fractions has to ensure that the sum of the first $K-1$ minerals is smaller than 1, i.e., 
the support of $\widehat{f}_{d+1,\dots,d+K-1}^{\mathrm{c}}$ has to be bounded by the set $\{m=(m_1,\dots,m_{K-1}) \in \R^{K-1} \colon m_i \geq 0 \text{ for each } i=1,\dots,K-1, \|m\|\leq 1\}$.
Furthermore, similarly to the approach presented in Section~\ref{pre.min.sub} the distribution of the CT-based particle descriptors for particles almost purely composed of the $i$-th mineral can be modeled by $d$-variate probability densities $\widehat{f}^{(i)}$ for each  $i=1,\dots,K$. 
In  (\ref{eq.dec.two}) the prediction model for the case $K=2$ is presented with respect to the median value of the conditional probability density $\widehat{f}^{\mathrm{c}}_{d+1,\dots, d+K-1 \mid x}$. 
 However, instead of the median value other characteristic values of the distribution like the mean or mode can be considered. A possible generalization of the prediction model $\predictionModel \colon \R^d \to [0,1]^K$ for $K\geq 2$ different minerals which considers the mean value is given by
\begin{equation}
    \predictionModel(x)=
    \left\{
    \begin{array}{ll}
        e_i, & \text{if }  \frac{n_{i}}{n} \widehat{f}^{(i)}(x) \geq \frac{n_\mathrm{c}}{n} \widehat{f}_{1,\dots,d}^\mathrm{c}(x)
        \text{ and }  \frac{n_{i}}{n} \widehat{f}^{(i)}(x) >   \max\limits_{j\in \{1,\dots,K\}\setminus\{i\} }
      \frac{n_{j}}{n}\widehat{f}^{(j)}(x), \\  
     \left(\phi(x), 1-\|\phi(x)\|\right), & \text{otherwise},
    \end{array}
    \right.
\end{equation}
for each $x\in\R^d$, where $n_i$ is the number of particles mainly composed of the $i$-th mineral, $n_\mathrm{c}$ is the number of composite particles, $n=n_\mathrm{c}+\sum_{i=1}^K n_i$ is the total number of  particles and $e_i\in \{0,1\}^K$ is the unit vector for which the $i$-th entry is equal to $1$. Moreover,
$
\phi(x)=
        \bigl(\phi_1(x), \ldots, \phi_{K-1}(x)\bigr),
$
denotes the mean vector of the $(K-1)$-variate conditional probability density $\widehat{f}^{\mathrm{c}}_{d+1,\dots, d+K-1 \mid x}$ for each CT-based particle descriptor vector $x\in \R^d$ and $\|\phi(x)\|=\sum_{i=1}^{K-1} \phi_i(x)$.

\section{Conclusions}\label{sec:conclusions}
In this paper we describe an approach for deriving prediction models which can estimate the \composition{} of particles solely based on information gathered from CT image data. The methods were developed using CT image data of a particle system consisting of zinnwaldite, quartz, topaz and muscovite composites---but, they can be adapted to further particle systems.

In a first step, CT image data was segmented using an adapted 3D U-net architecture. For the training of the 3D U-net, we calibrated a loss function which focuses on the separation of touching particles for improved segmentation results. Moreover, the loss function can handle partially labeled data 
in order to reduce the amount of hand-labeled data necessary to train the network. More precisely, for training just sparsely annotated slices are required to achieve a 3D particle-wise segmentation.
From the segmented CT data individual particles were extracted such that vectors of descriptors describing the size, shape and texture of particles can be computed (CT-based descriptor vectors). 
Additionaly, for some particles SEM-EDS information was available, such that the CT-based descriptor vectors could be extended with a descriptor of their \composition{}. Then, in the stochastic modeling step, we utilized R-vine and Archimedean copulas to fit multivariate probability densities to the computed descriptor vectors,  achieving an in-depth probabilistic characterization of the particle system.

Moreover, the building blocks  $\widehat{f}^\nzin,\widehat{f}^\zin,\widehat{f}^\czin$ of the seven-variate probability density $\widehat{f}$ introduced in Eq.~\ref{eq:multivaraiteCharacterizationDensity}
 provide, as a  ``by\textcolor{blue}{-}product", conditional probability densities which can be used to construct a  prediction model $\predictionModel$ given in Eq.~(\ref{eq.dec.two}). This model can estimate the \composition{} of a particle quantitatively by means of  CT-based descriptor vectors. Furthermore, it turned out that the prediction model $\predictionModel$ performs significantly better when utilizing R-vine copulas instead of Archimedean copulas.

\section*{Acknowledgements}

The financial support provided by the German Research Foundation (DFG) for  the research projects PE1160/22-2, SCHM997/27-2  and
SCHM997/45-1 within the priority programs SPP 2045 ``Highly specific and multidimensional fractionation of fine particle systems with technical relevance'' and SPP 2315 ``Engineered artificial minerals (EnAM) - A geo-metallurgical tool to recycle critical elements from waste streams''  is gratefully
acknowledged.

\bibliography{Bibliography}{}

\if\submission1
\section{Figure legends}
\begin{table}[ht]
        \centering

\begin{tabular}{p{0.24\textwidth}|p{0.24\textwidth}|p{0.3\textwidth}}

  & using  R-vine copulas  & using Archimedean copulas \\
\hline 
 log-likelihood $\mathbb{L}$& \textcolor[rgb]{0,0,0}{\hspace{1.2mm}\bf{3085.13}} & \hspace{0.8mm} 2295.62 \\
\hline 
 AIC & \bf{-6046.26} & -4583.24 \\
\hline 
 BIC & \bf{-5732.03} & -4562.97 \\
\hline 
 misclassification rate & \hspace{1.2mm} \bf{0.0477} & \hspace{0.8mm} 0.0579 \\
 
\end{tabular}

     \caption{Validation scores of the prediction model $g$ determined by means of R-vine  (left) and Archimedean copulas (right).}
    \label{tab:scores_approach1}

\end{table}

 \begin{table}[ht]
        \centering

\begin{tabular}{p{0.24\textwidth}|p{0.24\textwidth}|p{0.3\textwidth}}

  & using R-vine copulas & using Archimedean copulas \\
\hline 
 log-likelihood $\mathbb{L}$ & \textcolor[rgb]{0,0,0}{\hspace{1.2mm}\bf{4963.41}} & \hspace{0.8mm} 4034.87 \\
\hline 
 AIC & \bf{-9526.82} & -7867.74 \\
\hline 
 BIC & \bf{-8486.58} & -7342.42 \\
 \hline 
  MAE & \hspace{1.05mm}\bf{0.0990} & \hspace{1.05mm}0.1304 \\
\hline 
  MSE & \hspace{0mm} \bf{0.0622} & \hspace{0mm} 0.0707\\
 
\end{tabular}

        \caption{
        Validation scores of the prediction model $\predictionModel$ calibrated by means of vine   (left) and  Archimedean (right) copulas, where the scores were computed on the data set $D$ of particle descriptor vectors.        }
    \label{tab:scores_all}
\end{table}

 \begin{table}[ht]
        \centering
     
\begin{tabular}{p{0.24\textwidth}|p{0.24\textwidth}|p{0.3\textwidth}}

  & using R-vine copulas & using Archimedean copulas \\
\hline 
 log-likelihood  $\mathbb{L}_\czin$& \textcolor[rgb]{0,0,0}{\hspace{1.2mm}\bf{1315.06}} & 725.51 \hspace{0.8mm}  \\
\hline 
 $\mathrm{AIC}_{\czin}$ & \bf{-2454.12
} & -1377.02
 \\
\hline 
 $\mathrm{BIC}_{\czin}$ & \bf{-2093.88} & -1225.55 \\
 \hline 
   $\mathrm{MAE}_\czin$ & \hspace{1.05mm}\bf{0.1378} & \hspace{1.05mm}0.2153 \\
\hline 
 $\mathrm{MSE}_\czin$ & \hspace{0mm} \bf{0.0631} & \hspace{0mm} 0.0952 \\
 
\end{tabular}

        \caption{Validation scores for the prediction model $\predictionModel$ calibrated by means of vine   (left) and  Archimedean (right) copulas, where the scores were computed on the data set $D_\czin\subset D$ composed of descriptor vectors for composite particles. 
        }
    \label{tab:scores_mix}
\end{table}

\newpage

\newpage

\begin{figure}[ht]
\centering

\includegraphics[width=0.95\textwidth]{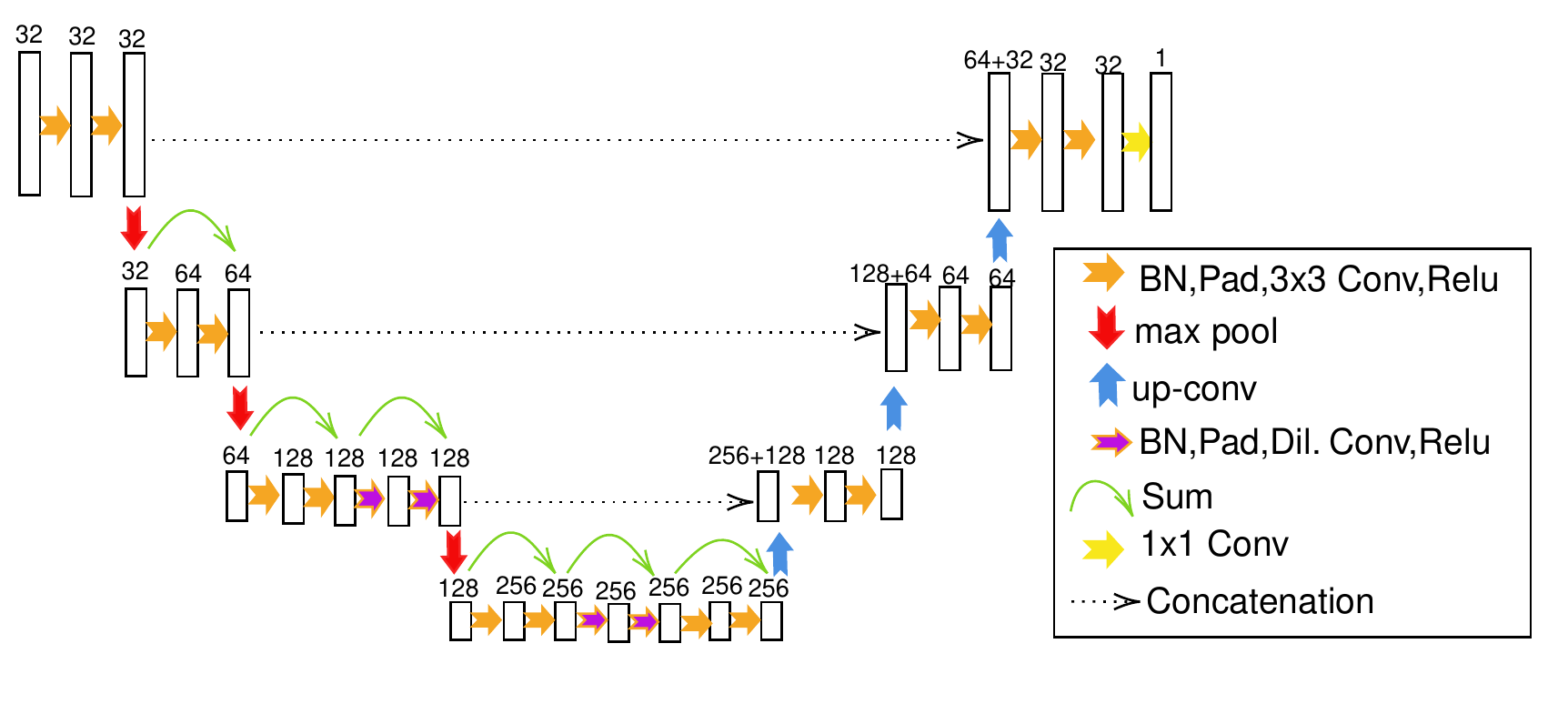}

\caption{U-net architecture used for the initial segmentation task. Each $3\times3\times3$ convolutional layer ($3\times3$ Conv) and each dilated convolutional layer (Dil. Conv)  is preceded by a batch normalization layer (BN) and a symmetric padding layer (Pad) of suitable width. }
\label{Fig:u-net}      
\end{figure}

\newpage

  \begin{figure}[ht]
   
  \includegraphics[width=1\linewidth]{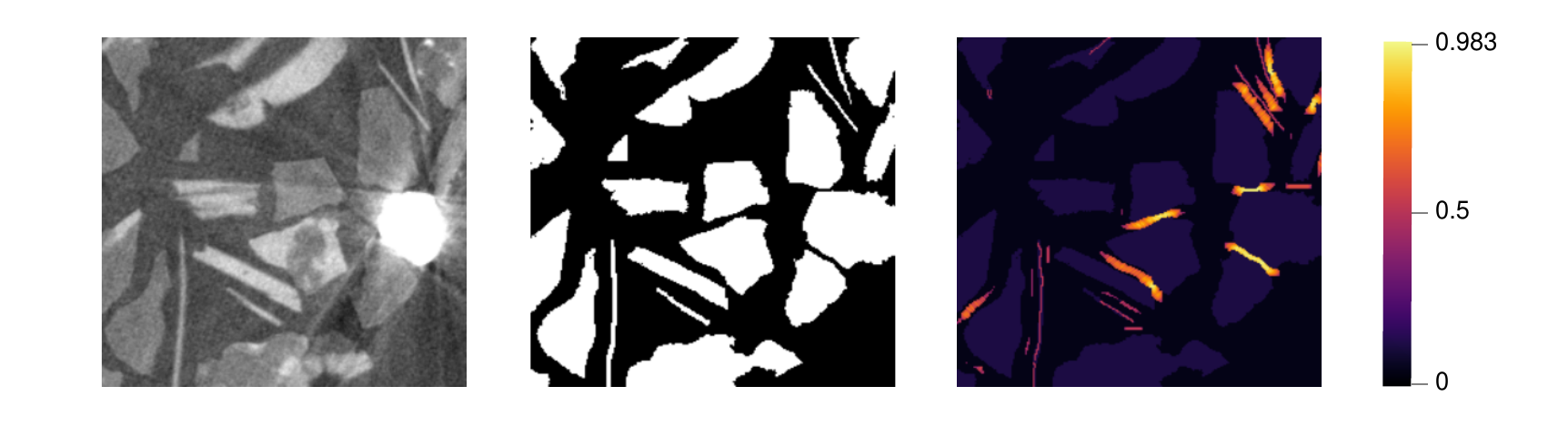}

  \caption{ (a) 2D cutout of raw CT data showing  a highly reflective particle, (b)  the corresponding ground truth labels  of particles (white) and background (black), and (c) the corresponding weight map, where brighter areas indicate higher weights.
Note that the weights of voxels corresponding to platelike particles are increased to avoid oversegmention of such particles.
 \label{Fig:2D_gt}
}
 \end{figure}

\begin{figure}[ht]

\begin{subfigure}{.7\textwidth}

  \centering
  \includegraphics[width=0.7\linewidth]{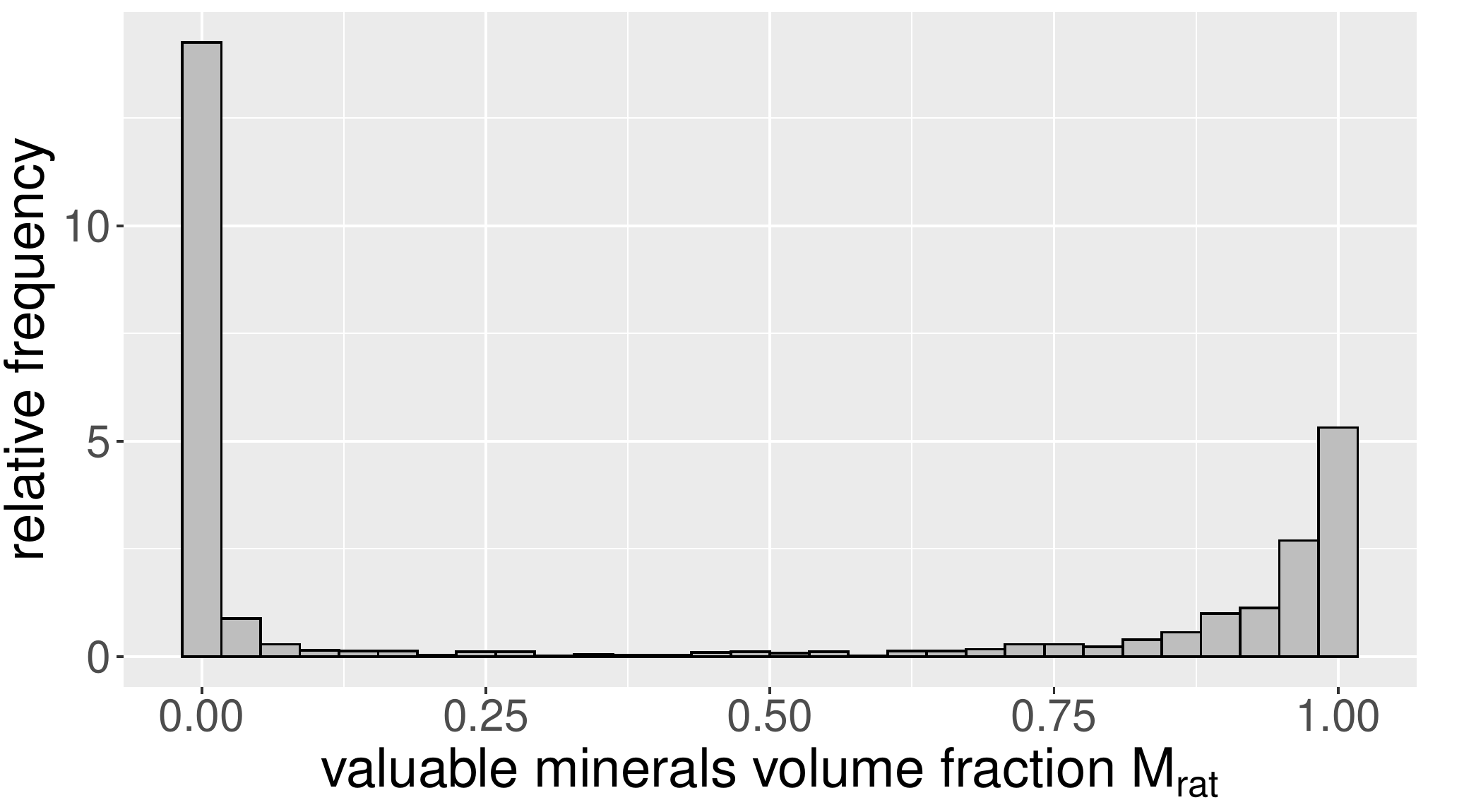}
\end{subfigure}

\caption{Histogram of the VFVM of particles intersecting with a SEM-EDS slice.  }
\label{Fig:fractions}
\end{figure}

\newpage

\begin{figure}[ht]

\tikzset{every picture/.style={line width=0.75pt}} 
\begin{tikzpicture}[x=0.75pt,y=0.75pt,yscale=-1,xscale=1]

\draw (161.52,91.83) node  {\includegraphics[width=231.46pt,height=115.25pt]{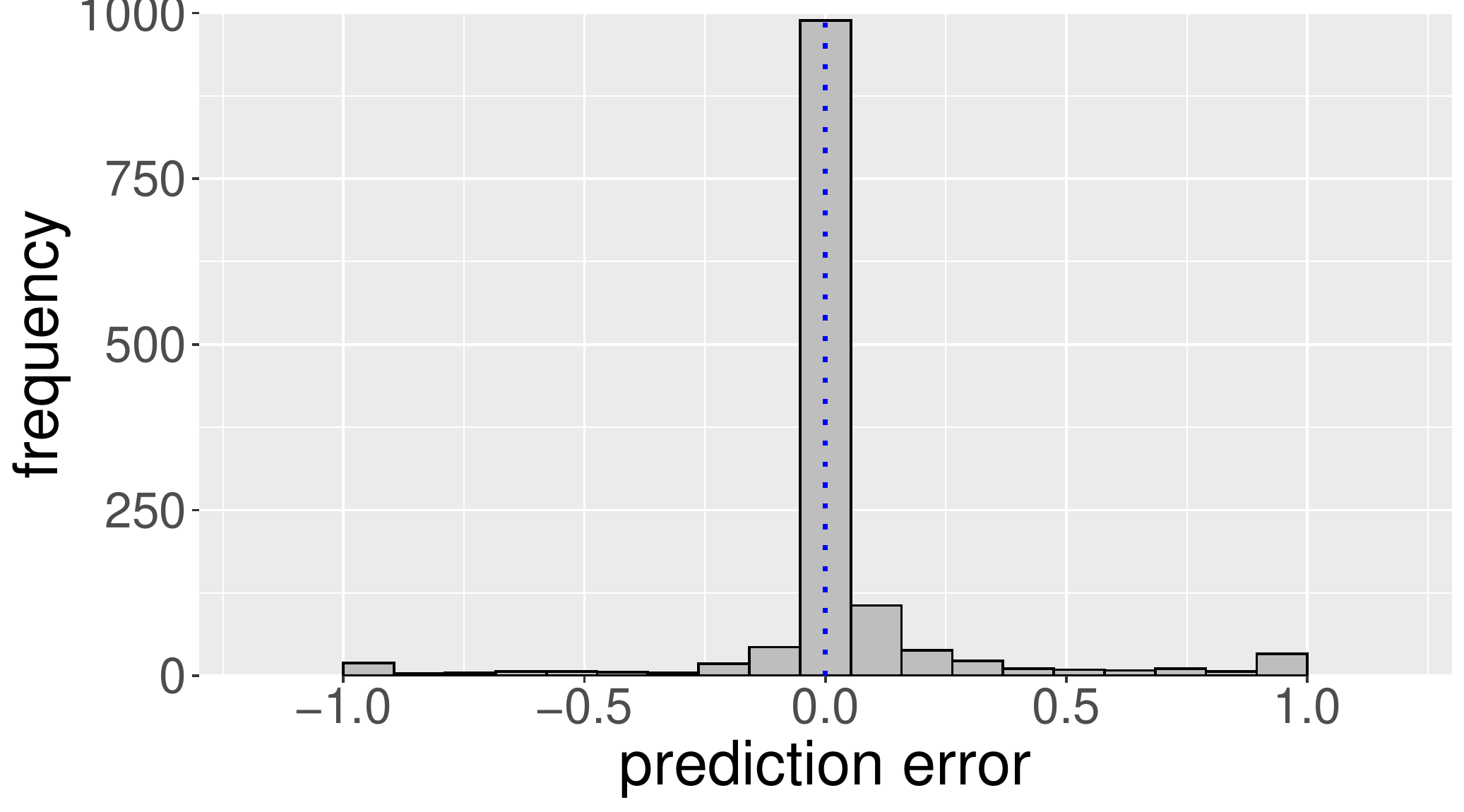}};

\draw (460.7,91.5) node  {\includegraphics[width=231.46pt,height=115.25pt]{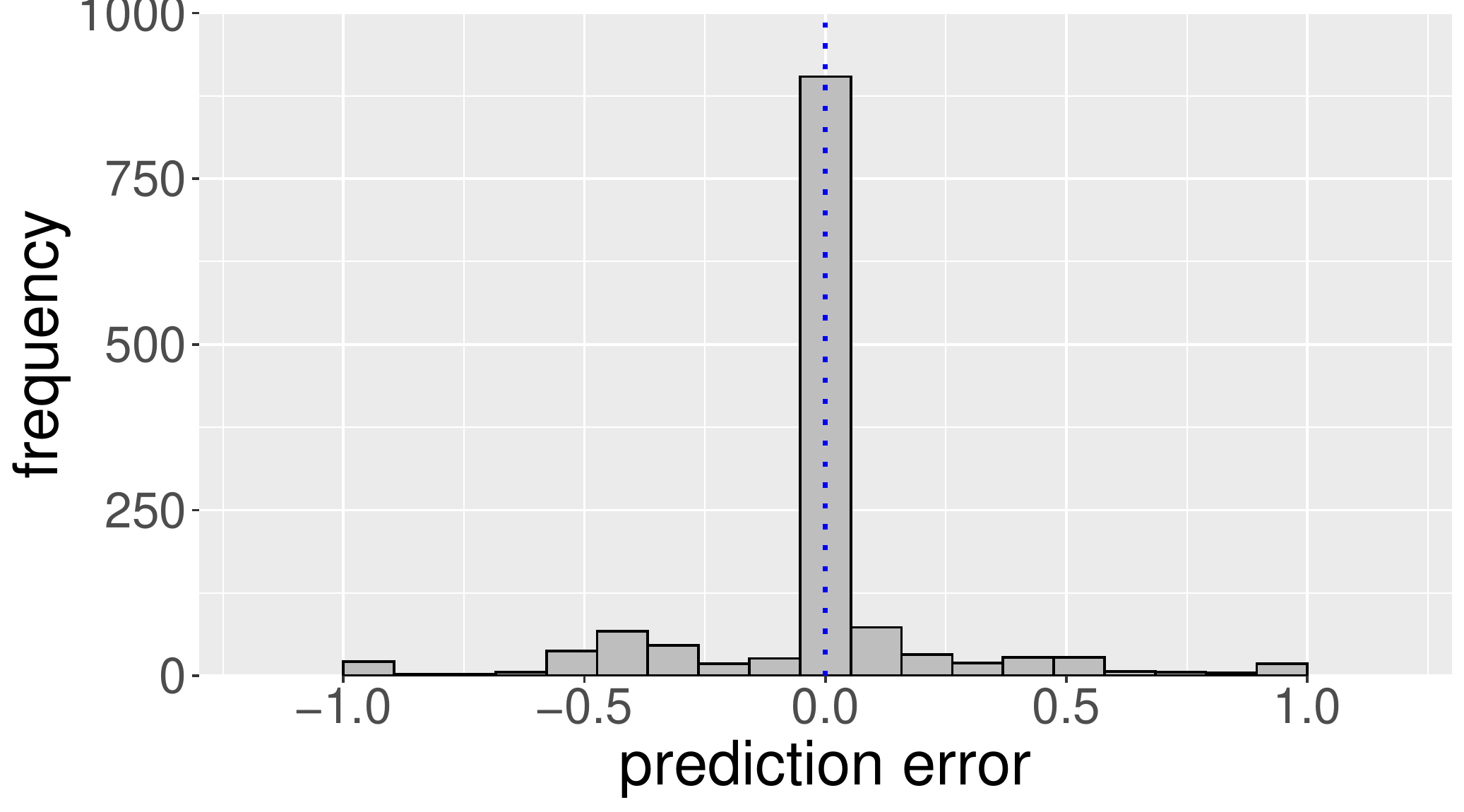}};

\end{tikzpicture}

 \caption{
 Histograms of prediction errors
  $g_{D_\czin\setminus\{x\}}(x_{1,\dots,6})-x_7$  resulting from the use of R-vine  (left) and Archimedean  (right) copulas, computed for each particle descriptor vector $x=(x_{1,\dots,6},x_7)\in D$.
 }
    \label{fig:pred_all}

\end{figure}

\tikzset{every picture/.style={line width=0.75pt}} 
\begin{figure}[ht]

\begin{tikzpicture}[x=0.75pt,y=0.75pt,yscale=-1,xscale=1]

\draw (161.52,91.83) node  {\includegraphics[width=231.46pt,height=115.25pt]{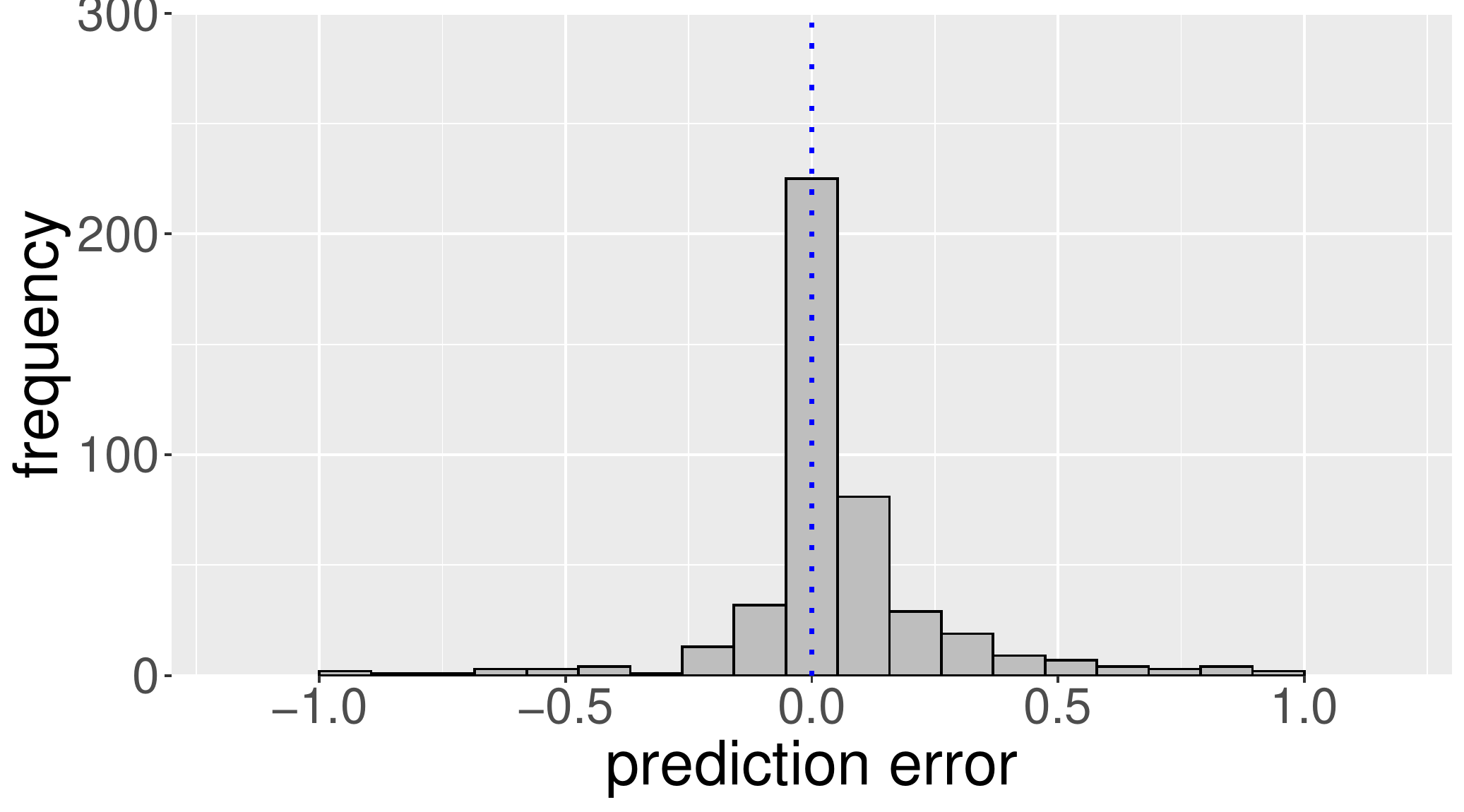}};

\draw (460.7,91.5) node  {\includegraphics[width=231.46pt,height=115.25pt]{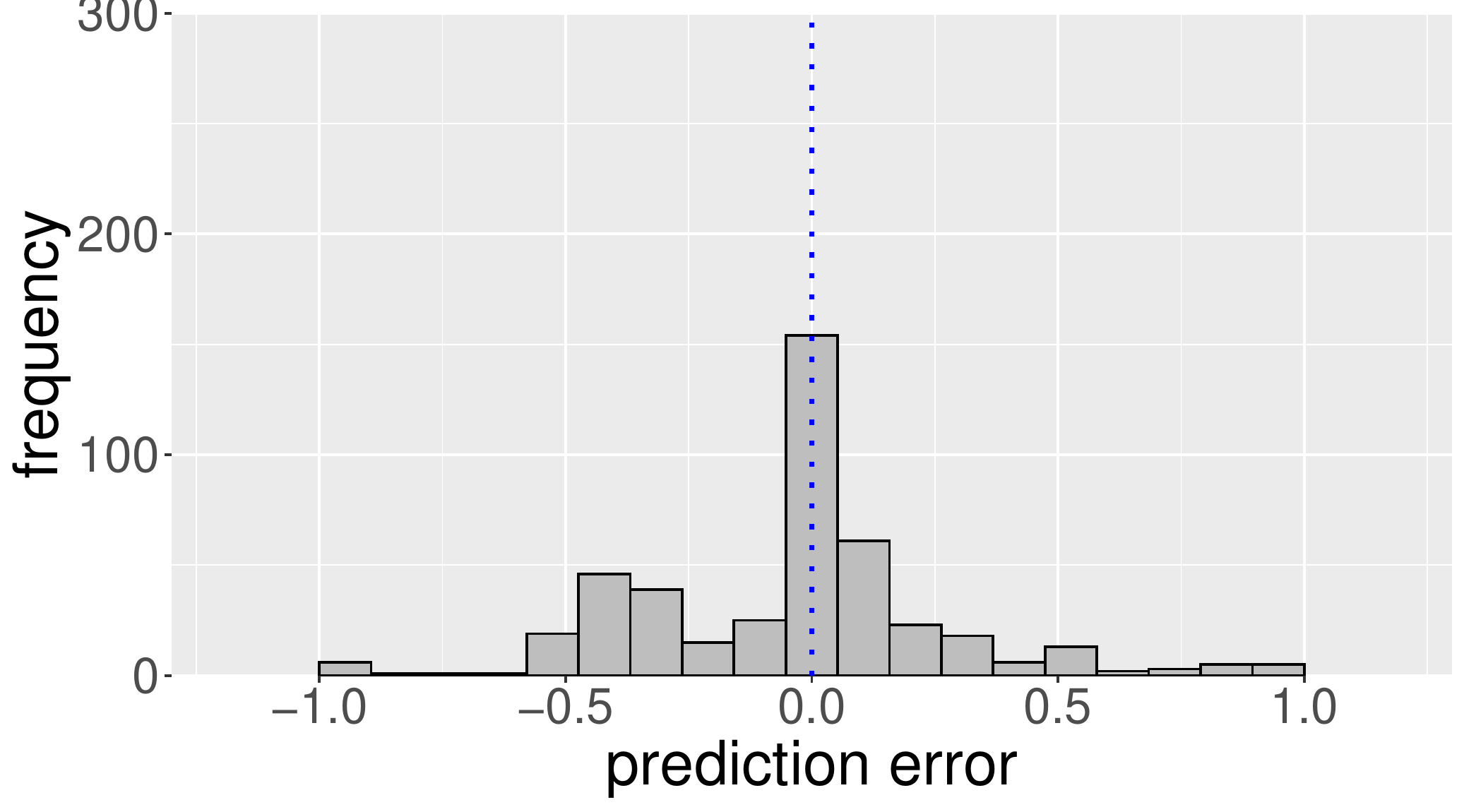}};
\end{tikzpicture}

 \caption{Histograms of prediction errors
  $\predictionModel_{D\setminus\{x\}}(x_{1,\dots,6})-x_7$  resulting from the use of R-vine  (left) and Archimedean  (right) copulas, computed for each particle descriptor vector $x=(x_{1,\dots,6},x_7)\in D_{\czin}$.   }
    \label{fig:pred_mix}

\end{figure}

\fi

\end{document}